\documentclass[article,english,12pt]{article}

\usepackage{amscd,amsfonts,amsmath,amssymb,amsthm,bbm,bm,latexsym,mathrsfs}
\usepackage{epsfig,graphics,graphicx,natbib,subfigure,longtable}
\usepackage[english]{babel}
\usepackage[usenames]{color}
\usepackage{bookmark}
\usepackage{sgame}
\usepackage{gensymb}
\usepackage{epstopdf}
\usepackage[top=1.7in, bottom=1.7in, left=0.9in, right=0.9in]{geometry}

\usepackage{natbib}
\makeatother

\theoremstyle{remark}

\theoremstyle{plain}

\def\simind{\stackrel{\mbox{\scriptsize{ind}}}{\sim}}
\def\simiid{\stackrel{\mbox{\scriptsize{iid}}}{\sim}}

\begin{document}

\title{\vspace{-50pt}Bayesian Nonparametric Conditional Copula Estimation of Twin Data\thanks{Email Addresses for correspondences:  \href{mailto:fabrizio.leisen@gmail.com}{fabrizio.leisen@gmail.com} (Fabrizio Leisen); \href{mailto:luca.rossini@unive.it}{luca.rossini@unive.it} (Luca Rossini);\href{mailto:luciana.dallavalle@plymouth.ac.uk}{luciana.dallavalle@plymouth.ac.uk} (Luciana Dalla Valle).}}

\author{
\hspace{-20pt} Luciana Dalla Valle \textsuperscript{1} \hspace{15pt}
Fabrizio Leisen\textsuperscript{2} \hspace{15pt} Luca Rossini\textsuperscript{3}
        \\
        \vspace{5pt}
        \\
        {\centering {\small \textsuperscript{1}
        University of Plymouth, UK \hspace{18pt}
        \textsuperscript{2}
            University of Kent, UK }} \vspace{5pt} \\
          {\centering {\small  \textsuperscript{3}
             Ca' Foscari University of Venice, Italy
             }}}

\date{\today}
\maketitle



\begin{abstract}
Several studies on heritability in twins aim at understanding the different contribution of environmental and genetic factors to specific traits.
Considering the National Merit Twin Study, our purpose is to correctly analyse the influence of the socioeconomic status on the relationship between twins' cognitive abilities.
Our methodology is based on conditional copulas, which allow us to model the effect of a covariate driving the strength of dependence between the main variables.
We propose a flexible Bayesian nonparametric approach for the estimation of conditional copulas, which can model any conditional copula density. Our methodology extends the work of \cite{WuWaWa15} by introducing dependence from a covariate in an infinite mixture model.
Our results suggest that environmental factors are more influential in families with
lower socio-economic position. \\


\end{abstract}

%


\maketitle


%


\section{Introduction}
\label{Intro}

The literature on heritability of traits in children often focusses on twins, due to the shared environmental factors and the association of genetical characteristics.
Among studies on the heritability of diseases, \cite{WaGuHeZh11} applied an efficient estimation method to mixed-effect models to analyze disease inheritance in twins.

One of the main purposes of studies on heritability is to estimate the different contribution of genetic and environmental factors to traits or outcomes (see, for example, the latent class twin method of \cite{Baker16}).
\cite{BaLeWe13} studied the interactions between environmental and genetic effects to intelligence in twins, showing that higher socioeconomic status is associated with higher intelligence scores.
Bioecological theory states that environmental factors may significantly influence the heritability of certain characteristics, such as cognitive ability, which is the readiness for future intellectual or educational pursuits.
Several studies have found that cognitive ability is more pronounced and evident among children raised in higher socioeconomic status families. 
Such families can offer greater opportunities to children, due to their socioeconomic wealth status, and represent stimulating environments where children's 
inherited capabilities may become more manifest.

\begin{figure}[h!]
 \centering
     \subfigure[\scriptsize{Mother's education}]
   {\includegraphics[width=6cm]{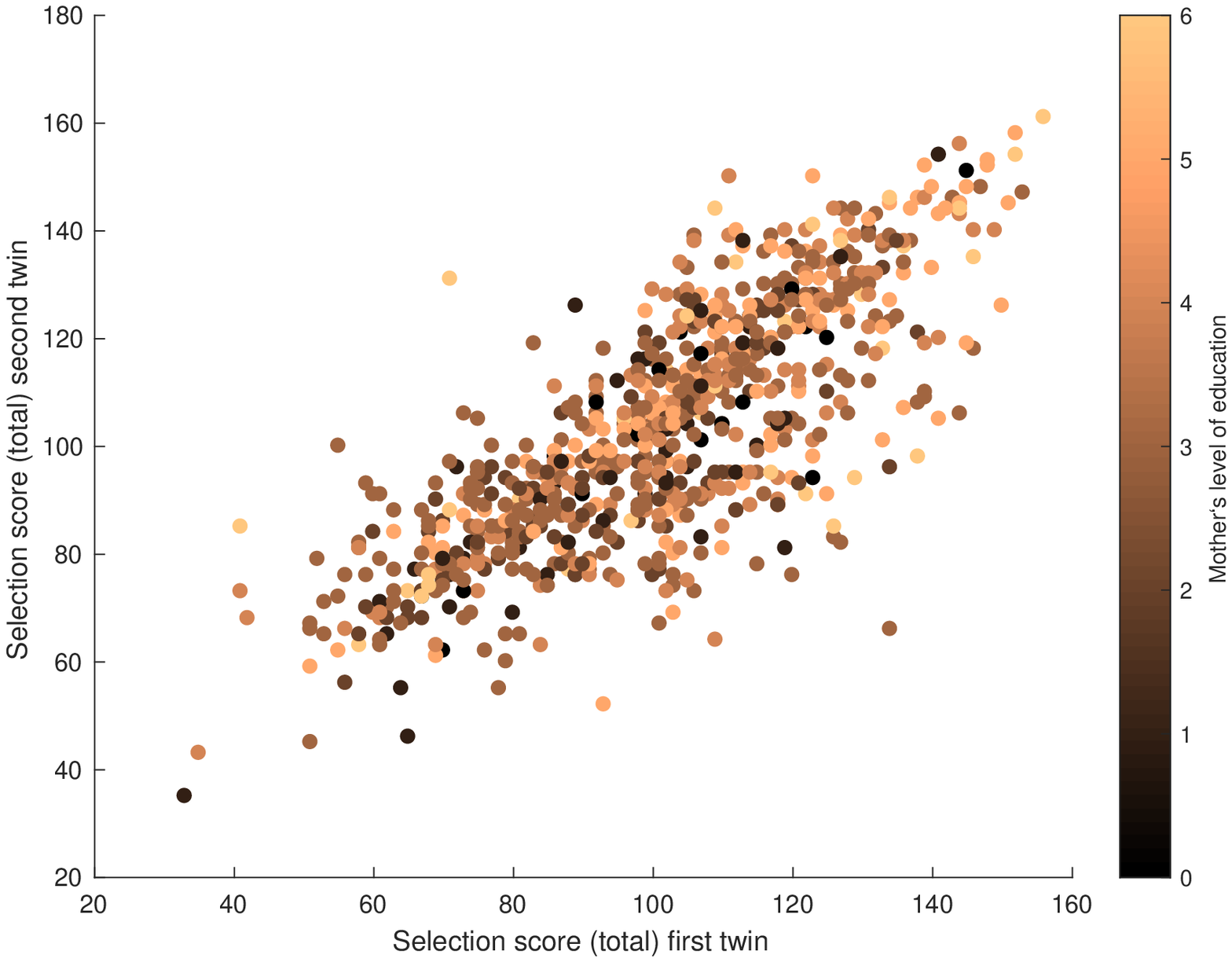}}
    \subfigure[\scriptsize{Father's education}]
   {\includegraphics[width=6cm]{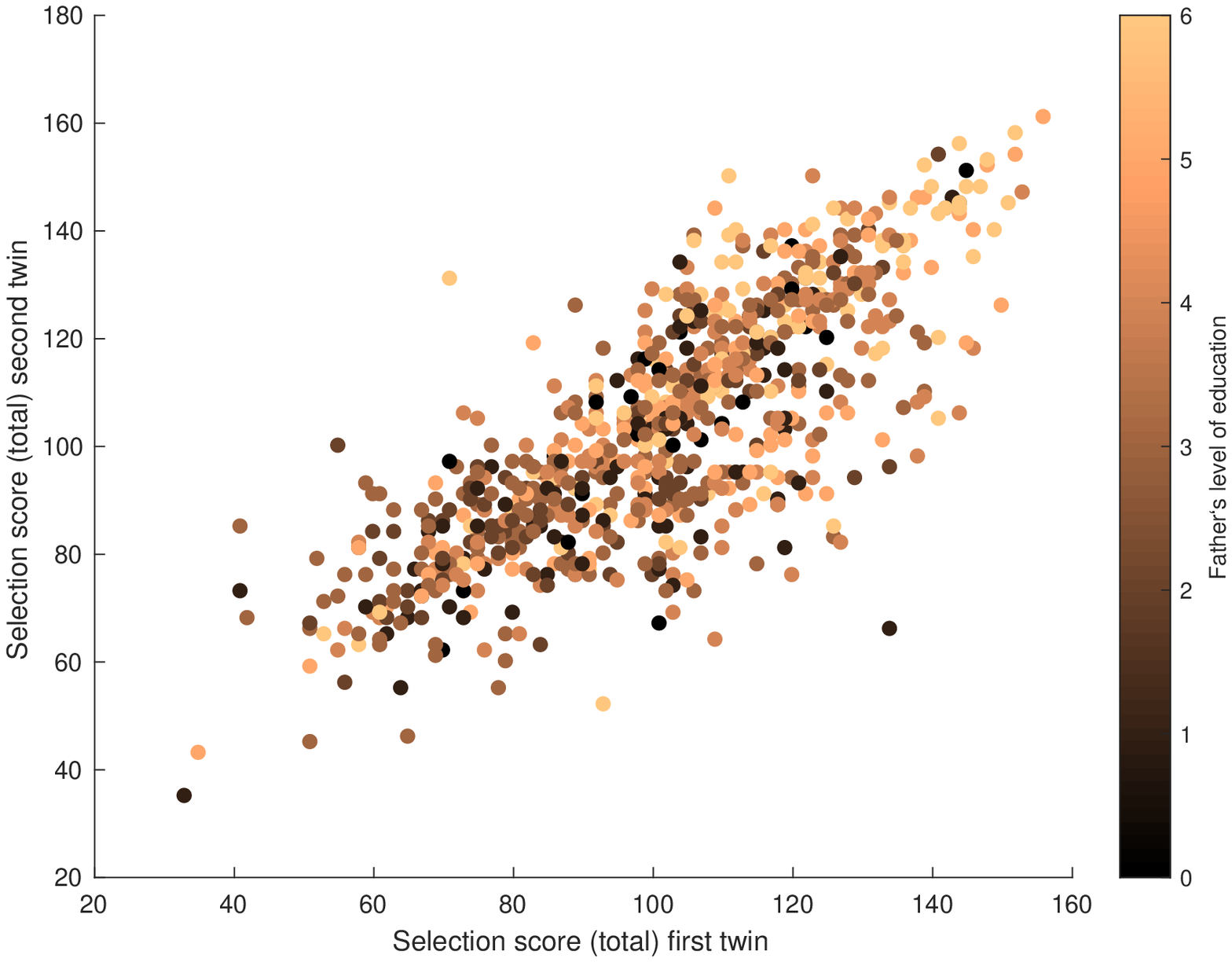}}
    \subfigure[\scriptsize{Family income}]
   {\includegraphics[width=6cm]{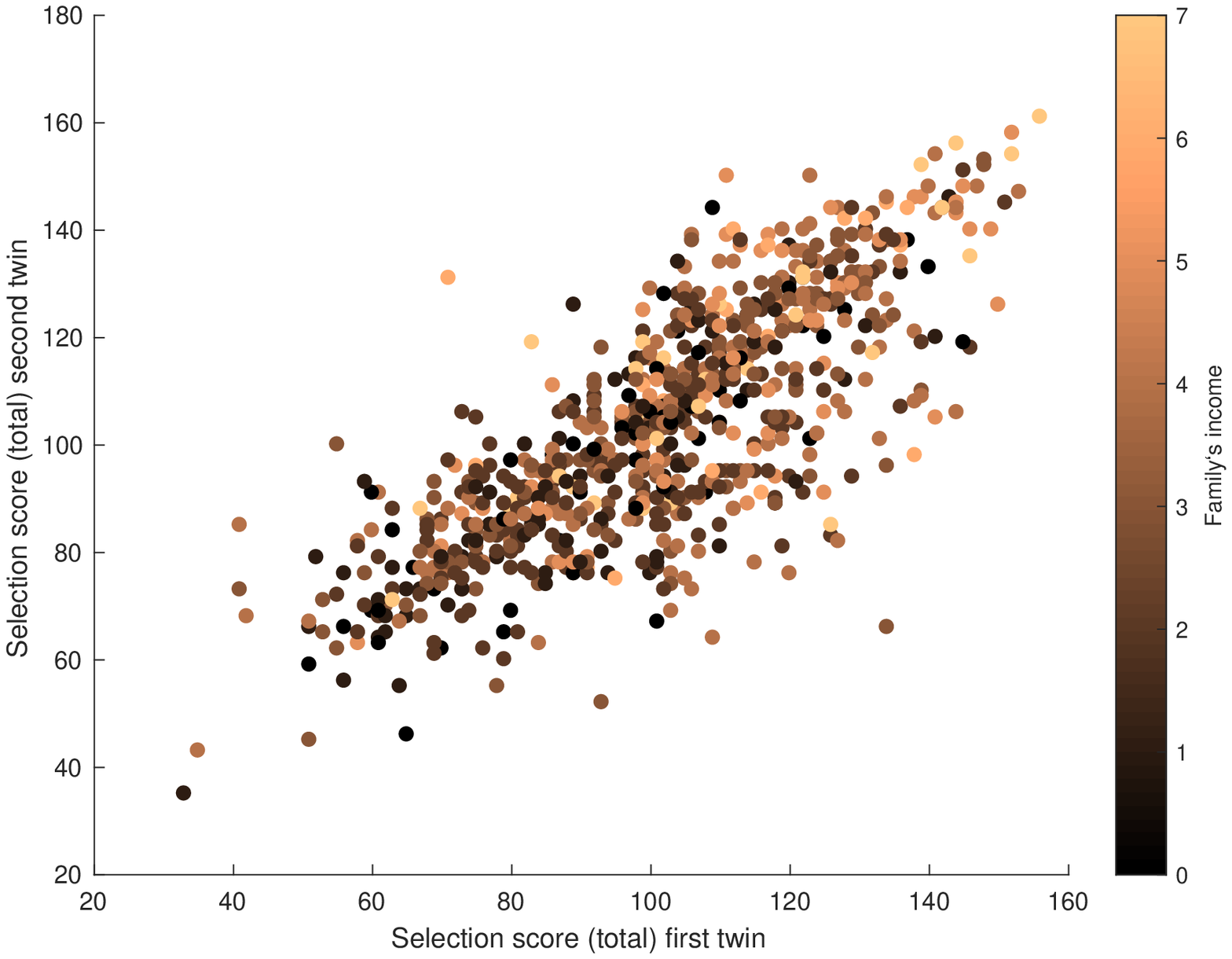}}
\caption{\small{Scatterplots of the twins overall scores with respect to the mother's (panel (a)) and father's level of education (panel (b)) and family income (panel (c)).}}
\label{DATA}
\end{figure}

The aim of this paper is to correctly analyse the effect of socioeconomic factors on the relationship between twins' cognitive abilities.
From a sample of $839$ US adolescent twin pairs who completed the National Merit Scholarship Qualifying Test, we consider each twin's overall school performance (measured by a total score including English, Mathematics, Social Science, Natural Science and Word Usage), the mother's and father's education level and the family income.
The data are plotted in Figure \ref{DATA}, which shows the scatterplots of the twins' school performances, on each axis, against the socioeconomic variables, whose values are in different colours (dark brown denotes low values, while light brown denotes high values).
Figure \ref{DATA} indicates that the twins' school performances are positively correlated and their dependence is influenced by the values of the socioeconomic variables (the mother's (panel (a)), the father's level of education (panel (b)) and the family income (panel (c))).
Indeed, most of the light brown dots (denoting high values of the covariates) are grouped in the upper right corner, while the dark brown dots (denoting low values of the covariates) lie in the bottom left corner of each plot. 
Hence, the higher the parents' education or family income, the higher the twins' school performance.
This means that the twins' performance scores are functions of each covariate and they vary according to the values of the covariates.

\begin{figure}[h!]
 \centering
  \includegraphics[width=4cm]{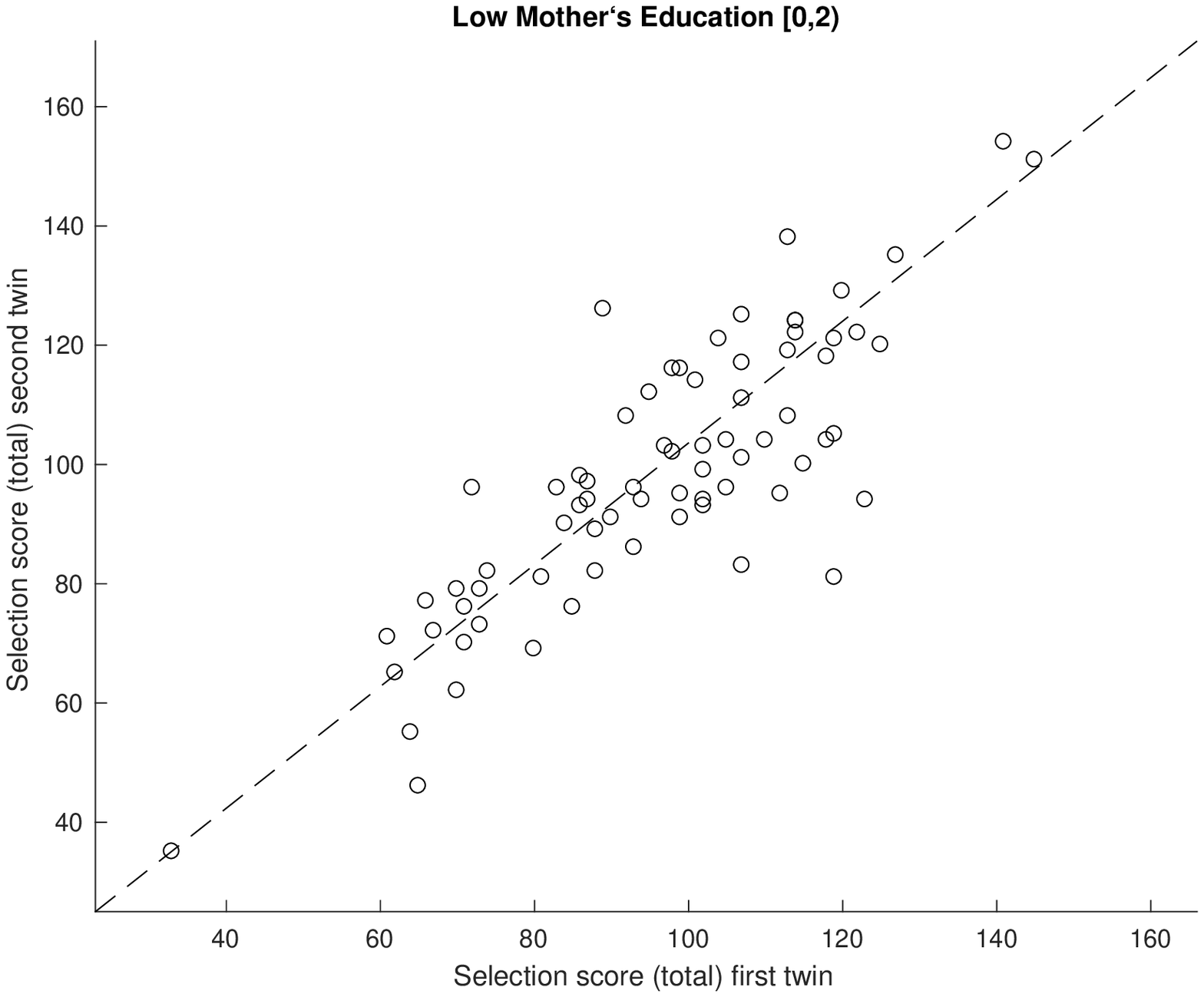} 
    \includegraphics[width=4cm]{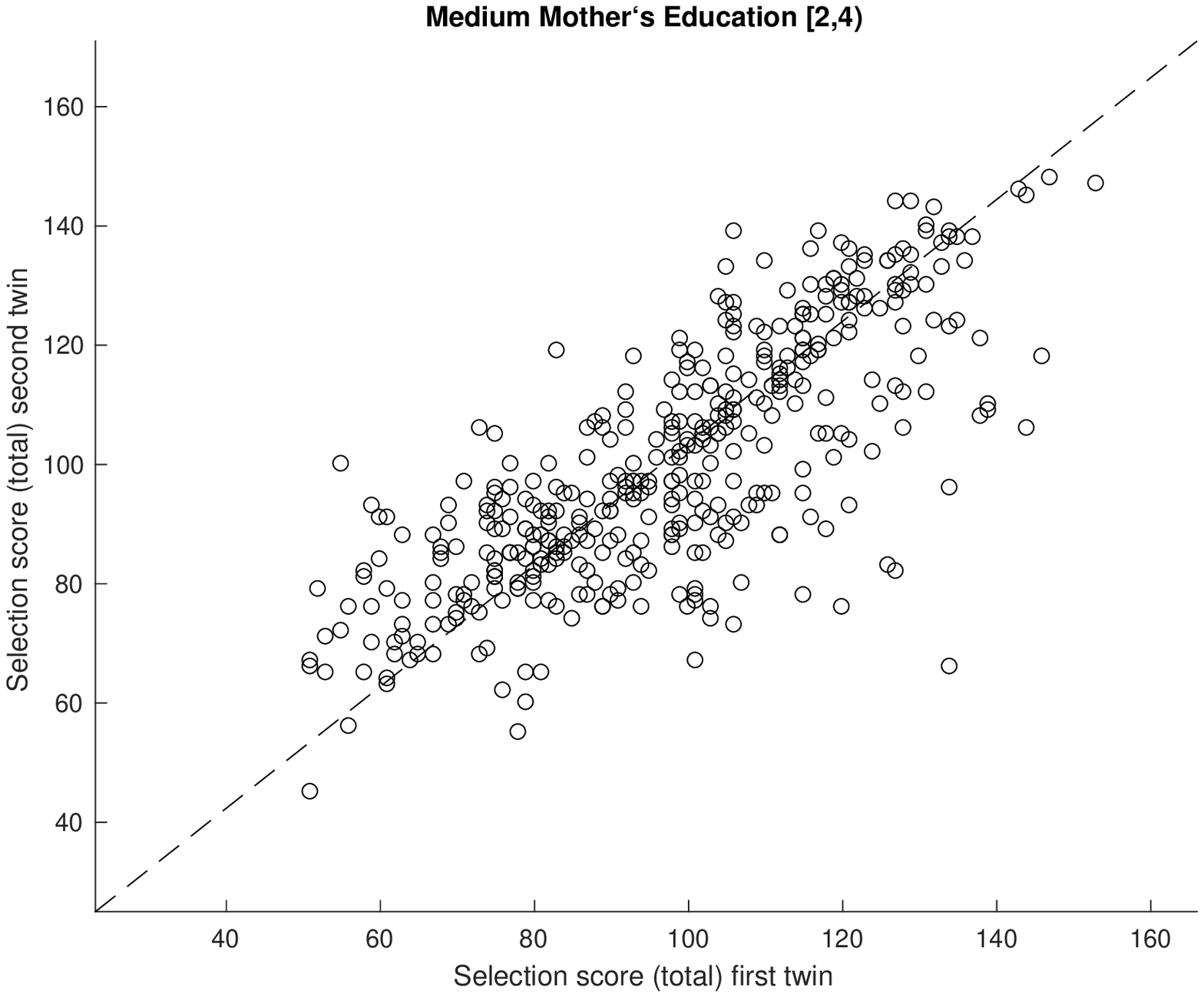} 
   \includegraphics[width=4cm]{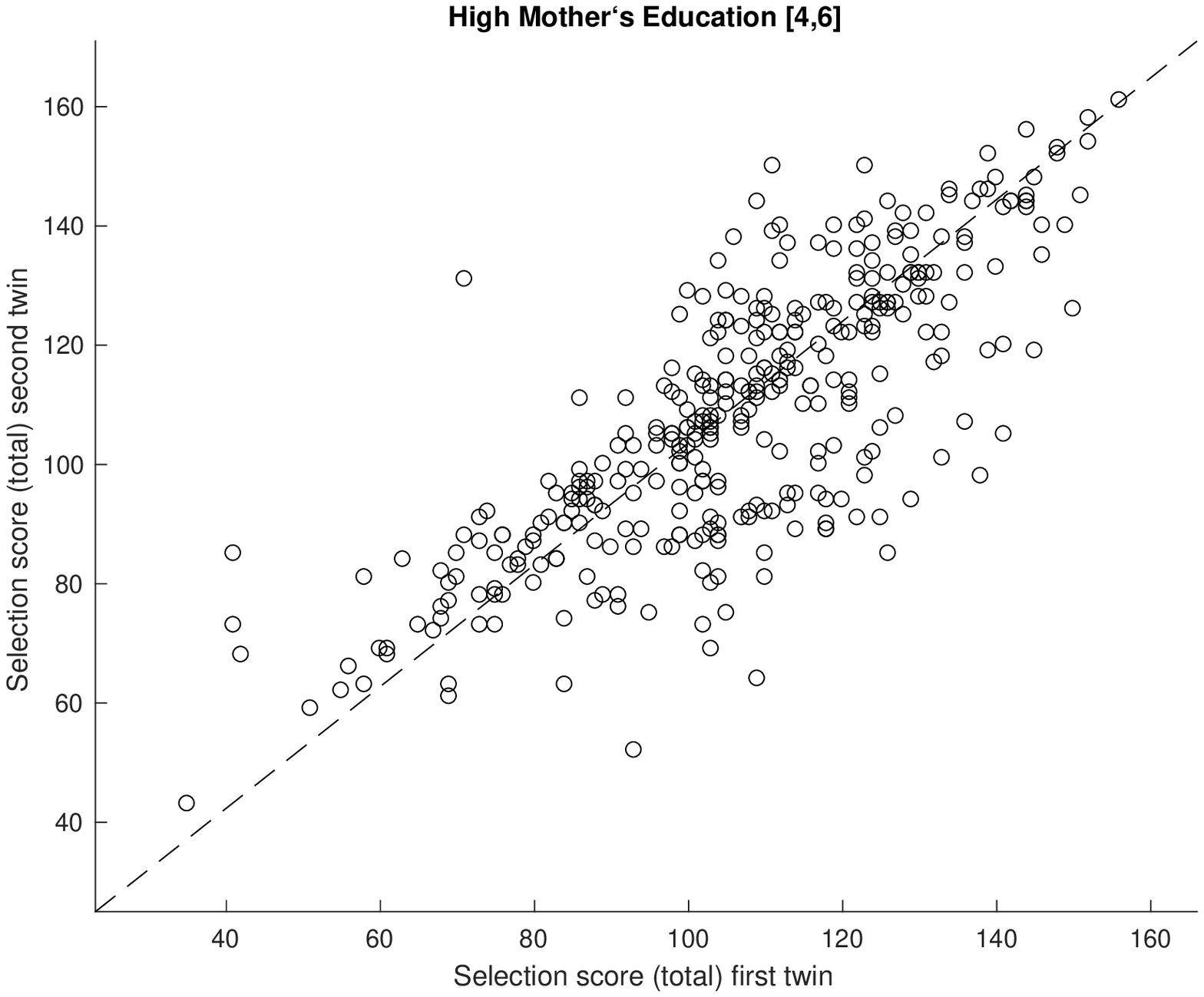}
   \\
    \includegraphics[width=4cm]{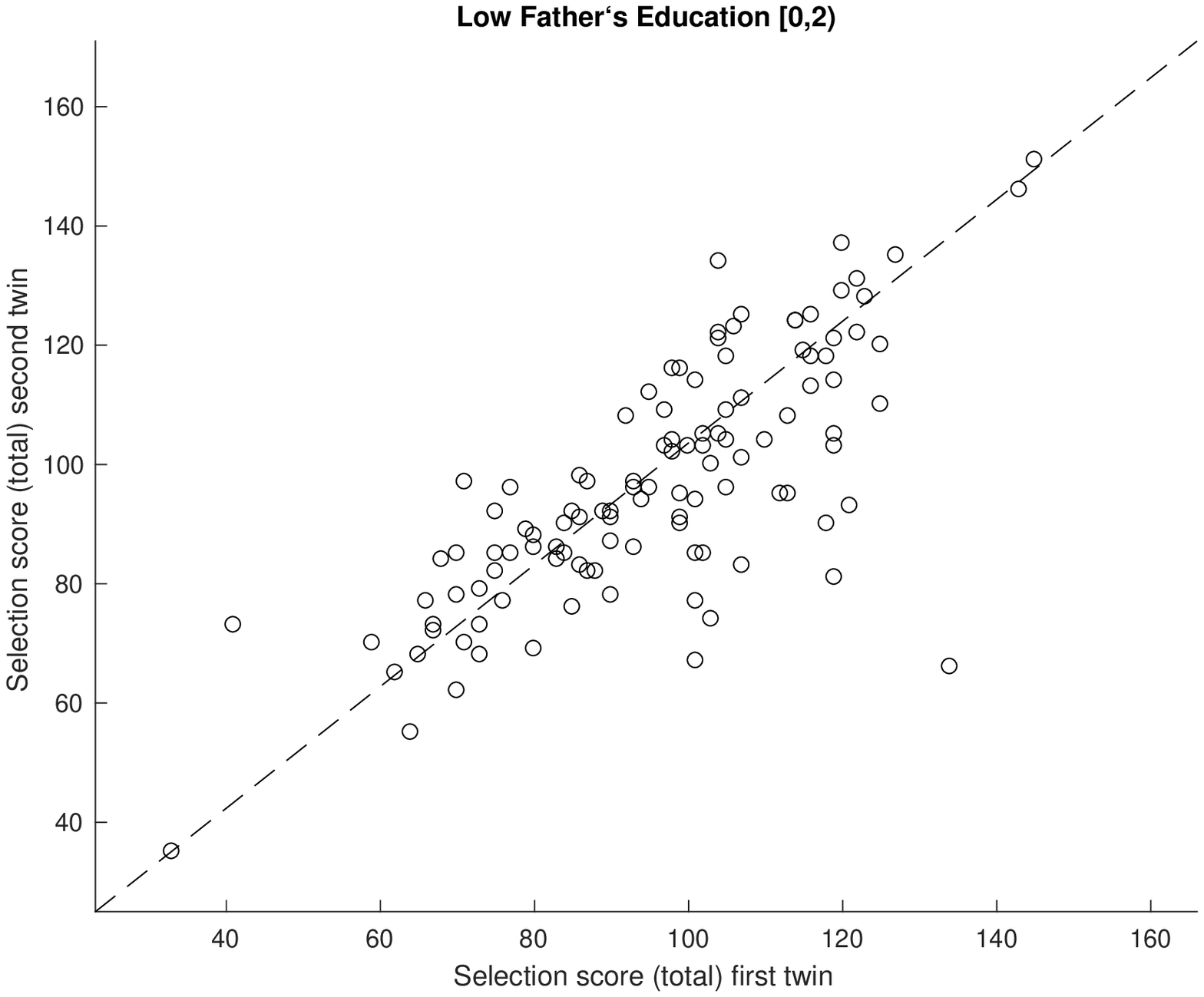} 
    \includegraphics[width=4cm]{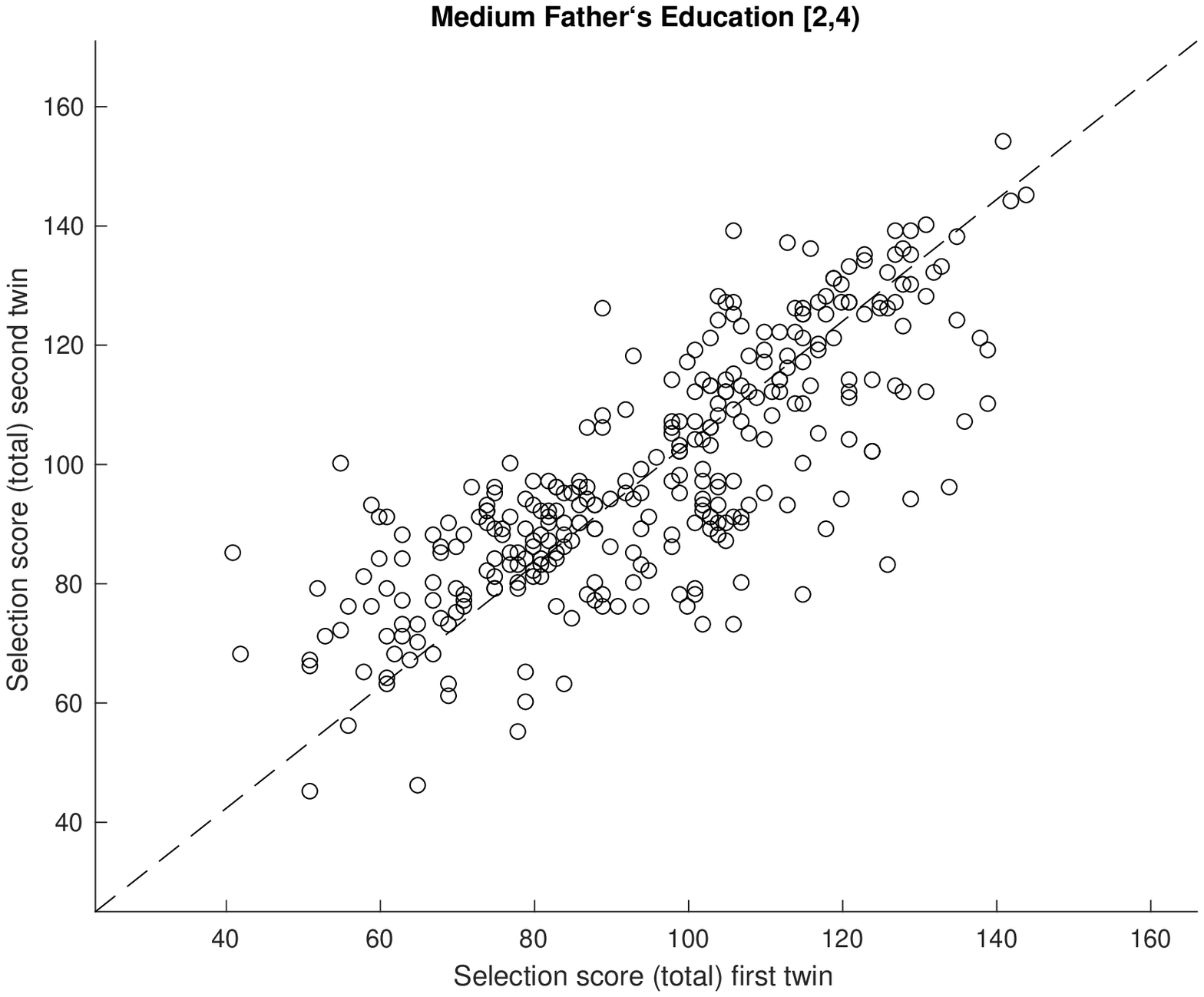}
     \includegraphics[width=4cm]{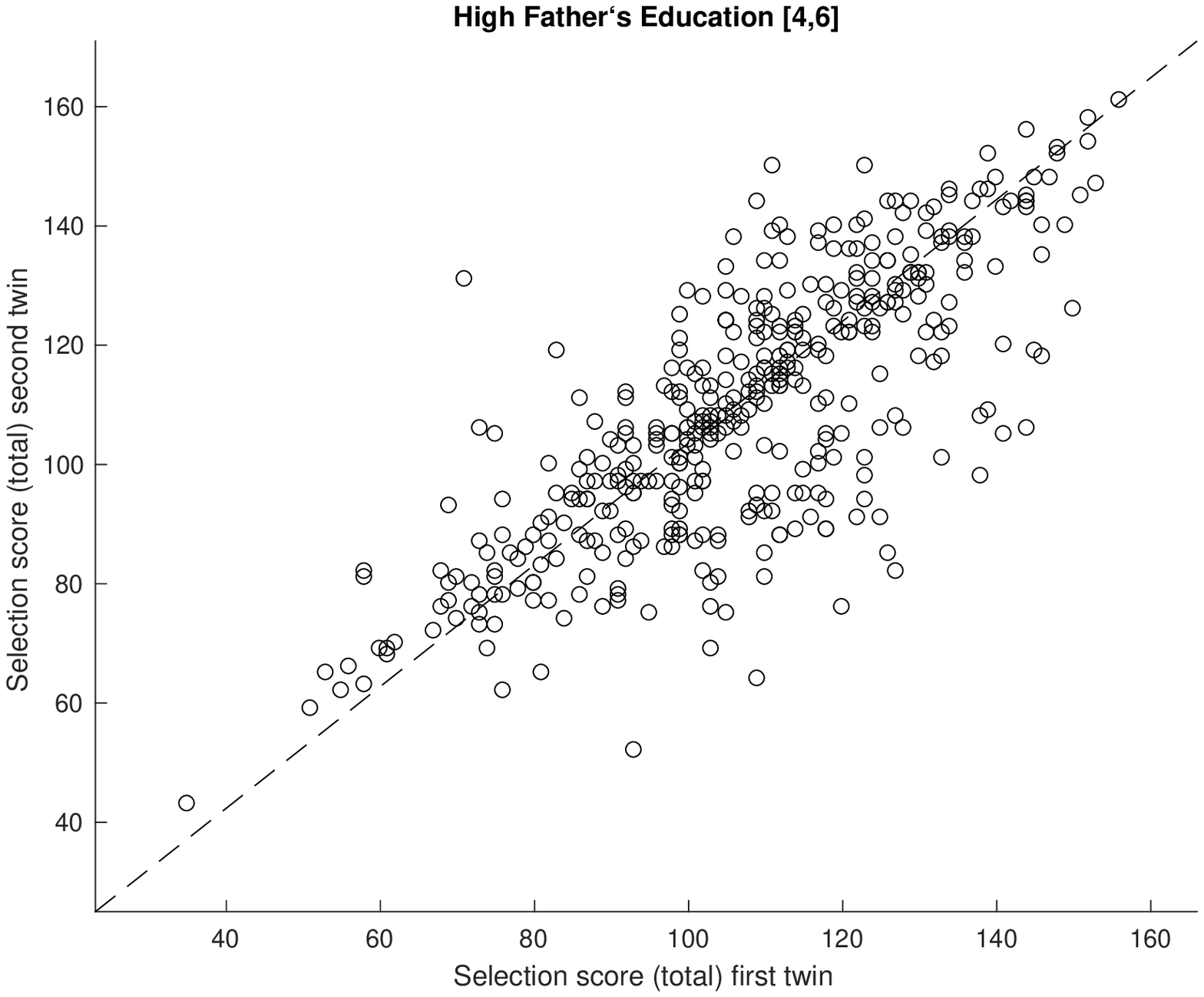}
 \\ 
   \includegraphics[width=4cm]{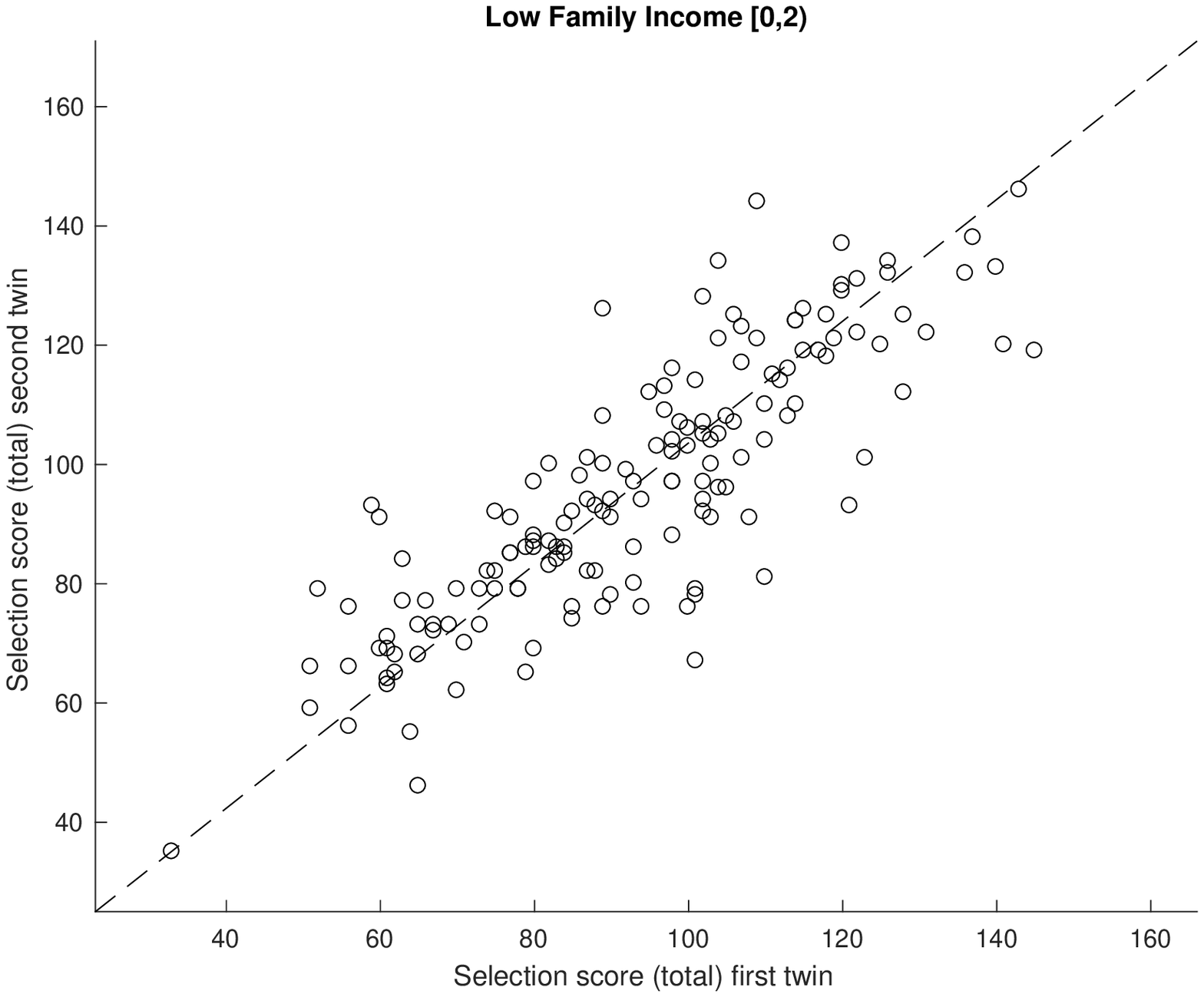}
   \includegraphics[width=4cm]{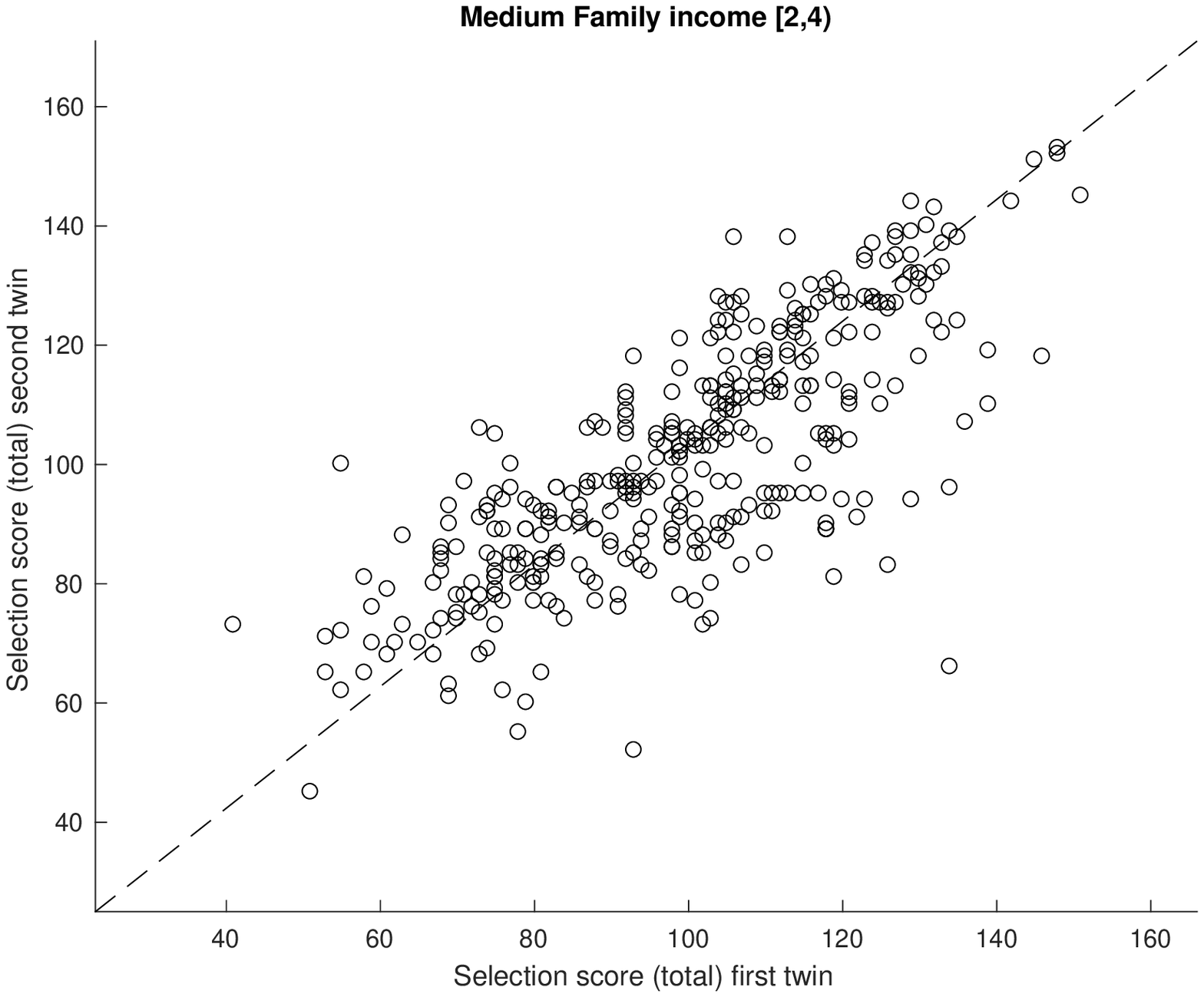}
    \includegraphics[width=4cm]{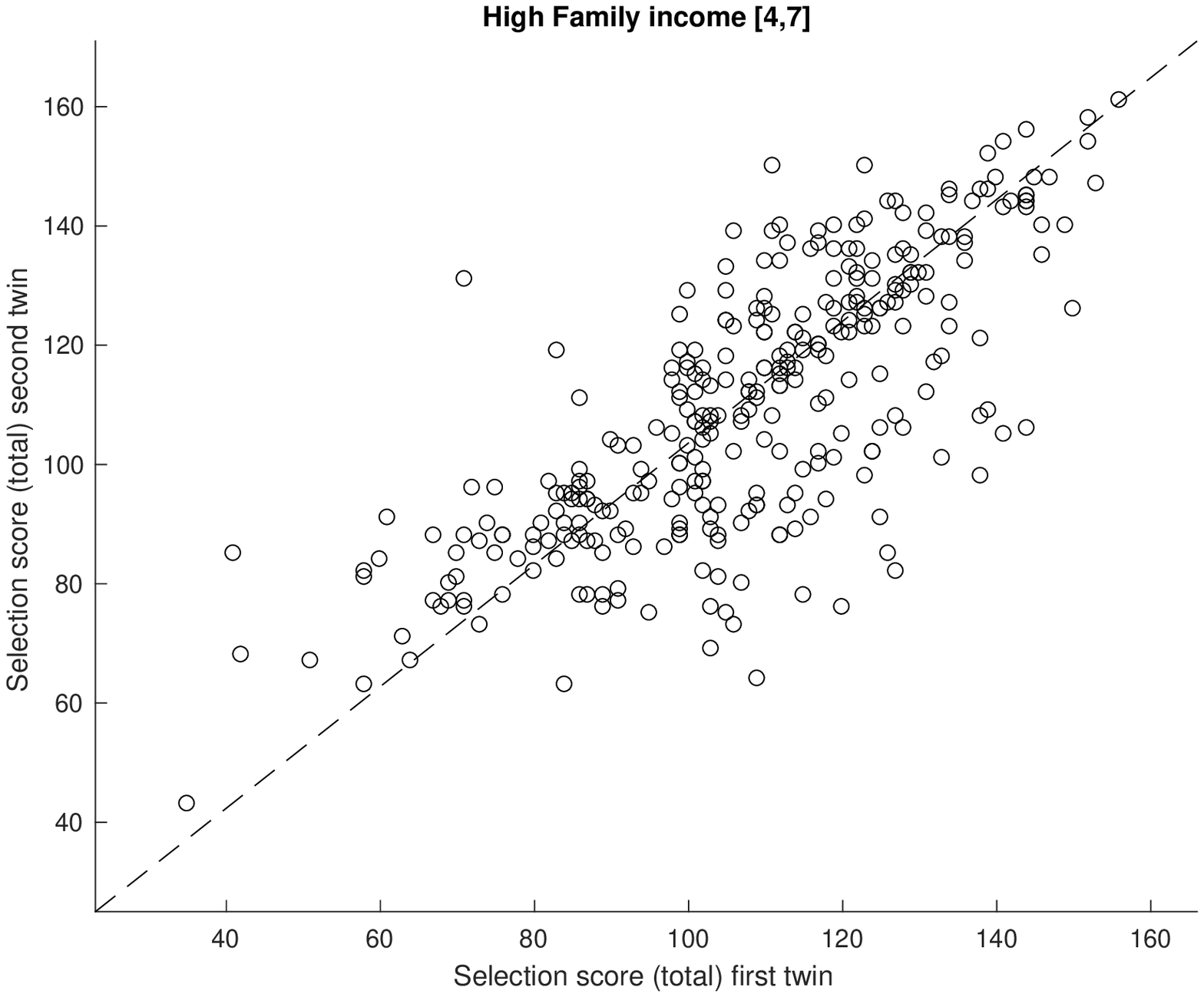}
\caption{\small{Scatterplots of the twins overall scores with respect to the mother's (top panels) and father's level of education (middle panels) and family income (bottom panels).
From left to right, the plots correspond to low [0,2), medium [2,4) and high [4,6] ([4,7] for the family income) levels of the covariates. The black dashed line corresponds to the 45 degrees diagonal.}}
\label{ZoCov}
\end{figure}

In Figure \ref{ZoCov} we discretized the covariates into low, medium and high values and we produced scatterplots of the twins' school performance scores. 
The top plots refer to the mother's level of education, the central plots refer to the father's level of education and the bottom plots refer to the family income.
From left to right, the plots correspond to low [0,2), medium [2,4) and high [4,6] ([4,7] for the family income) levels of the covariates. 
As we move from low to high levels of the covariates, the point clouds tend to change shape around the diagonal and move to the upper right corner. 
Therefore, in all three cases, as already pointed out, low values of covariates correspond to low performance scores, while high values of covariates correspond to high performance scores.
In addition, the different shapes of the scatterplot points corresponding to low, medium and high level of covariates suggest that the dependence structure between the twins' school outcomes changes according to the levels of the covariates.
Therefore, a flexible model, able to capture the effect of a covariate on the dependence between the kids' performance scores is necessary.


In order to model the dependence structure between the twins' school performances, we use copulas, which are popular modeling approaches in multivariate statistics allowing the separation of the marginal components of a joint distribution from its dependence structure.
More precisely, \cite{Sklar59} proved that a $d$-dimensional distribution $H$ of the random variables $Y_1, \ldots, Y_d$ can be fully described by its marginal distributions and a function $C:[0,1]^d\rightarrow [0,1]$, called copula, through the relation $H(y_1,\dots,y_d)=C(F_1(y_1),\dots,F_d(y_d))$.
In the literature, copulas have been applied to model the dependence between variables in a wide variety of fields (see \cite{KolAnjMen06} and \cite{CheLucVec04}). In particular, applications of copula models involved lifetime data analysis (\cite{Ander2005}), survival analysis of Atlantic halibut (\cite{BraVer05}) and transfusion-related AIDS and cancer analysis (\cite{EmuWang12}, \cite{HuaZha08} and \cite{OwJuSen07}).

The introduction of covariate adjustments to copulas has attracted an increased interest in recent years. 
\cite{CraSab12} propose a conditional copula approach in regression settings where the bivariate outcome can be continuous or mixed. \cite{Patton06} introduces time-variation in the dependence structure of ARMA models (see also \cite{JonRock06} and \cite{BaTaWa07} for other applications of time-series analysis to dependence modelling).
The paper of \cite{AcarCra10} provides a nonparametric procedure to estimate the functional relationship between copula parameters and covariates, showing that the gestational age drives the strength of dependence between the birth weights of twins.
\cite{AbGijVer12} and \cite{GijOmVer12} propose semiparametric and nonparametric methodologies for the  estimation of conditional copulas, establishing consistency and asymptotic normality results for the estimators. The methodology is then applied to examine the influence of the gross domestic product (GDP), in USD per capita, on the life expectancy of males and females at birth.

In a similar vein, parametric models such as Bayesian regression copulas allow the specification of Bayesian marginal regressions for a set of outcomes, linking the marginals to covariates, and combining them via a copula to form a joint model.
The general framework of Bayesian Gaussian regression copulas with discrete, continuous or mixed outcomes is presented by \cite{pitt2006efficient} and allows to handle a multivariate regression with Gaussian and non-Gaussian marginal distributions.  
\cite{yin2009bayesian} adopt a Bayesian regression copula model in cancer
clinical trials for dose finding to account for the synergistic effect of combinations of 
multiple drugs.
A copula constructed from the skew t distribution is employed by \cite{smith2012modelling}  to capture asymmetric and extreme dependence between variables modelled via Bayesian marginal regressions.
While most Bayesian regression copula models focus on covariate adjustments for the marginals, recently \cite{KleinKneiss2016} proposed simultaneous
Bayesian inference for both the marginal distributions
and the copula.  Other contributions along the same lines are \cite{taglioni2016bayesian}, \cite{stander2015a} and \cite{stander2015b}. 
However, the authors selected the copula family by using the deviance information criterion, which may suffer from limitations, as discussed for example by \cite{Plum08}. Indeed, the choice of the copula family may be controversial and it is still an open problem (see \cite{Joe14}).  
The literature offers a rich range of copula families, such as elliptical copulas (e.g. Gaussian and Student's t) and archimedean copulas (e.g. Frank, Gumbel, Clayton and Joe) to accommodate various dependence structures.
In this paper, we adopt a Bayesian nonparametric approach which allows us to overcome the issue of the choice of copula and we adopt a conditional copula approach to model the effect of a covariate on the dependence between variables.  
Our methodology builds on \cite{WuWaWa15}, who propose a Bayesian nonparametric procedure to estimate any unconditional copula density function. The authors combine the well-known Gaussian copula density with the modeling flexibility of the Bayesian nonparametric approach, proposing to use an infinite mixture of Gaussian copulas. \cite{BurProk14} propose to use nonparametric univariate Gaussian mixtures for the marginals and a multivariate random Bernstein polynomial copula for the link function under the Dirichlet process prior.  Our paper extends the work of \cite{WuWaWa15} to the conditional copula setting, by proposing a novel methodology which combines the advantages of a conditional copula approach with the modeling flexibility of Bayesian nonparametrics.
In particular, we include a conditional covariate component to explain the variable dependence structure, allowing us further flexibility to the copula density modelling. Up to our knowledge, this is the first Bayesian nonparametric proposal in the conditional copulas literature.

The outline of the paper is the following. In Section \ref{BNPCondCopula} we briefly review the literature about conditional copulas and Bayesian nonparametric copula estimation. In Section \ref{NP} we introduce our novel Bayesian nonparametric conditional copula setting. Section \ref{Gibbs} provides an algorithm for estimating the posterior parameters and Section \ref{Simul} illustrates the performance of the methodology. Section \ref{RealData} is devoted to the application of our methodology to the analysis of the National Merit Twin Study. Concluding remarks are given in Section \ref{Conc}.

\section{Preliminaries}
\label{BNPCondCopula}

In this Section, we review some preliminary notions about conditional copulas and illustrate the Bayesian nonparametric copula density estimation introduced in \cite{WuWaWa15}. In what follows, we focus on the bivariate case for
simplicity, however the arguments can be easily extended to more than two dimensions.

\subsection{The conditional copula}

Let $Y_1$ and $Y_2$ be continuous variables of interest and $X$ be a covariate that may affect the dependence between $Y_1$ and $Y_2$. Following \cite{GijOmVer12}, \cite{AbGijVer12} and \cite{AcarCra10}, we suppose that the conditional distribution of $(Y_1,Y_2)$ given $X=x$ exists and we denote the corresponding conditional joint distribution function by
$$H_x(y_1,y_2)=P(Y_1\leq y_1,Y_2\leq y_2|X=x).$$
If the marginals of $H_x$, denoted as
$$F_{1x}(y_1)=P(Y_1\leq y_1|X=x),\qquad F_{2x}(y_2)=P(Y_2\leq y_2|X=x),$$
are continuous, then according to Sklar's theorem there exists a unique copula $C_x$ which equals
\begin{equation}\label{CondCop}
C_x(u,v)=H_{x}(F_{1x}^{-1}(u),F_{2x}^{-1}(v))
\end{equation}
where $F_{1x}^{-1}(u)=\inf\lbrace y_1: F_{1x}\geq u\rbrace$ and $F_{2x}^{-1}(v)=\inf\lbrace y_2: F_{2x}\geq v\rbrace$ are the conditional quantile functions and $u=F_{1x}(y_1)$ and $v=F_{2x}(y_2)$ are called pseudo-observations. The conditional copula $C_x$ fully describes the conditional dependence structure of $(Y_1,Y_2)$ given $X=x$. An alternative expression for \eqref{CondCop} is
\begin{equation}\label{jointCondCop}
H_x(y_1,y_2)=C_x(F_{1x}(y_1), F_{2x}(y_2)).
\end{equation}

\subsection{Bayesian nonparametric copula density estimation}

Let $\Phi_{\rho}(y_1,y_2)$ denote the standard bivariate normal distribution function with correlation coefficient $\rho$. Then, $C_{\rho}$ is the copula corresponding to $\Phi_{\rho}$, taking the form:
\begin{equation}
C_{\rho}(u,v)=\Phi_{\rho}(\Phi^{-1}(u),\Phi^{-1}(v)) \label{cop}
\end{equation}
where $\Phi$ is the univariate standard normal distribution function. The Gaussian copula density is:
\begin{equation}
\begin{footnotesize}
c_{\rho}(u,v)=|\Sigma|^{-\frac{1}{2}} \exp{\biggl\{ -\frac{1}{2} (\Phi^{-1}(u),\Phi^{-1}(v)) (\Sigma^{-1}-\textbf{I})\begin{pmatrix} \Phi^{-1}(u) \\ \Phi^{-1}(v) \end{pmatrix}\biggr\}}
\end{footnotesize}
\label{Gaus}
\end{equation}
where the correlation matrix is:
\begin{equation}
\Sigma= \begin{bmatrix} 1 & \rho \\ \rho & 1 \end{bmatrix}.\notag
\end{equation}
\cite{WuWaWa15} proposed to use an infinite mixture of Gaussian copulas for the estimation of a copula density, as follows
\begin{equation}
c(u,v)=\sum_{j=1}^{\infty} w_{j}c_{\rho_j}(u,v)\label{mixGau}
\end{equation}
where the weights $w_j$'s sum up to $1$ and the $\rho_j$'s vary in $(-1,1)$. Given a set of $n$ observations $(u_1,v_1), \dots, (u_n,v_n)$, their model can be described through a hierarchical specification, i.e. 
\begin{align}\label{BnpCopula}
\begin{split}
   (u_i,v_i) \mid \rho_i \, \simind & \quad c_{\rho_i}(u_i,v_i), \qquad i = 1, \ldots, n, \\
   \rho_i \mid G   \, \simiid & \quad G, \\
 G \, \sim & \quad DP(\lambda,G_0),
  \end{split}
\end{align}
where $G$ is a Dirichlet Process prior with total mass $\lambda$ and base measure $G_0$. 
This proposal is motivated by the fact that bivariate density functions on the real plain can be arbitrarily well approximated by a mixture of a countably infinite number of bivariate normal distributions of the form
$$f(y_1,y_2)=\sum_{j=1}^{\infty} w_{j}N((y_1,y_2)|(\mu_{1j},\mu_{2j}),\Sigma_j)$$
where $N((y_1,y_2)|(\mu_{1j},\mu_{2j}),\Sigma_j)$ is the joint bivariate normal density with mean vector $(\mu_{1j},\mu_{2j})$ and correlation matrix $\Sigma_j$ (see \cite{Lo84} and \cite{Fer83}). Roughly speaking, the authors are mimicking the Dirichlet process mixture model in the copula setting (see \cite{Escobar94} and \cite{EscobarWest1995}). The sampling strategy follows the slice sampler of \cite{walker2007} and \cite{walker2011}.
The authors show that the Gaussian mixture is flexible enough to accurately approximate any bivariate copula density.


\section{Conditional copula estimation with Dirichlet process priors}
\label{NP}

The data object of study requires a model which can take into account the effect of a covariate. We build on the model introduced by \cite{WuWaWa15} and illustrated in the previous section. The idea is to replace the Gaussian copula with a conditional version where the correlation is a function of the covariate, i.e. 
$$c_{\rho}(u,v|x)=c_{\rho(x)}(u,v).$$ 
The function $\rho(x)$ can be modelled as preferred, for instance, with a generalized linear model or with a non-linear function. In any case, we have that $\rho(x)$ will depend on a vector of parameters $\bm{\beta}$, so that 
$$c_{\rho(x)}(u,v)=c_{\rho(x|\bm{\beta})}(u,v).$$
We assume a Dirichlet process prior on the vector of parameters $\bm{\beta}=(\beta_1,\dots,\beta_d)$. Following the model description provided in equation \eqref{BnpCopula}, we can summarize our model as follows, 
\begin{align}\label{BnpCondCopula}
\begin{split}
   (u_i,v_i) \mid \rho(x_i|\bm{\beta}_i) \, \simind & \quad c_{\rho(x_i|\bm{\beta}_i)}(u_i,v_i), \qquad i = 1, \ldots, n, \\
   \bm{\beta}_i \mid G   \, \simiid & \quad G, \\
 G \, \sim & \quad DP(\lambda,G_0),
  \end{split}
\end{align}
where $G$ is a Dirichlet process prior with total mass $\lambda$ and base measure $G_0$. As in \cite{WuWaWa15}, our model can be described as an infinite mixture of Normal distributions,
\begin{equation}\label{InfCondCop}
c_{\rho}(u,v|x)=\sum_{j=1}^{\infty} w_{j}c_{\rho(x|\bm{\beta}_j)}(u,v),
\end{equation}
and hence suitable for implementing a slice sampling algorithm, as explained in the next section. 

\medskip

\noindent In order to model the function $\rho(x|\bm{\beta})$, we would like to follow some standard approaches in the literature. \cite{AbGijVer12} model the dependence of the parameter of interest, with respect to the covariate,    through a \textit{calibration function} $\theta(x|\bm{\beta})$. It is important to highlight that in many copula families the parameter space is restricted. In contrast, the calibration function $\theta(x|\bm{\beta})$ can assume any value on the real line. In our case, the parameter is restricted to the interval $(-1,1)$ and we need a transformation which can link the calibration function $\theta(x|\bm{\beta})$ to $\rho(x|\bm{\beta})$. In this paper, we adopt the following transformation, 
\begin{align*}
\rho(x|\bm{\beta})&=\frac{2}{|\theta(x|\bm{\beta})|+1}-1.
\end{align*}
In our simulated and real data examples we focus on two particular calibration functions studied in the literature, which are 
\begin{align*}
\theta(x|\bm{\beta})&= \beta_1 +\beta_2 x^2 \\
\theta(x|\bm{\beta})&= \beta_1+\beta_2 x +\beta_3\exp{(-\beta_4 x^2)}
\end{align*} 
respectively, such that $\theta(x|\bm{\beta})\in (-\infty,+\infty)$ and, consequently, $\rho(x|\bm{\beta}) \in (-1,1)$.

\section{Posterior sampling algorithm}
\label{Gibbs}

The observations $(y_{1i},y_{2i})$, for $i=1,\dots,n$, are transformed into the corresponding pseudo-observations $(u_i,v_i)$ using a nonparametric estimation approach, as in \cite{GijVerOmel11}. The pseudo-observations are then plugged into the copula.
Following equation \eqref{InfCondCop}, given $(u_i,v_i)$ for $i=1,\dots,n$, and the conditional variable $x_i$, the conditional copula density function for each pair $(u_i,v_i)$ can be written as an infinite mixture of conditional Gaussian copulas, such that:
\begin{equation}
c(u_i,v_i|x_i)=\sum_{j=1}^{\infty} w_j c_{\rho(x_i|\bm{\beta}_j)}(u_i,v_i) \label{InfMix}
\end{equation}
where $w_{j}$'s are the stick-breaking weights, i.e.
$$w_{j}=\pi_{j}\prod_{l=1}^{j-1} (1-\pi_{l})$$ where the $\pi_{j}$ are distributed as a $\mathcal{B}e(1,\lambda)$, $\lambda>0$. In order to sample from the infinite mixture displayed in equation \eqref{InfMix}, we use the slice sampling algorithm for mixture models proposed by \cite{walker2007} and \cite{walker2011}. To reduce the dimensionality of the problem, the authors introduce a latent variable $z_i$ for each $i$ which allows us to write the infinite mixture model as follows:
\begin{equation}
c(u_i,v_i,z_i|x_i)=\sum_{j=1}^{\infty} \mathbb{I}(z_i< w_{j}) c_{\rho(x_i|\bm{\beta}_j)}(u_i,v_i). \label{Slice}
\end{equation}
The introduction of the slice variable $z_i$ reduces the sampling complexity to the analogous of a finite mixture model. In particular, letting
\begin{equation}
A_{w}=\{j: z_i< w_j\}, \label{set}
\end{equation}
then it can be proved that the cardinality of the set $A_{w}$ is almost surely finite. Consequently, there is a finite number of parameters to be estimated.
By iterating the data augmentation principle further, we introduce another latent variable $d_i$, which is called allocation variable, allowing us to allocate each observation to one component of the mixture model. Then, the conditional copula density $c(u_i,v_i,z_i,d_i|x_i)$ takes the form:
\begin{equation}
c(u_i,v_i,z_i,d_i|x_i)=\mathbb{I}(z_i<w_{d_i}) c_{\rho(x_i|\bm{\beta}_{d_i})}(u_i,v_i) \label{alloc}
\end{equation}
where $d_i\in\{1,2,\dots\}$. Hence, the full likelihood function of the conditional copula model is:
\begin{equation}
\prod_{i=1}^n c(u_i,v_i,z_i,d_i|x_i)=\prod_{i=1}^n \mathbb{I}(z_i<w_{d_i}) c_{\rho(x_i|\bm{\beta}_{d_i})}(u_i,v_i). \label{Lik}
\end{equation}
We use the notation $(U,V)=\{i=1,\dots,n: (u_{i},v_{i})\}$, $X=\{x_{1},\dots,x_{n}\}$ to describe the pseudo-observations and the covariate values, respectively. We denote with $\bm{\beta}=\{\bm{\beta}_{1},\bm{\beta}_{2},\dots\}$ the vector of parameters and $D=\{d_{1},\dots, d_{n}\}, \, Z=\{z_{1},\dots, z_{n}\}$ and $\bm{\pi}=\{\pi_{1},\pi_{2},\dots \}$ the new variables introduced so far.
\begin{figure}[h!]
 \centering
     \subfigure[\scriptsize{Simulated, $1^{st}$ cal.fun.}]
   {\includegraphics[width=3.5cm]{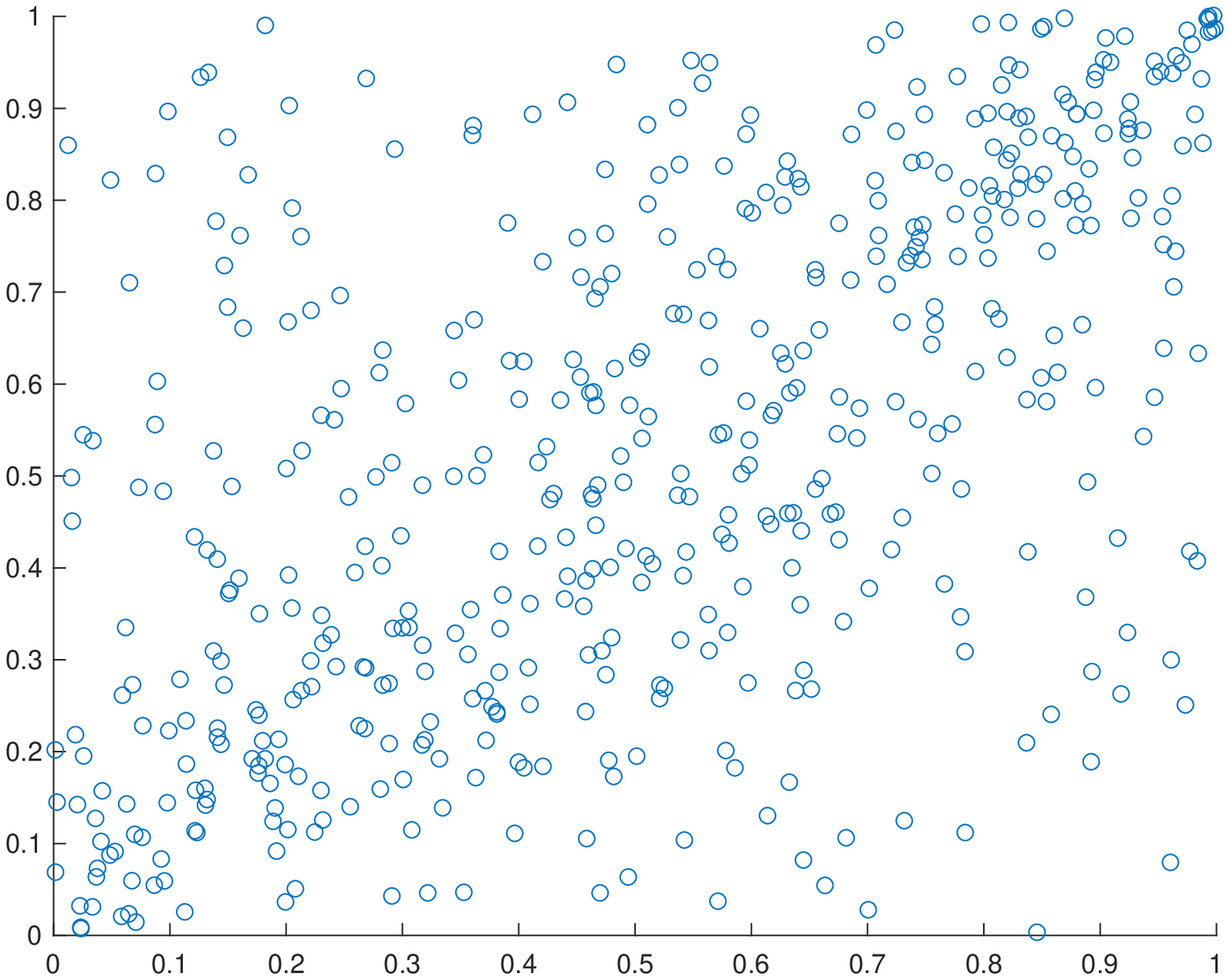}}
    \subfigure[\scriptsize{Simulated, $1^{st}$ cal.fun.}]
   {\includegraphics[width=3.5cm]{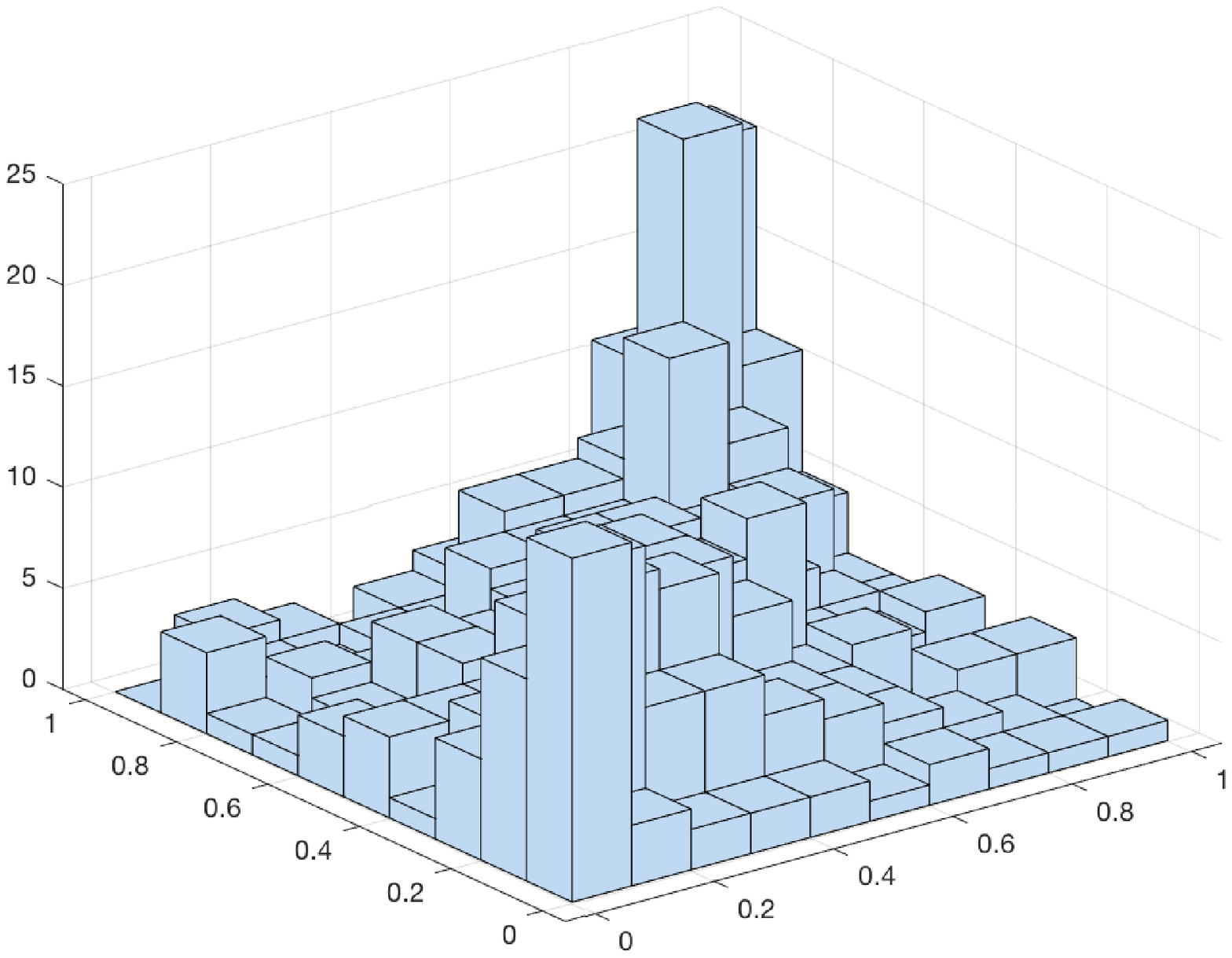}}
     \subfigure[\scriptsize{Predictive, $1^{st}$ cal.fun.}]
   {\includegraphics[width=3.5cm]{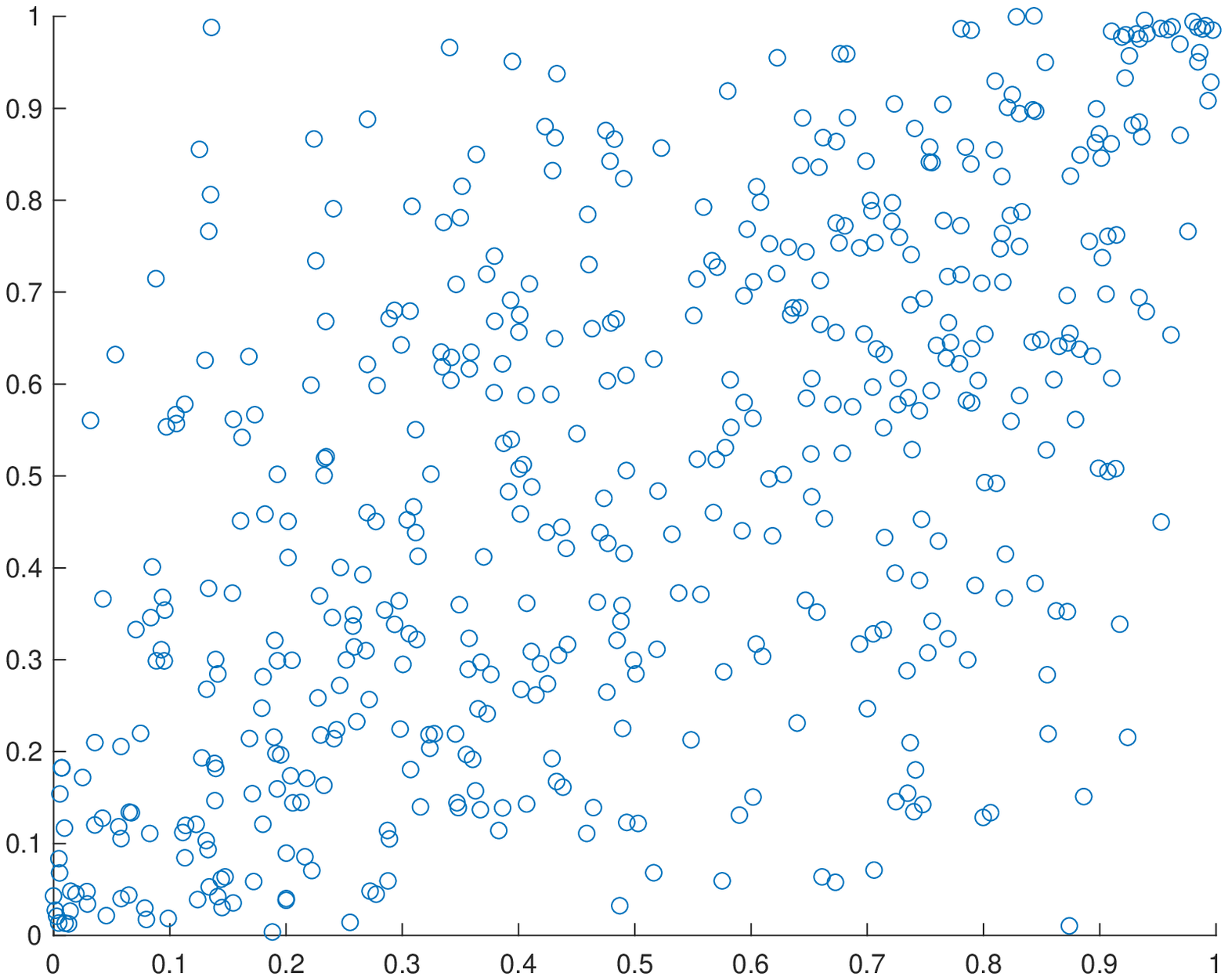}}
    \subfigure[\scriptsize{Predictive, $1^{st}$ cal.fun.}]
   {\includegraphics[width=3.5cm]{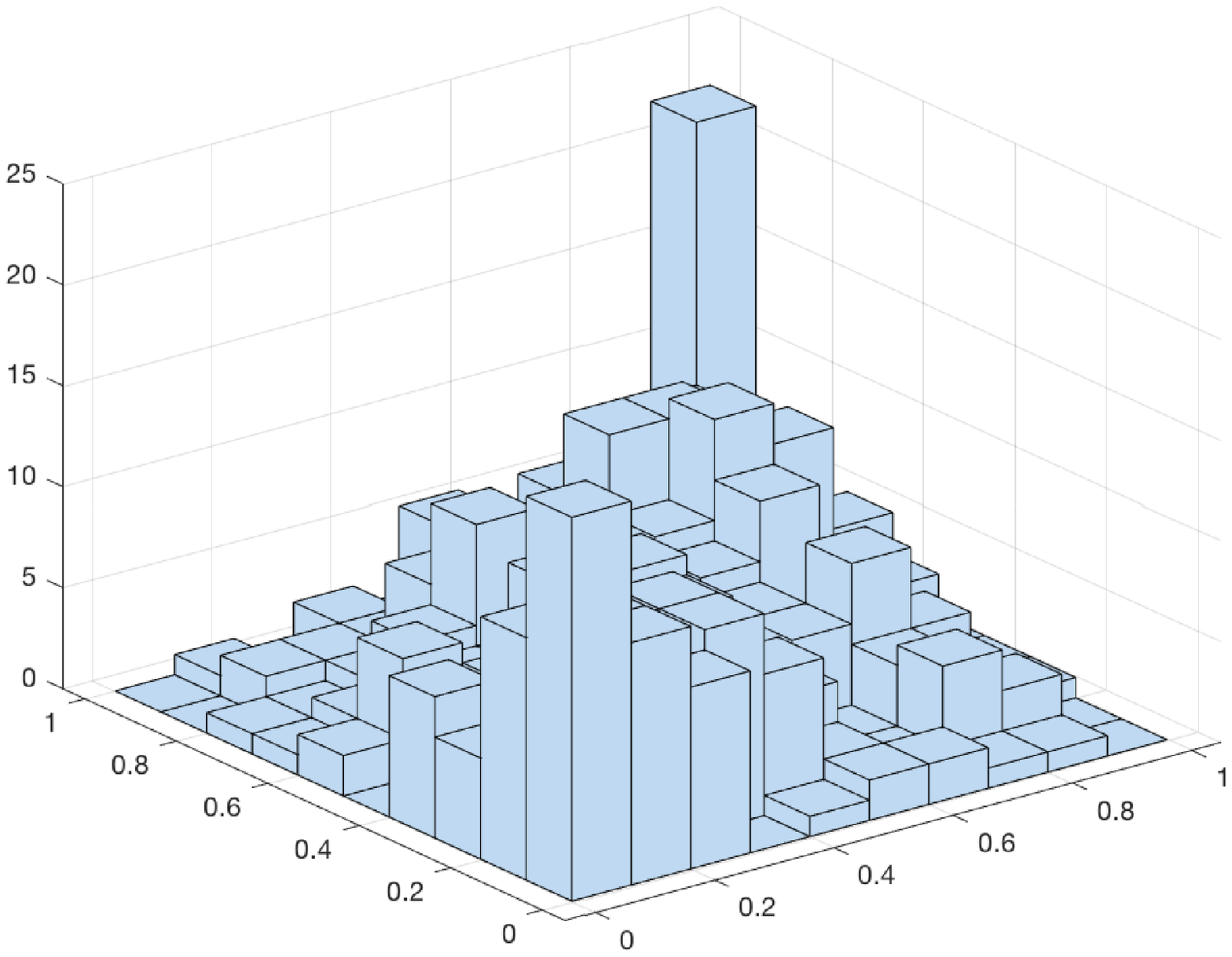}}
    \subfigure[\scriptsize{Simulated, $2^{nd}$ cal.fun.}]
   {\includegraphics[width=3.5cm]{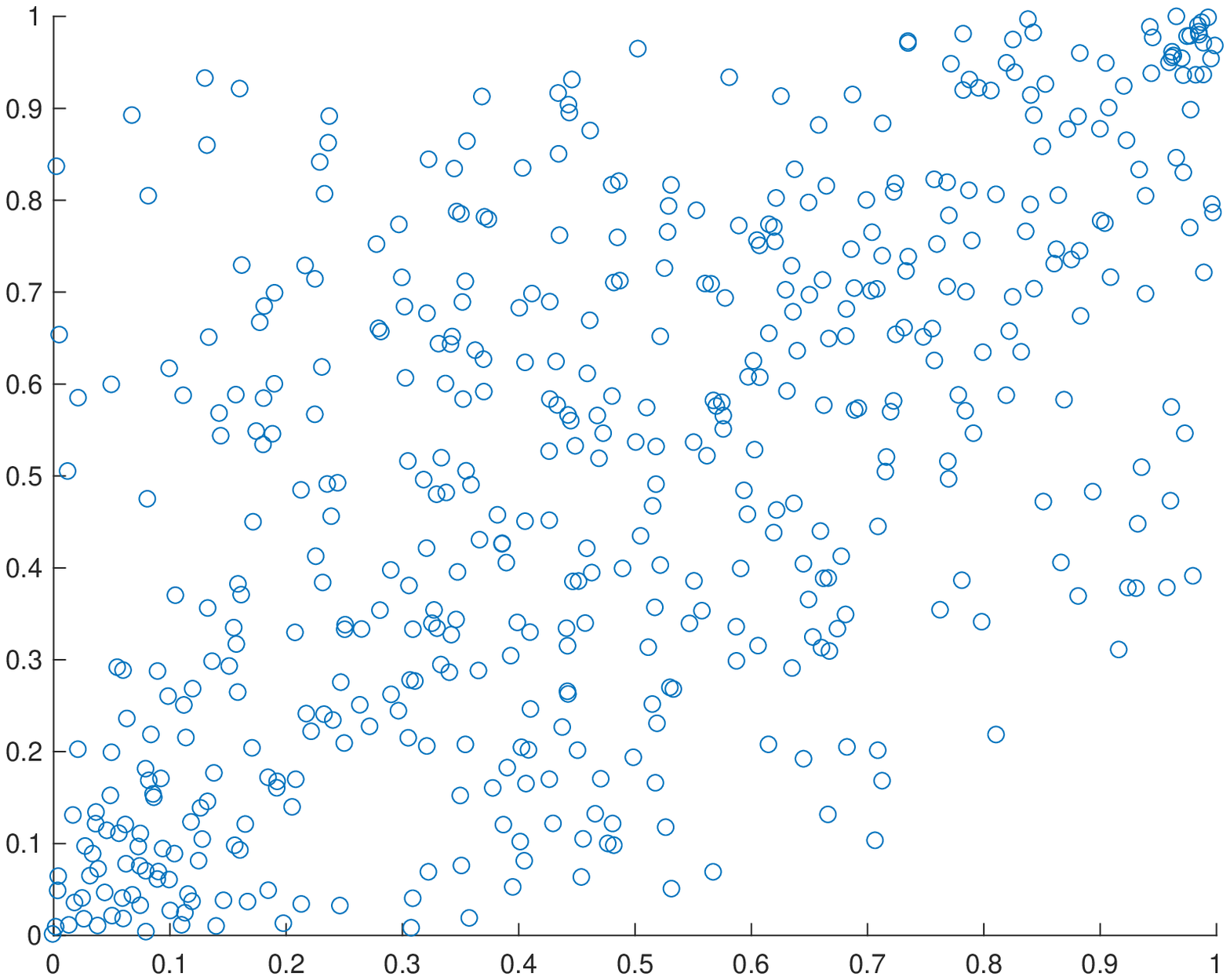}}
    \subfigure[\scriptsize{Simulated, $2^{nd}$ cal.fun.}]
   {\includegraphics[width=3.5cm]{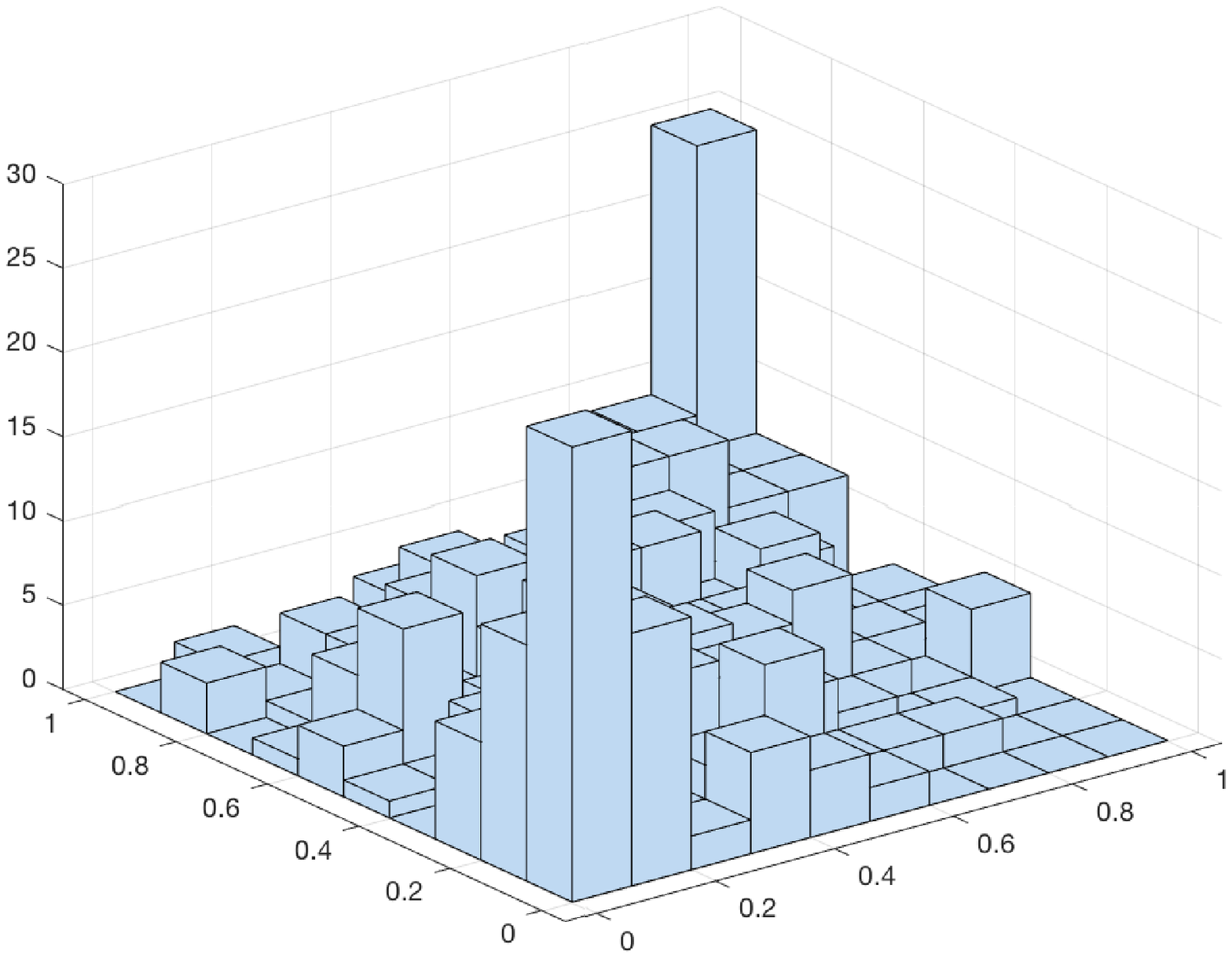}}
     \subfigure[\scriptsize{Predictive, $2^{nd}$ cal. fun.}]
   {\includegraphics[width=3.5cm]{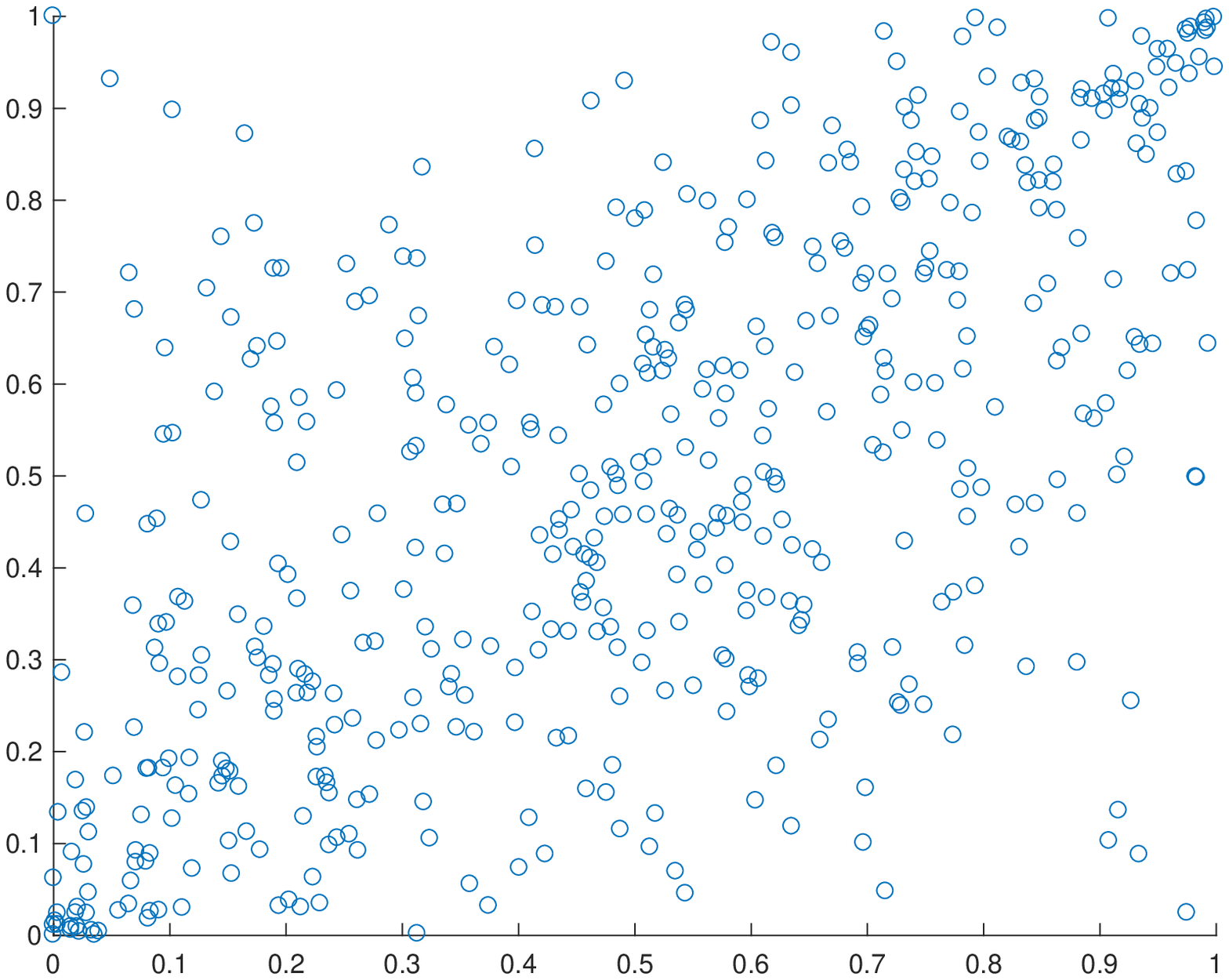}} \hspace{10pt}
    \subfigure[\scriptsize{Predictive, $2^{nd}$ cal. fun.}] 
   {\includegraphics[width=3.5cm]{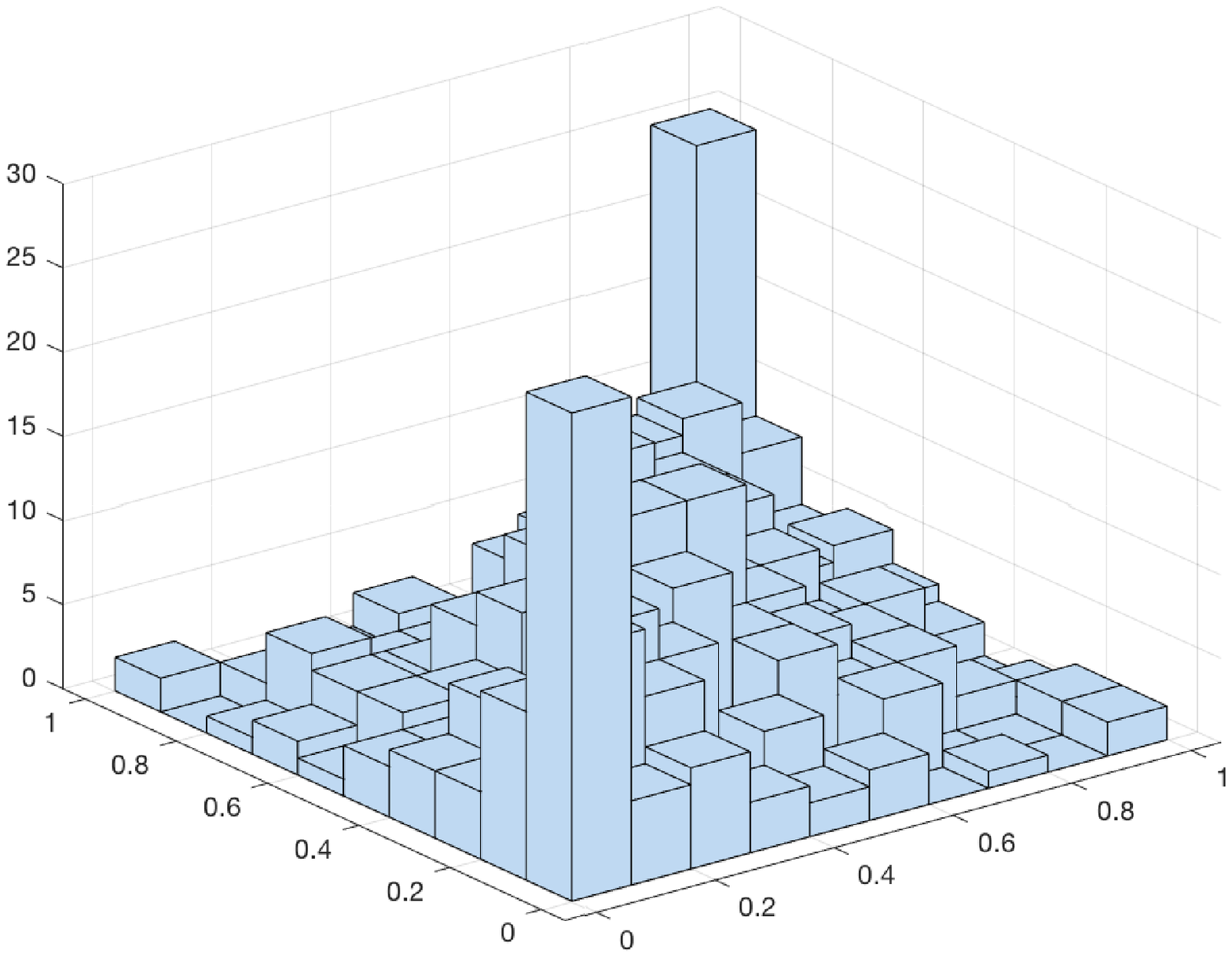}}
 \caption{\footnotesize{Gaussian copula with sample size $n=500$. Panels (a), (b), (c) and (d) depict the scatter plots and histograms, obtained with the first calibration function, of the simulated and predictive samples, respectively; panels (e), (f), (g) and (h) depict the scatter plots and histograms, obtained with the second calibration function, of the simulated and predictive sample, respectively.}}
\label{CondNorm500}
\end{figure}

Therefore, we used a Gibbs sampler to simulate iteratively from the posterior distribution function, according to the following steps:
\begin{enumerate}
\item The stick-breaking components $\bm{\pi}$ are updated given $[Z,D,\bm{\beta}, (U,V),X]$;
\item The latent slice variables $Z$ are updated given $[\bm{\pi},D,\bm{\beta},(U,V),X]$;
\item The allocation variables $D$ are updated given $[\bm{\pi},Z,\bm{\beta}, (U,V),X]$;
\item The vector of parameters $\bm{\beta}$ is updated given $[\bm{\pi},Z,D,(U,V),X]$.
\end{enumerate}
The Gibbs sampling details are explained in Appendix \ref{AppA}.

\begin{figure}[h!]
 \centering
     \subfigure[\scriptsize{Simulated, $1^{st}$ cal.fun.}]
   {\includegraphics[width=3.5cm]{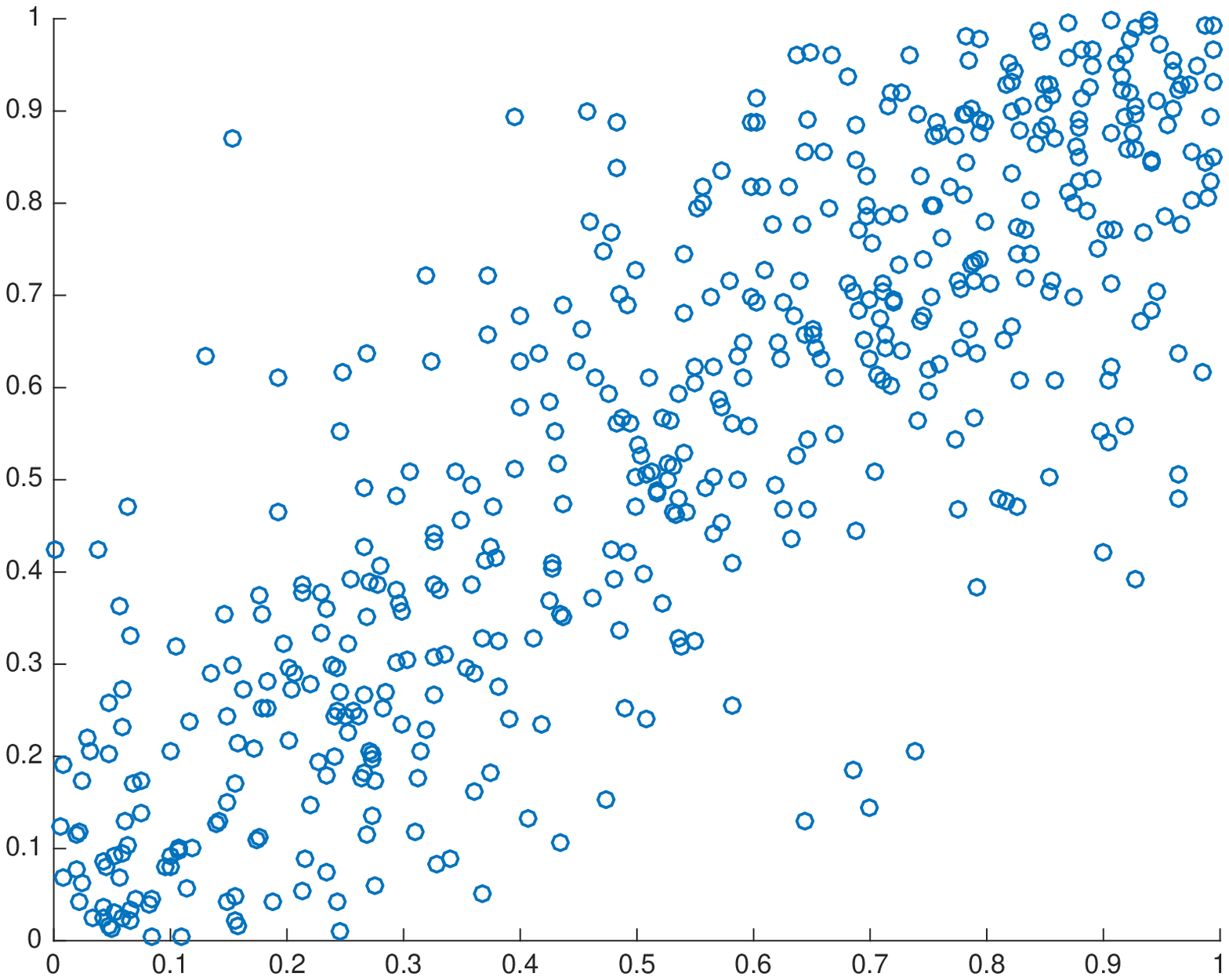}}
    \subfigure[\scriptsize{Simulated, $1^{st}$ cal.fun.}]
   {\includegraphics[width=3.5cm]{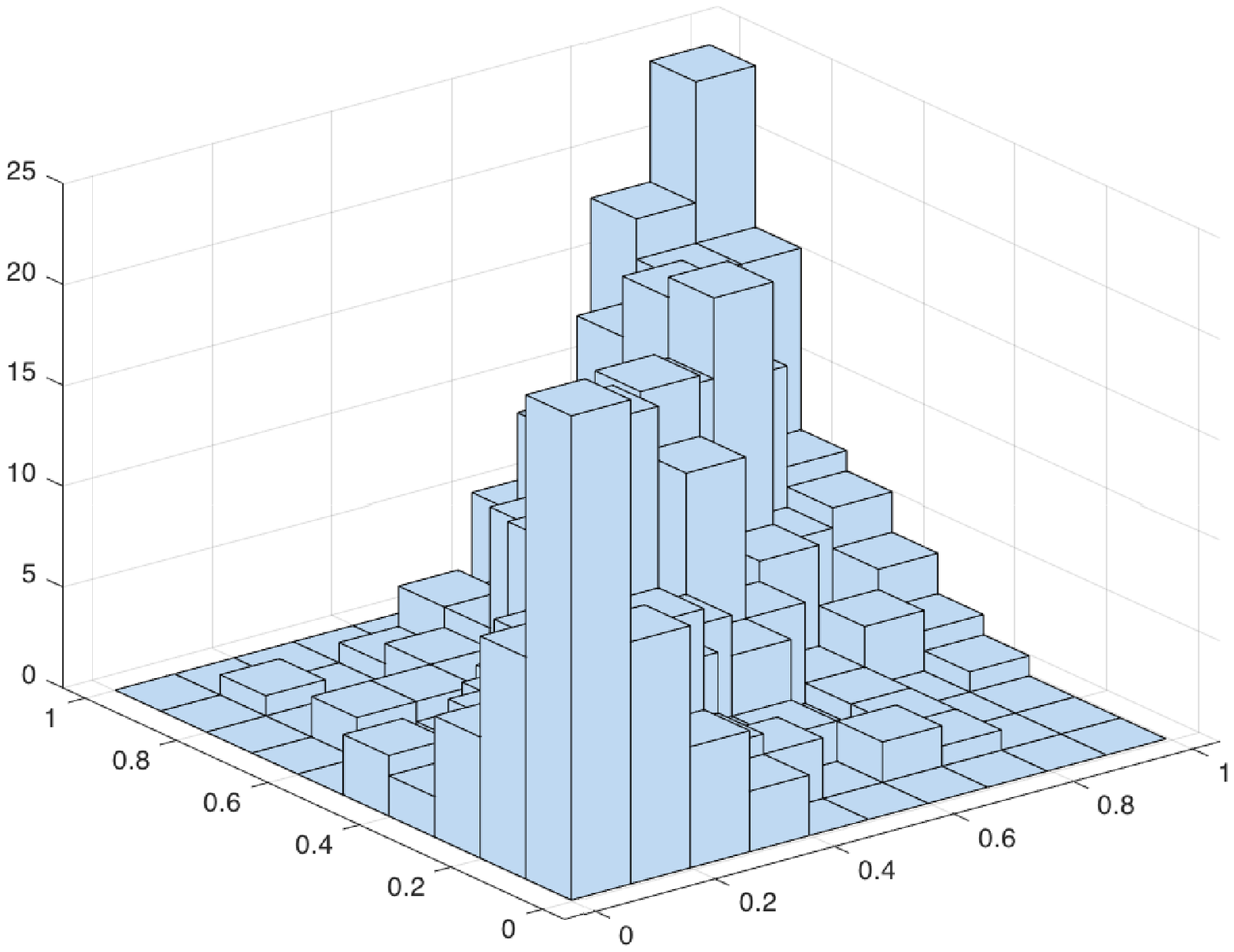}}
     \subfigure[\scriptsize{Predictive, $1^{st}$ cal.fun.}]
   {\includegraphics[width=3.5cm]{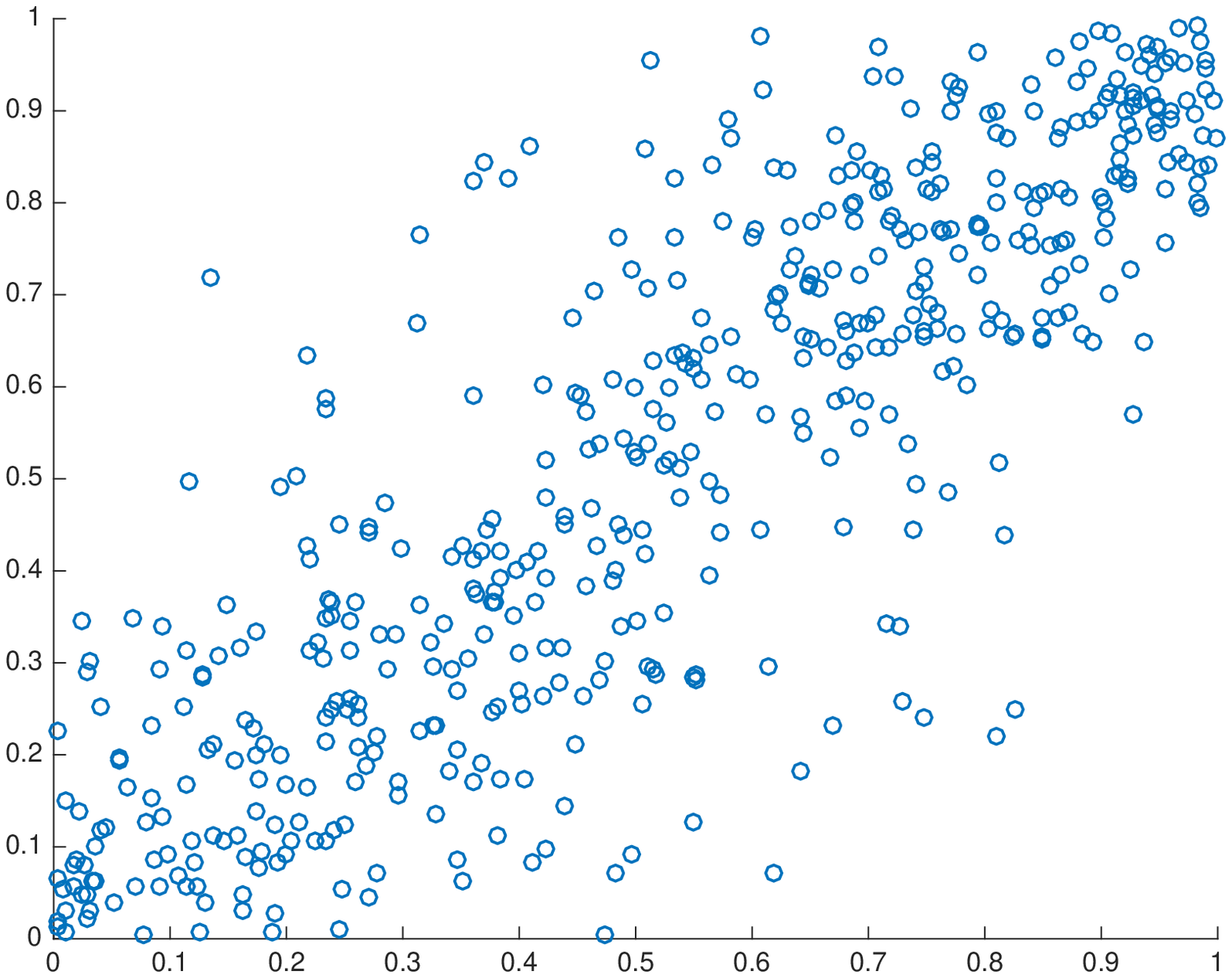}}
    \subfigure[\scriptsize{Predictive, $1^{st}$ cal.fun.}]
   {\includegraphics[width=3.5cm]{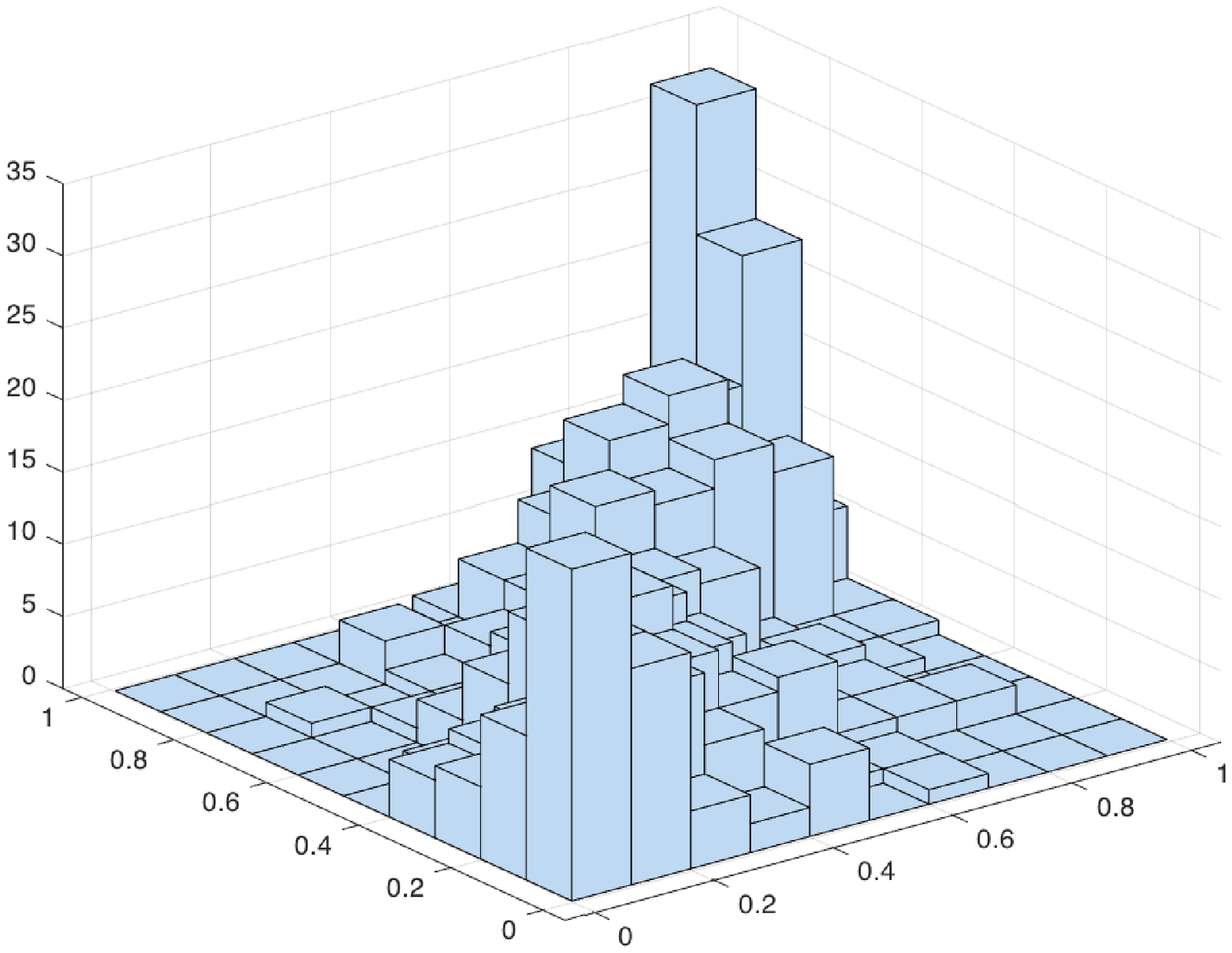}}
    \subfigure[\scriptsize{Simulated, $2^{nd}$ cal.fun.}]
   {\includegraphics[width=3.5cm]{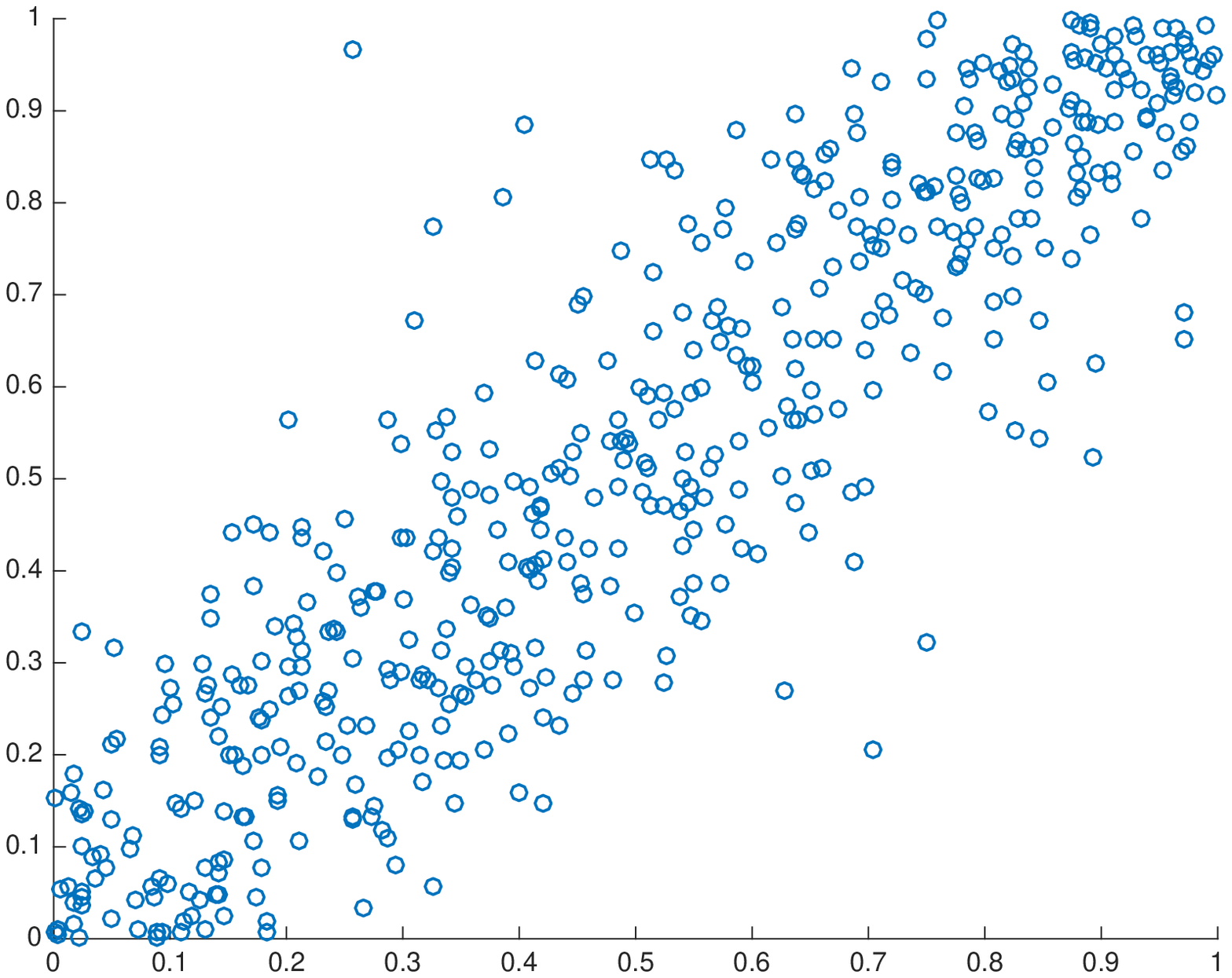}}
    \subfigure[\scriptsize{Simulated, $2^{nd}$ cal.fun.}]
   {\includegraphics[width=3.5cm]{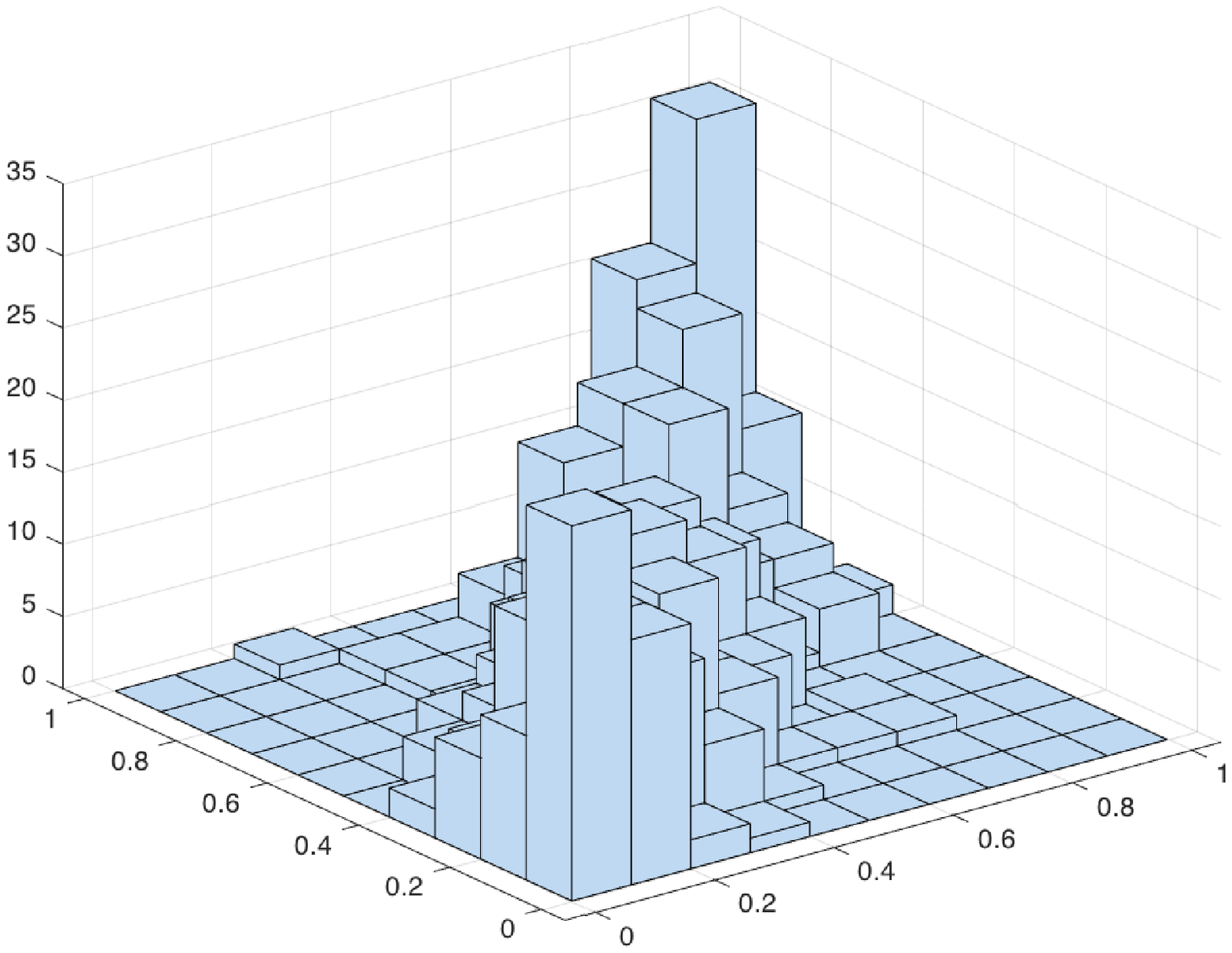}}
     \subfigure[\scriptsize{Predictive, $2^{nd}$ cal. fun.}]
   {\includegraphics[width=3.5cm]{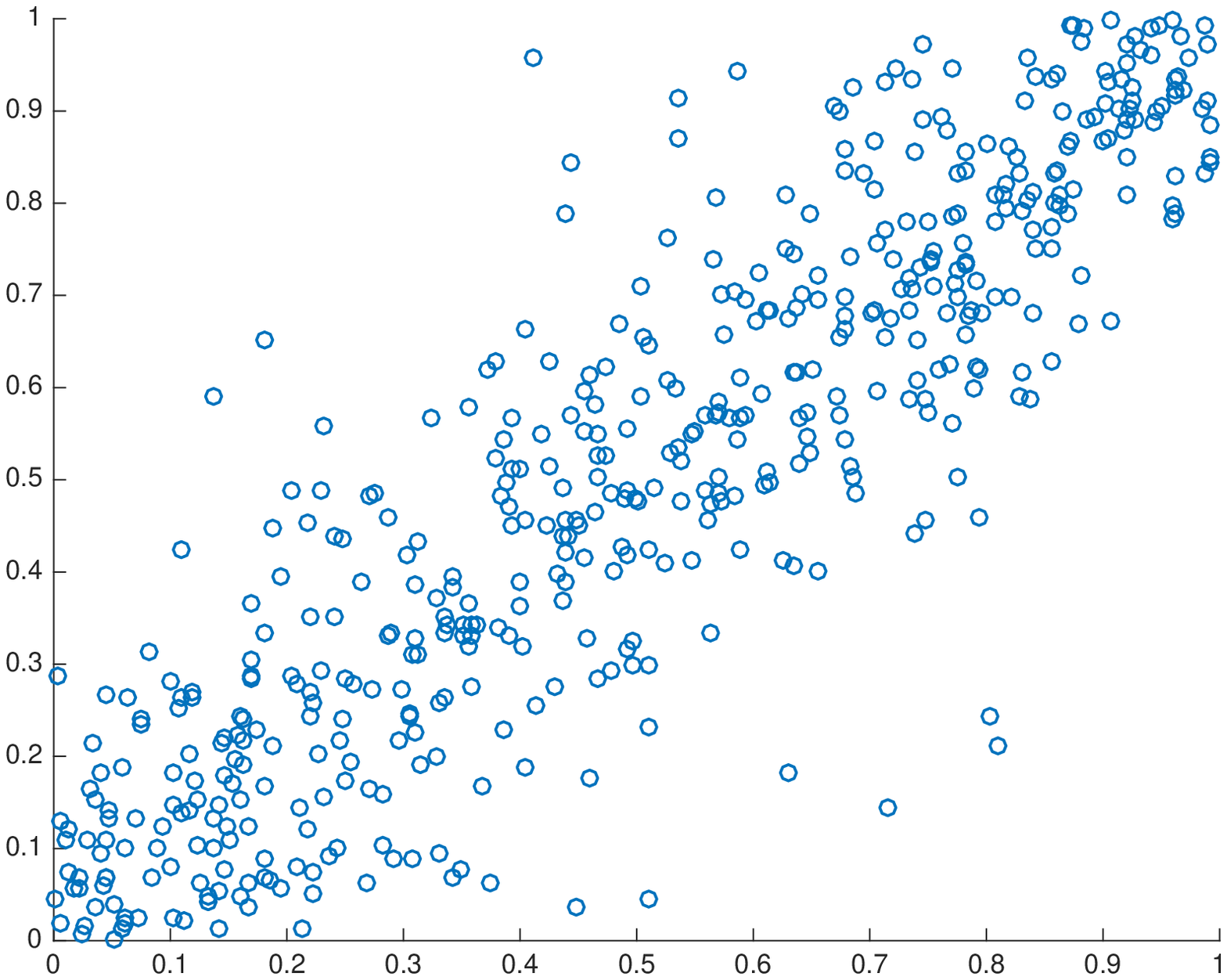}} \hspace{10pt}
    \subfigure[\scriptsize{Predictive, $2^{nd}$ cal. fun.}]
   {\includegraphics[width=3.5cm]{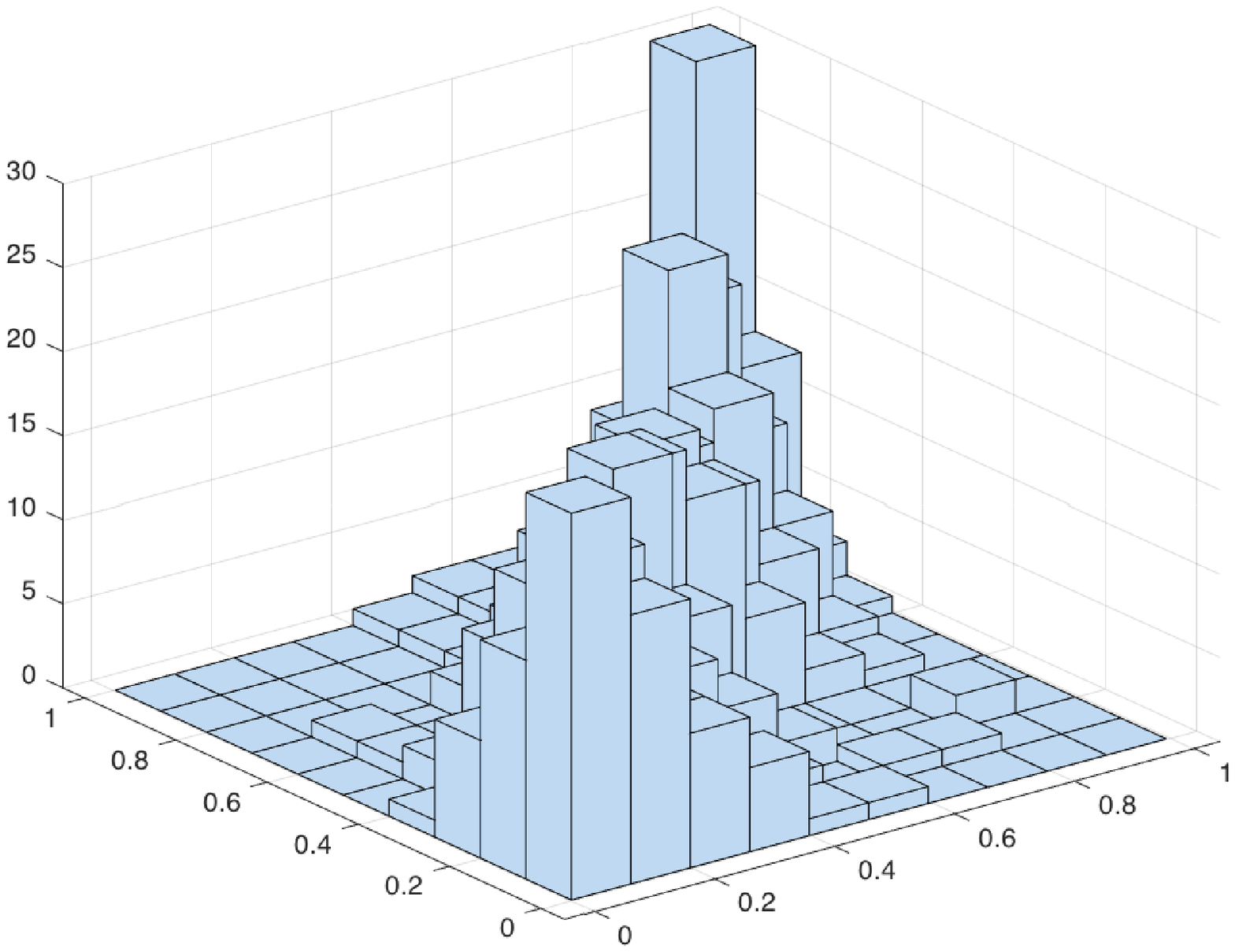}}
 \caption{\footnotesize{Frank copula with sample size $n=500$. Panels (a), (b), (c) and (d) depict the scatter plots and histograms, obtained with the first calibration function, of the simulated and predictive samples, respectively; panels (e), (f), (g) and (h) depict the scatter plots and histograms, obtained with the second calibration function, of the simulated and predictive sample, respectively.}}
\label{CondFrank500}
\end{figure}

\section{Simulation experiments}
\label{Simul}
This section illustrates the performance of the Bayesian nonparametric conditional copula model with simulated data. We generate datasets $(U,V)$ of sizes $n=250, 500~\mbox{and}~1000$ from different copula families, such as the Gaussian and Frank copulas. 
The copula dependence parameter is considered as a function of the exogenous variable $X$, which is simulated from an Uniform distribution in the interval $[-2,2]$.

For the Dirichlet process prior, we fix the total mass $\lambda=1$ and, for the base measure $G_0$, we adopt a bivariate normal distribution with zero mean vector and covariance matrix $\sigma^2 \bm{I}$, where $\sigma^2=100$. The following calibration functions are selected for $\theta(x|\bm{\beta})$,
\begin{align*}
\theta(x|\bm{\beta})&= \beta_1 +\beta_2 x^2 \\
\theta(x|\bm{\beta})&= \beta_1+\beta_2 x +\beta_3\exp{(-\beta_4 x^2)}.
\end{align*} 

As highlighted in Section \ref{NP}, we link the calibration functions $\theta(x|\bm{\beta})$ with $\rho(x|\bm{\beta})$ through the following transformation:
\begin{align*}
\rho(x|\bm{\beta})&=\frac{2}{|\theta(x|\bm{\beta})|+1}-1.
\end{align*}
This ensures that $\rho(x|\bm{\beta})$ assumes values between $(-1,1)$.

We run the Gibbs sampler algorithm described in Section \ref{Gibbs} for $4000$ iterations with (i) $500$ burn-in iterations and (ii) $3500$ burn-in iterations. Aiming at a parsimonious representation of the results, we focused on $3500$ burn-in iterations, since $500$ burn-in iterations gave very similar results.


\begin{figure}[h!]
\centering
\subfigure[Gaussian Copula]
{\includegraphics[width=6.5cm]{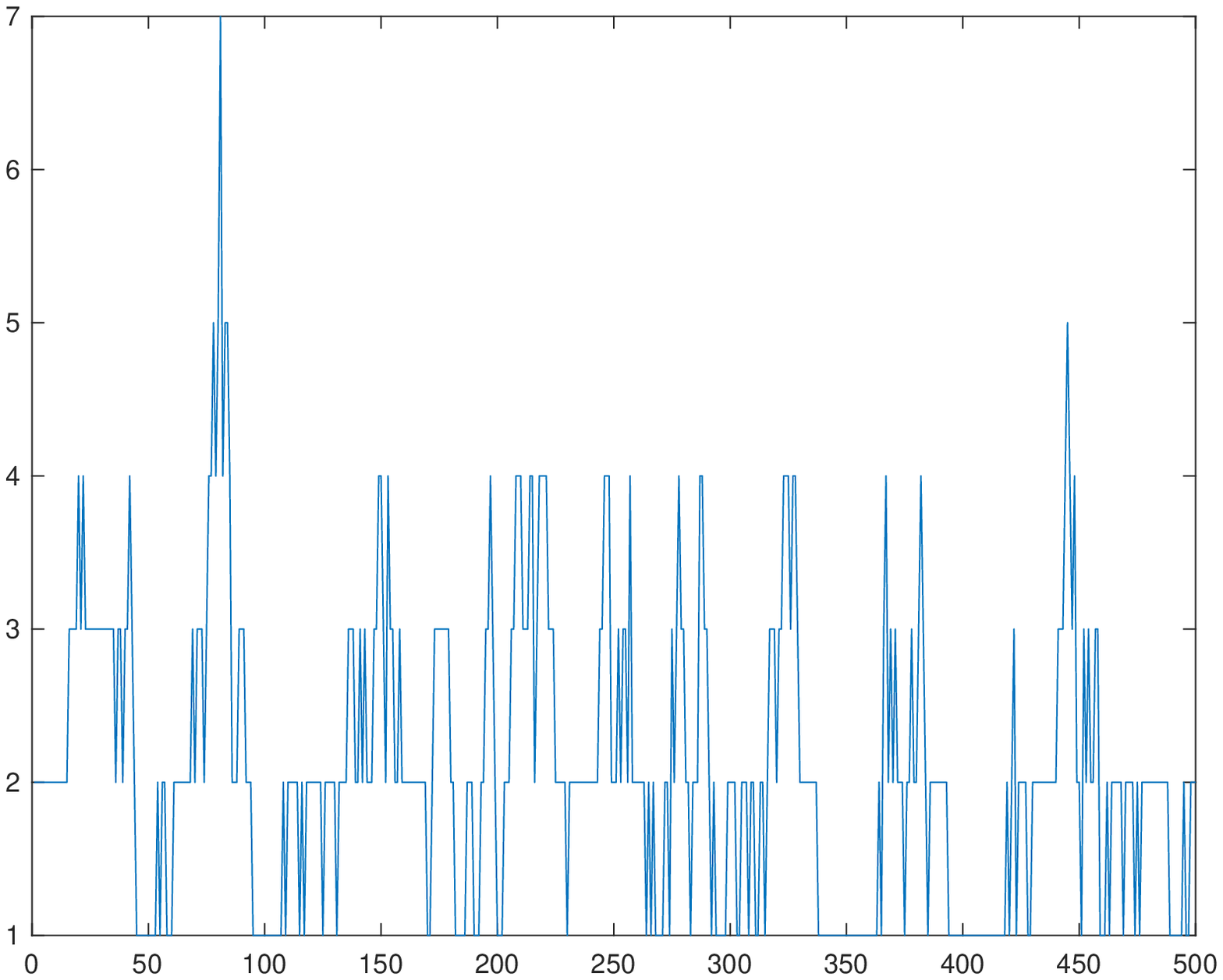}}
\subfigure[Frank Copula]
{\includegraphics[width=6.5cm]{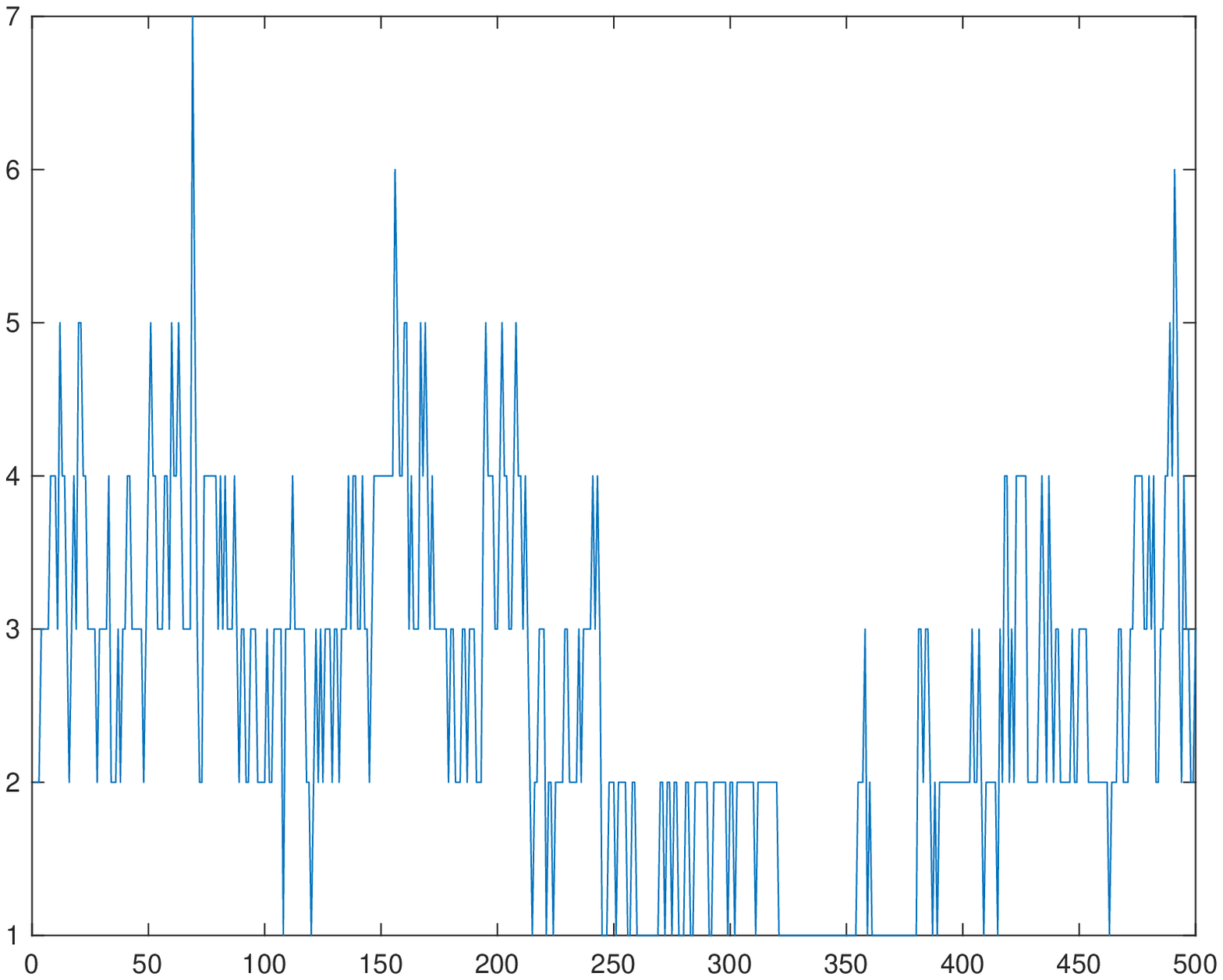}}
\caption{Number of components ($y$-axis) generated at each MCMC iteration ($x$-axis) for the Gaussian (left) and for the Frank copula (right) with sample size $n=500$.}
\label{NComSim}
\end{figure}

Figure \ref{CondNorm500}
illustrates the results of the application of the Bayesian nonparametric conditional copula model to data simulated from a Gaussian copula, with sample size $n= 500$. 
Figure \ref{CondFrank500}
illustrates similar results for the Frank copula. Since the performances of the model with sample sizes $n=250~\mbox{and}~1000$ for both copula families were analogous, here we omit the results.
In Figures \ref{CondNorm500} and \ref{CondFrank500}, panels (a), (b), (c) and (d) show the scatter plots and histograms of the simulated data and the predictive samples obtained using the first calibration function; while panels (e), (f), (g) and (h) show the scatter plots and histograms of the simulated data and the predictive sample obtained using the second calibration function.
The comparison between the simulated and predictive outputs highlights the excellent fit of the Bayesian nonparametric conditional copula model using either calibration function and for different copula families.
The model performance appears to be consistent across both copula families, demonstrating that the approach is suitable to model different dependence patterns and tail structures.
\begin{figure}[h!]
\centering
\subfigure[Gaussian Copula]
{\includegraphics[width=6.5cm]{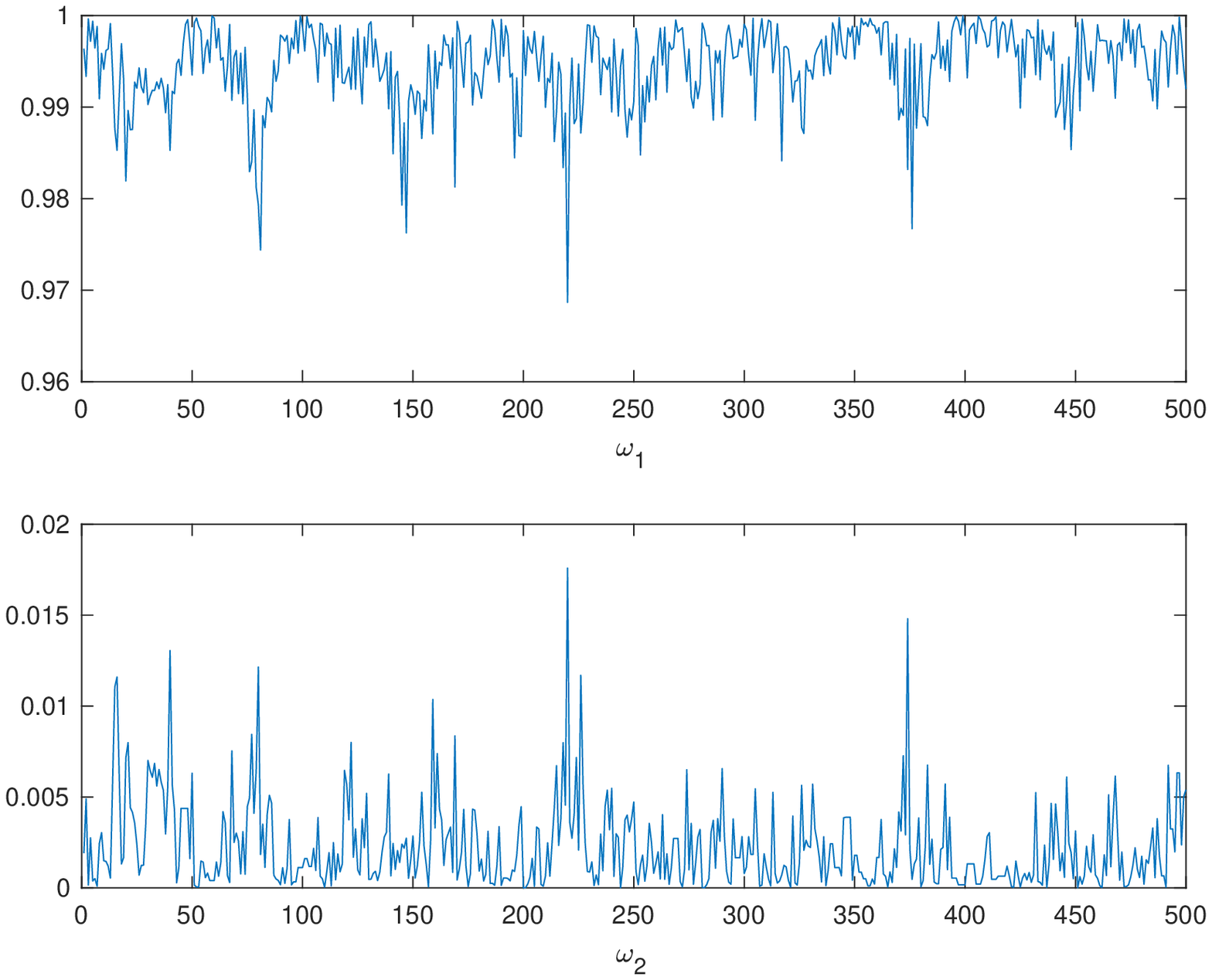}}
\subfigure[Frank Copula]
{\includegraphics[width=6.5cm]{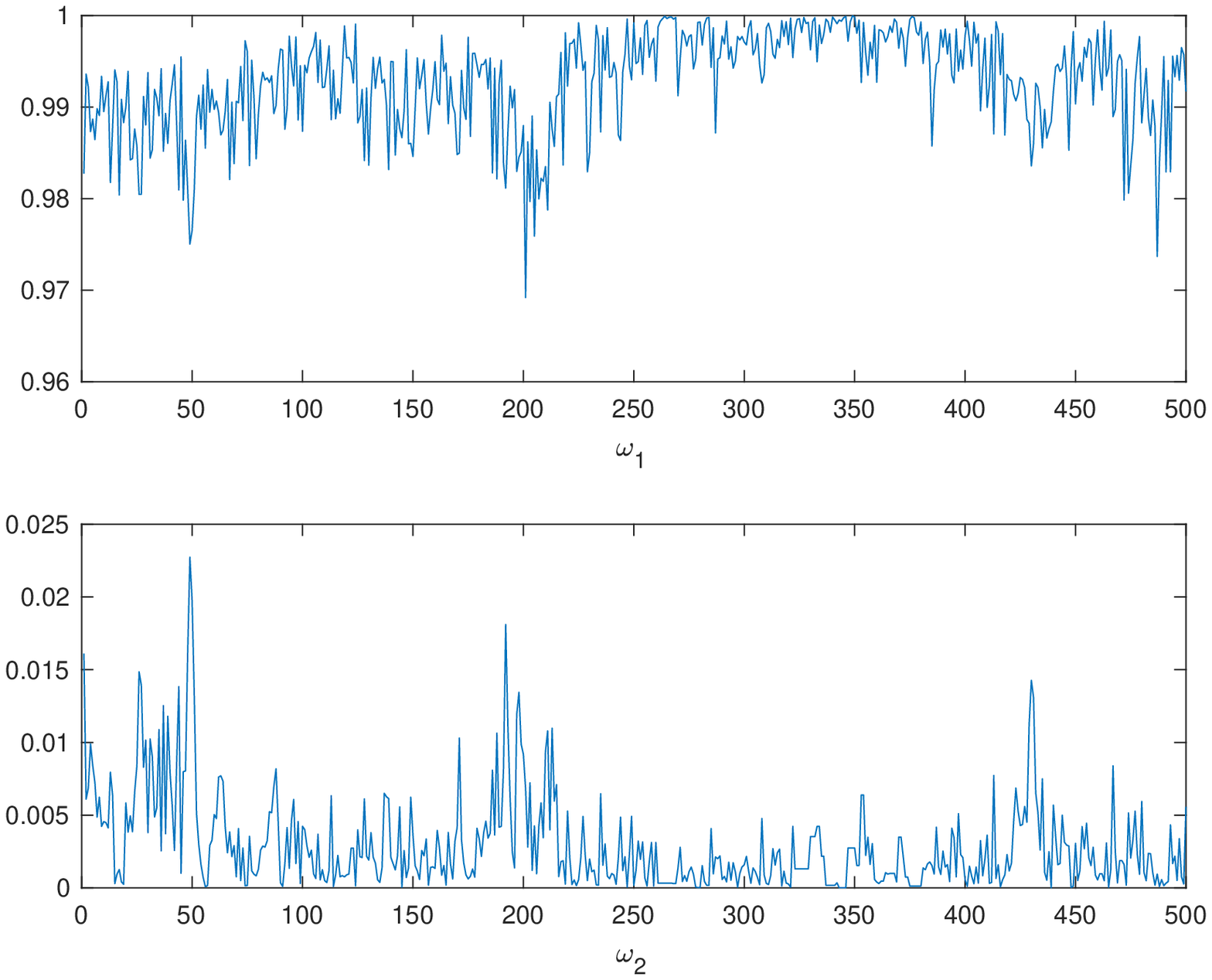}}
\caption{Plots of the values of the first two most significant weights ($y$-axis) generated at each MCMC iteration ($x$-axis) for the Gaussian (left) and for the Frank copula (right) with sample size $n=500$. The values of the first weight are plotted in the top panels, while the values of the second weight are plotted in the bottom panels.}
\label{WeiSim}
\end{figure}
Figure \ref{NComSim} shows the plot of the number of components generated at each MCMC iteration for both the Gaussian and the Frank copula. In Figure \ref{NComSim}, we focus on the first calibration function, since the second calibration function gave similar results. Table \ref{TabNC} shows the summary statistics of the number of components generated at each MCMC iteration for both copulas, indicating that the posterior median of the number of components is equal to $2$. For the two most significant components, we estimated the weights generated at each MCMC iteration for both copulas. In Figure \ref{WeiSim} we show the trace plots of the last $500$ iterations of the first two weights, as defined in Section (\ref{Gibbs}).
Figure \ref{WeiSim} suggests that the first weight is much more important than the second weight, since the first weight tends to take values close to $1$, while the second weight takes values close to $0$. 
For each of the two components we also estimated the posterior mean of the copula correlation coefficient $\rho_j$ (defined in eq. \ref{mixGau}), obtaining, for the Gaussian copula, a value of $0.6701$ for the first component and $-0.9860$ for the second component. On the other hand, for the Frank copula we obtained a posterior mean of $0.7941$ for the first component and $-0.9722$ for the second component, respectively.

\begin{table}[h!]
\centering
\begin{tabular}{lcccccc}
\hline
& Min. & 1st Quant. &  Median & Mean & 3rd Quant. & Max. \\
\hline
Gaussian & 1 & 1 & 2 & 2.102 & 3 & 7 \\
Frank & 1 & 2 & 2 & 2.546 & 3 & 7 \\ 
\hline
\end{tabular}
\caption{Summary statistics of the number of components generated at each MCMC iterations for the first calibration function.}
\label{TabNC}
\end{table}

\section{Real Data application}
\label{RealData}

We now apply the proposed Bayesian nonparametric conditional copula method to a sample of $839$ adolescent twin pairs, which is a subset of the National Merit Twin Study \citep{LoeNic09, LoeNic14}.
The dataset contains questionnaire data from 17 years old twins and their parents, where the twins were identified among $600.000$ US high school juniors who took part to the National Merit Scholarship Qualifying Test (NMSQT).

\begin{figure}[h!]
 \centering
   \subfigure[\scriptsize{Real data}]
   {\includegraphics[width=3.5cm]{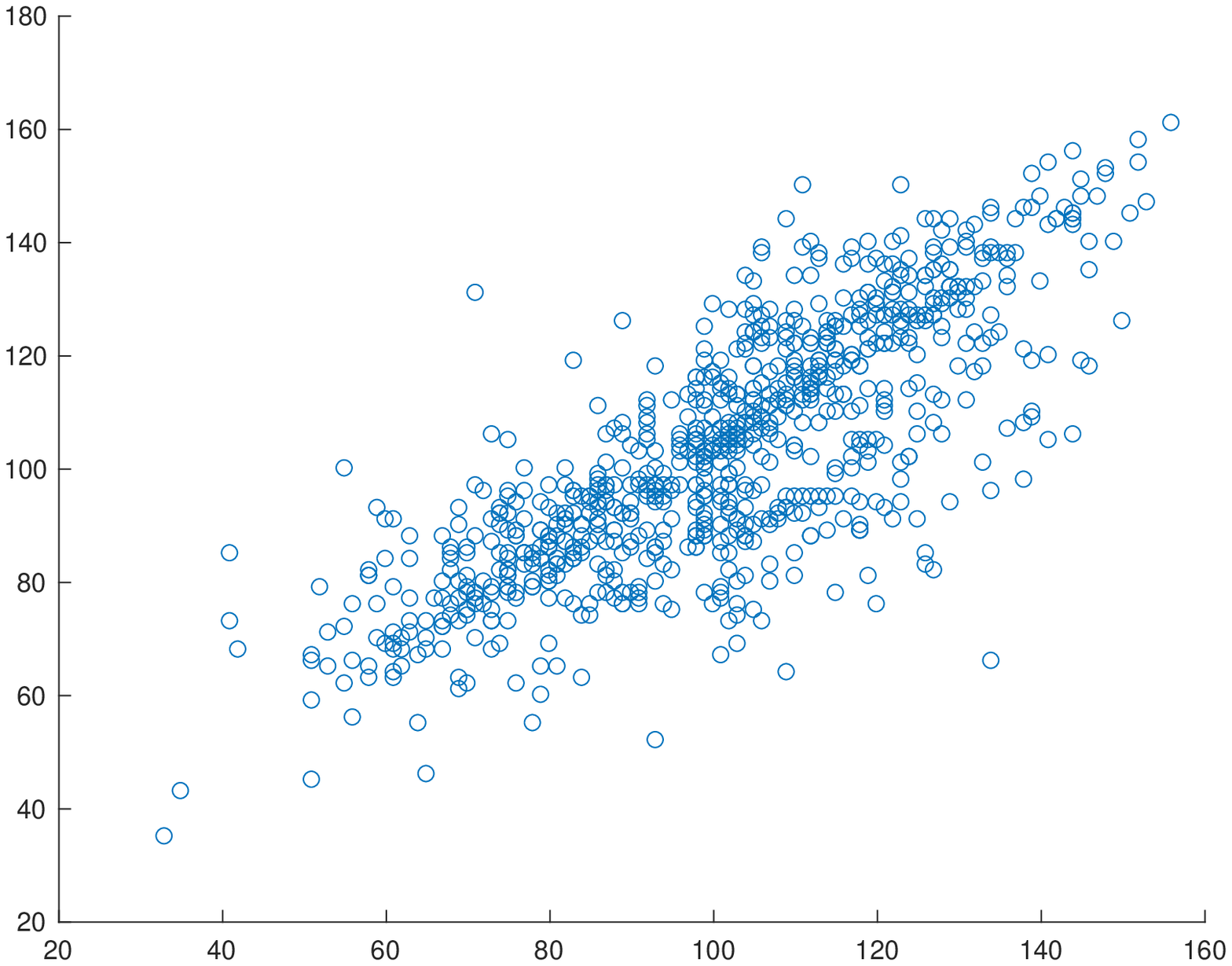}}
    \subfigure[\scriptsize{Transformed pseudo-observations}]
   {\includegraphics[width=3.5cm]{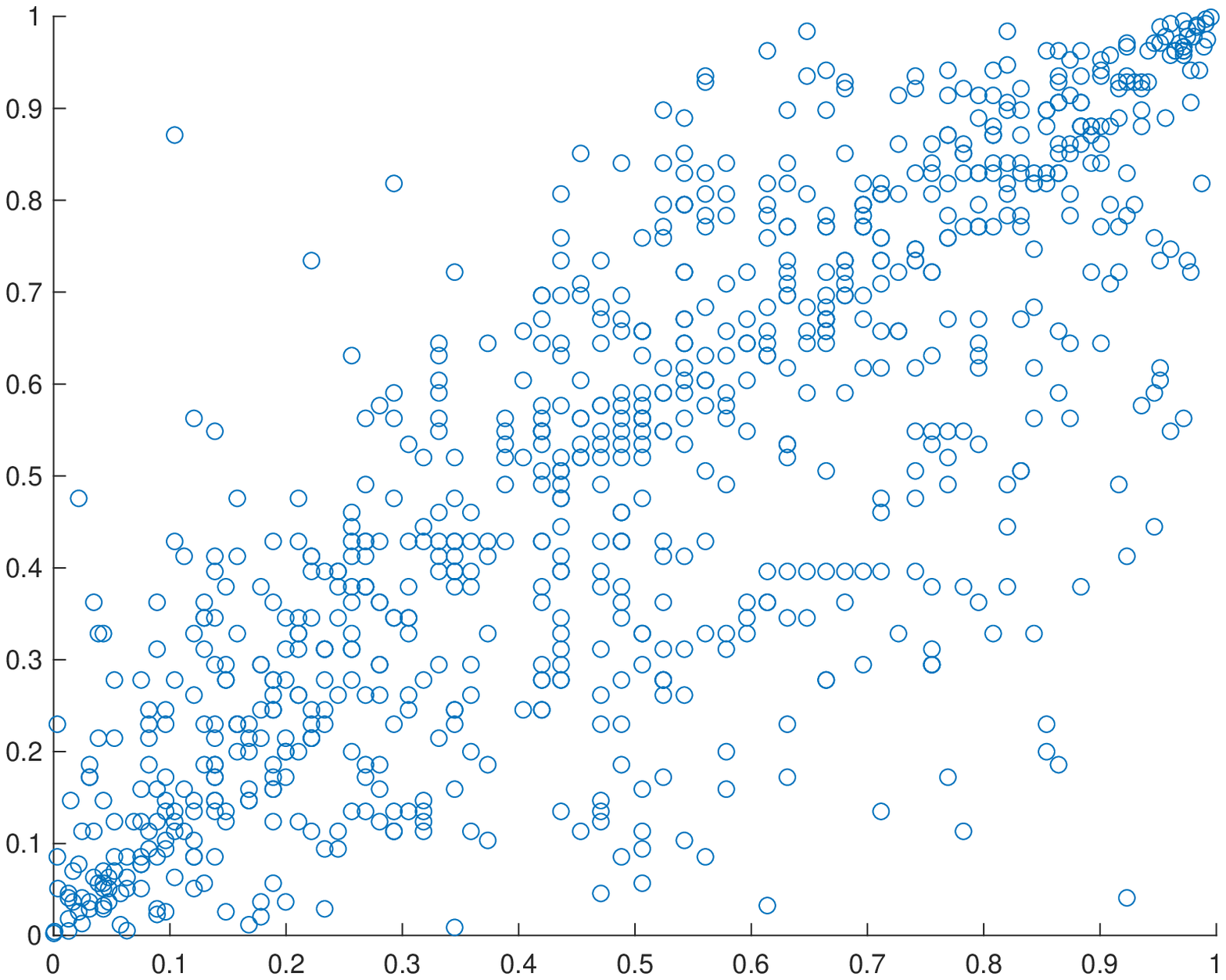}}
      \subfigure[\scriptsize{Predictive sample}]
   {\includegraphics[width=3.5cm]{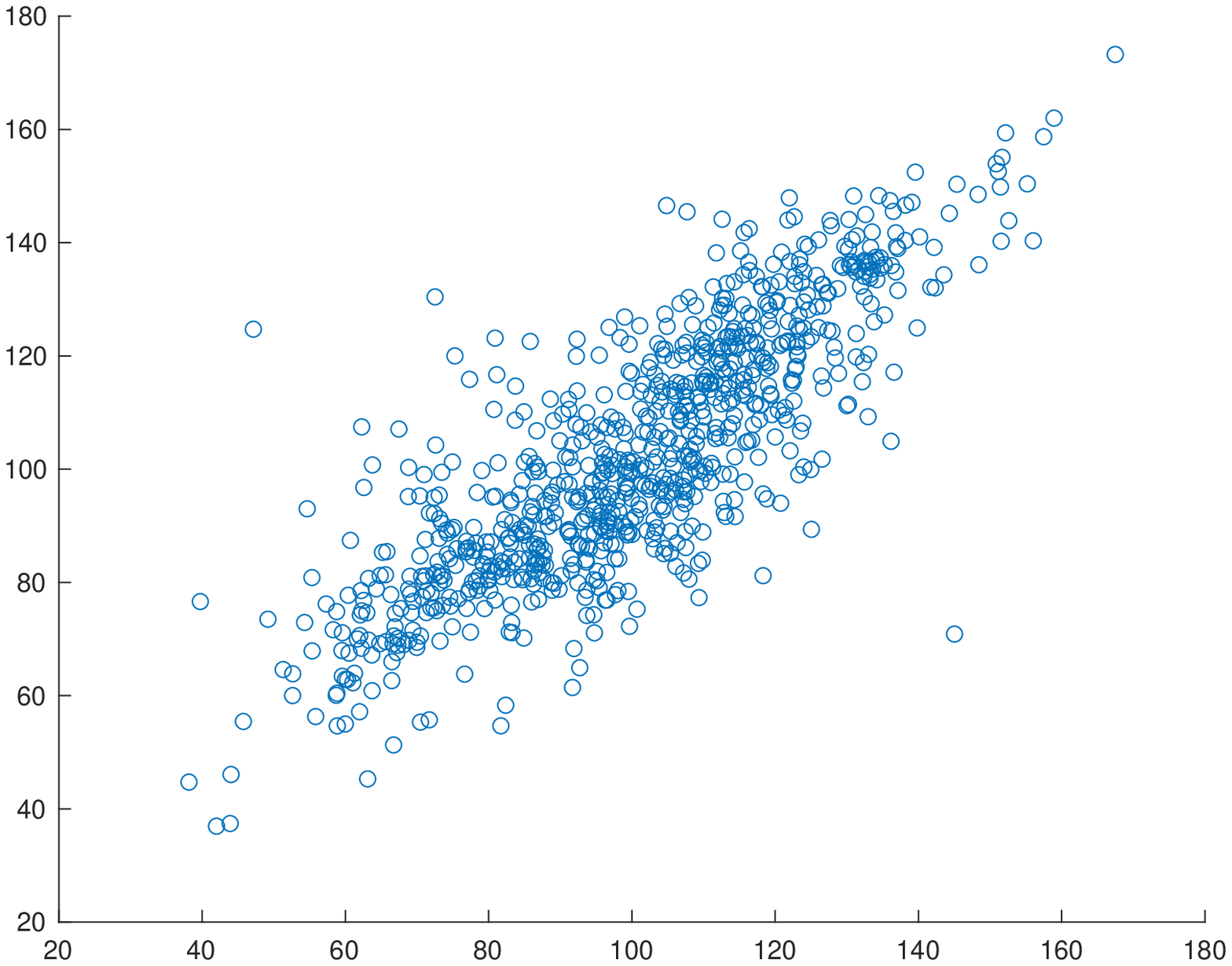}}
    \subfigure[\scriptsize{Predictive transformed sample}]
   {\includegraphics[width=3.5cm]{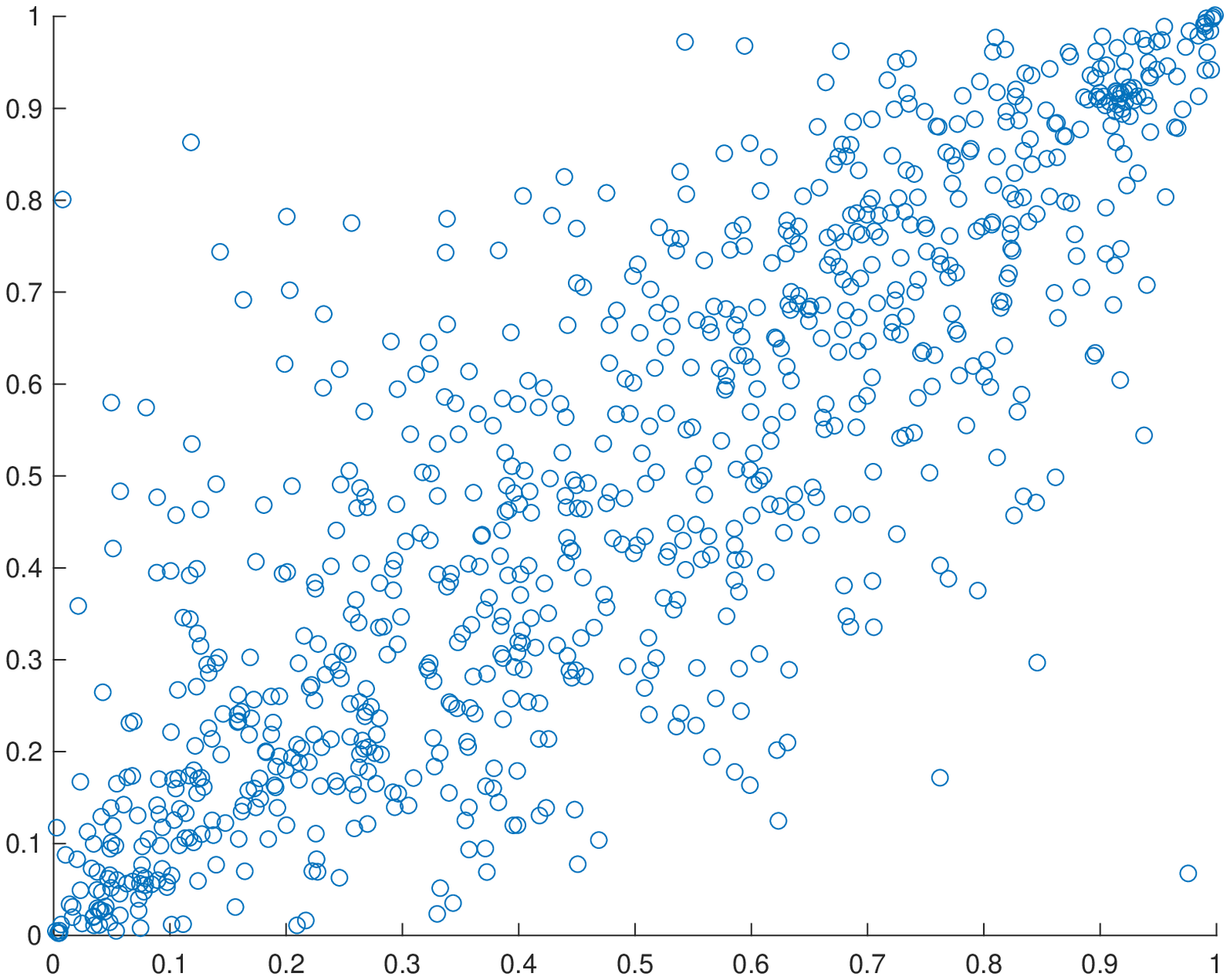}}
   \subfigure[\scriptsize{Histogram of real data}]
  {\includegraphics[width=3.5cm]{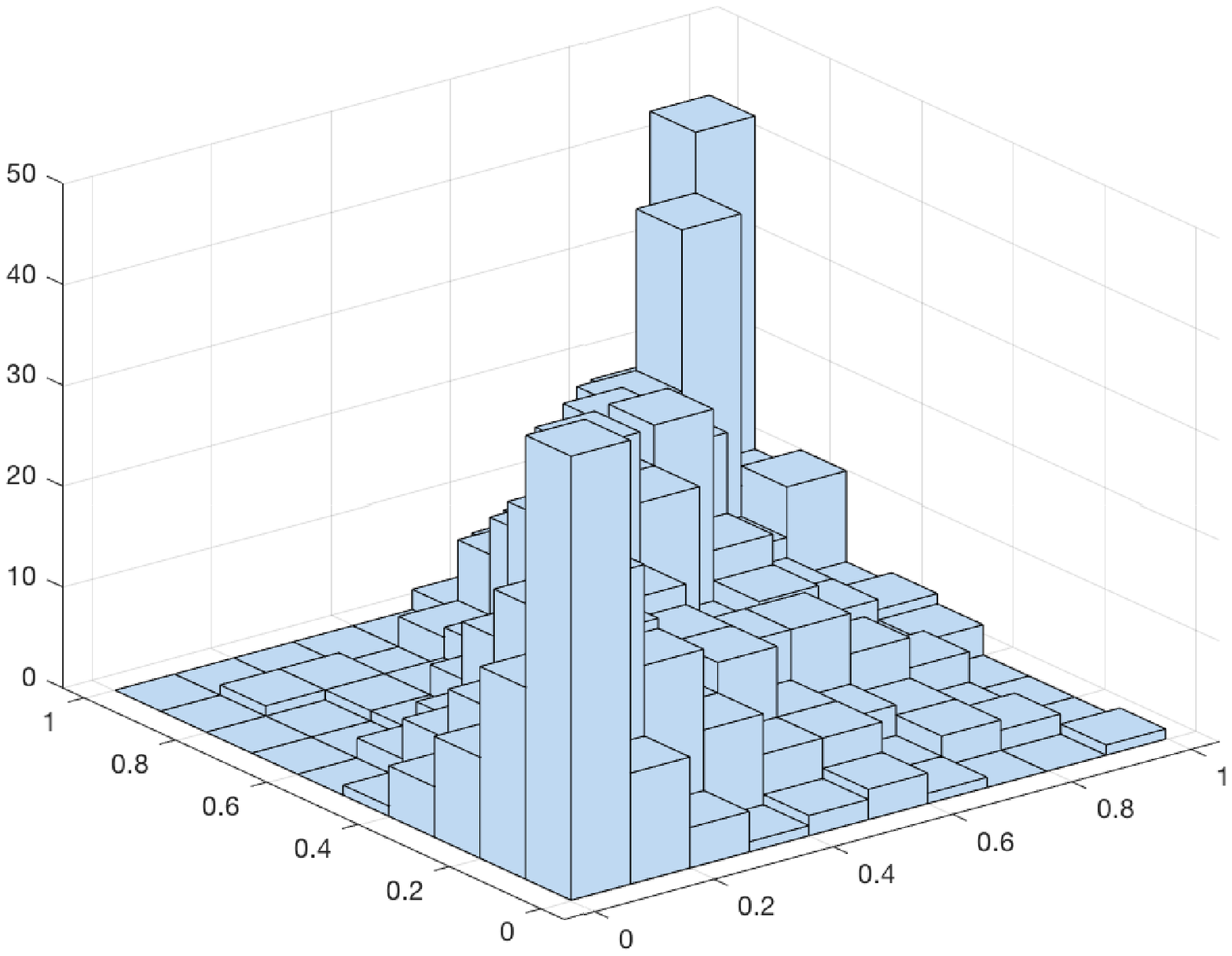}}
    \subfigure[\scriptsize{Predictive histogram}]
  {\includegraphics[width=3.5cm]{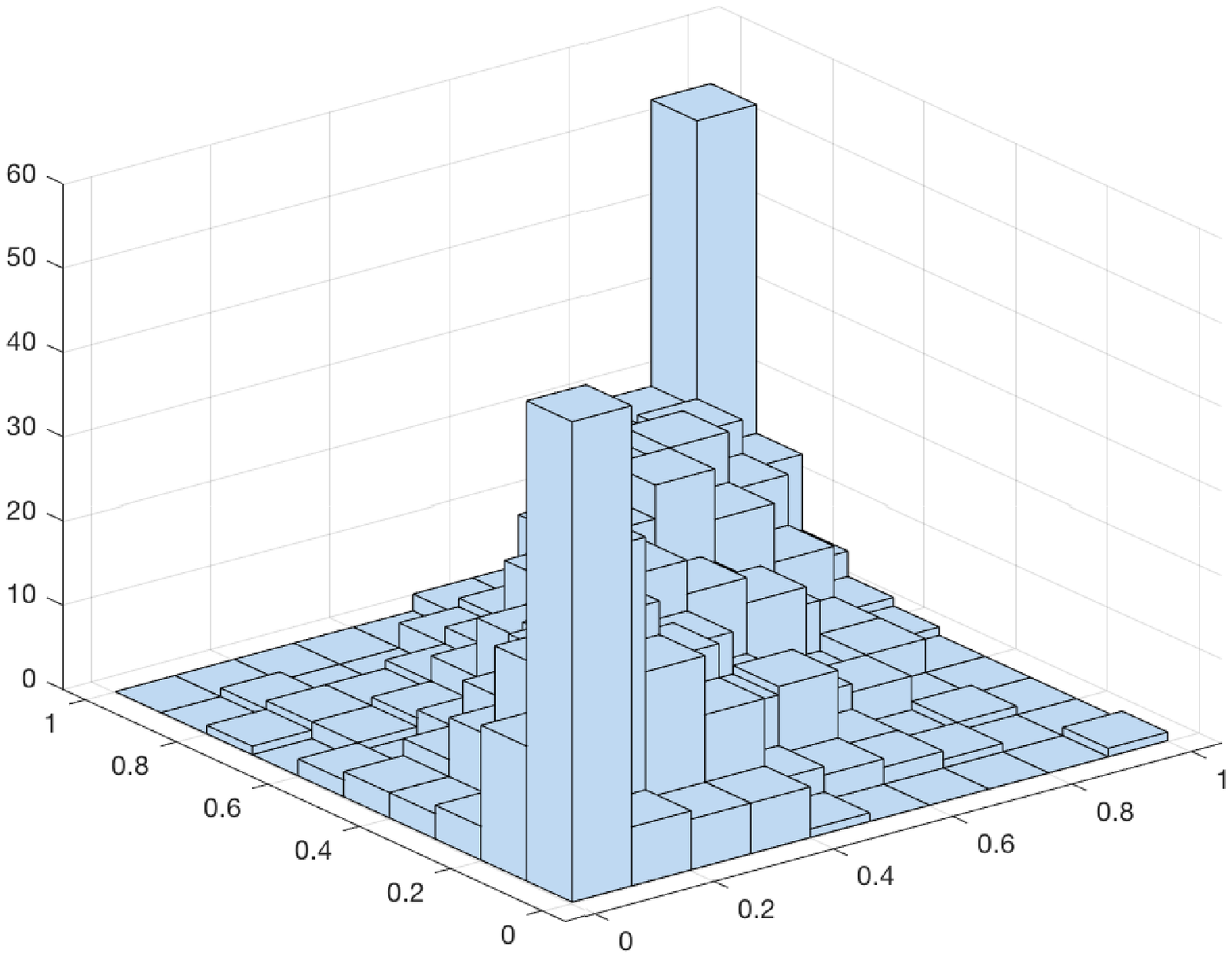}}
\caption{\footnotesize{Panels (a) and (b): scatterplots of the twins' overall scores for the real and pseudo-observations with respect to the mother's level of education; panels (c) and (d): scatterplots of the predictive and transformed predictive samples; panels (e) and (f): histograms of the real data and the predictive sample.}}
\label{DataEstimMother}
\end{figure}

The NMSQT was designed to measure cognitive aptitude, that is students' readiness for future intellectual or educational pursuits.
The participants to the test include identical twins and same-sex fraternal twins who were asked to fill in a complete questionnaire in order to understand their school performance and attitude.
Our purpose is to examine whether the relationship between twins' cognitive ability, measured by the NMSQT, is influenced by their socioeconomic status, measured by parent education and parental income.
The variables we considered from this study are the overall measures of each twin's performance at school (obtained as the sum of individual scores in English Usage, Mathematics Usage, Social Science Reading, Natural Science Reading and Word Usage/Vocabulary), the mother's and father's level of education and the family income.
The overall scores range from $30$ to $160$, the education covariates range from $0$ to $6$, while the family income covariate ranges from $0$ to $7$.
The levels of the education covariates correspond to: less than 8-th grade, 8-th grade, part high school, high school graduate, part college or junior college, college graduate, and graduate or professional degree beyond the bachelor's degree.
The levels of the income covariate correspond to values going from less than \$5000 per year to over than \$25000 per year.

\begin{figure}[h!]
 \centering
\subfigure[\scriptsize{Real data}]
   {\includegraphics[width=3.5cm]{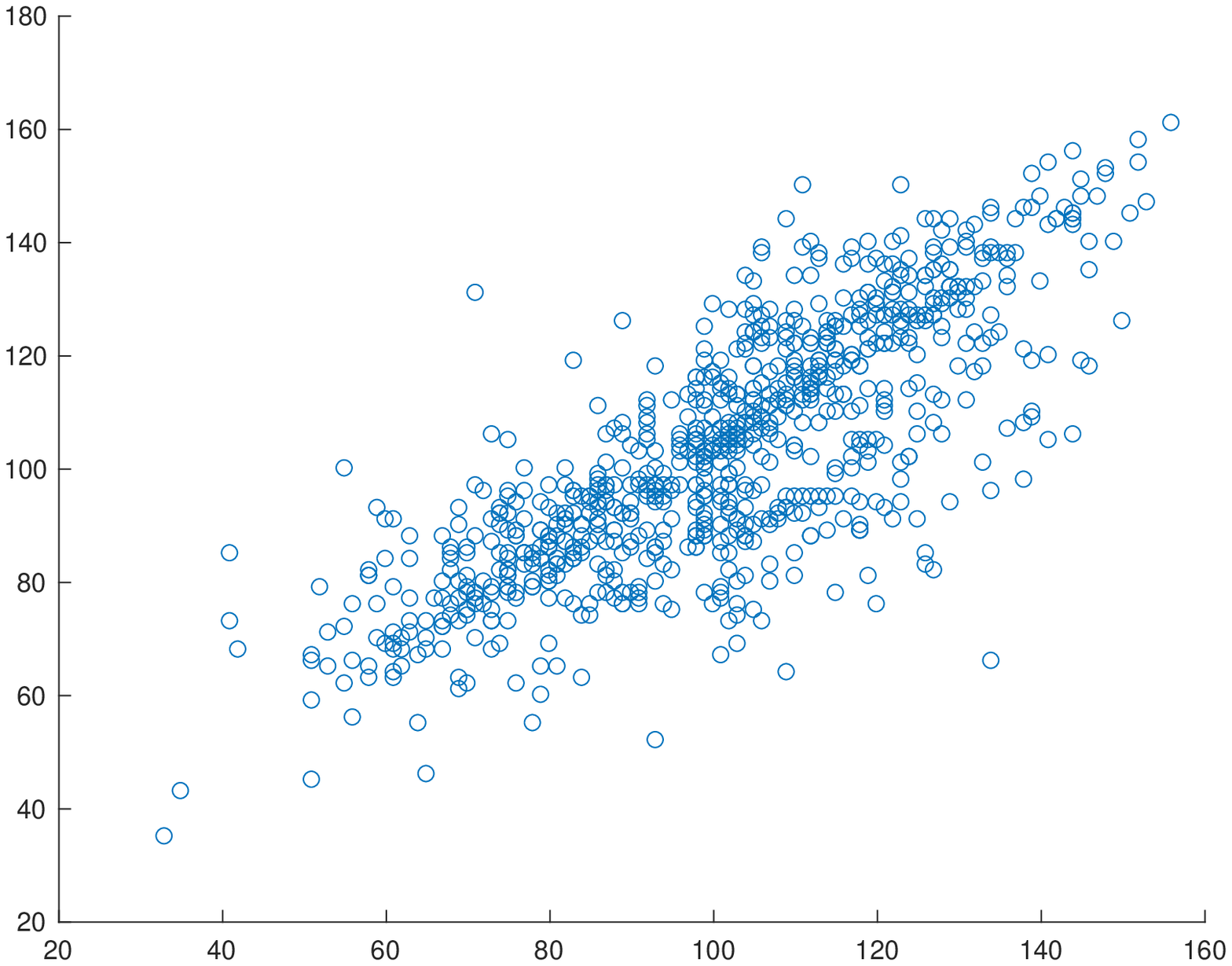}}
    \subfigure[\scriptsize{Transformed pseudo-observations}]
   {\includegraphics[width=3.5cm]{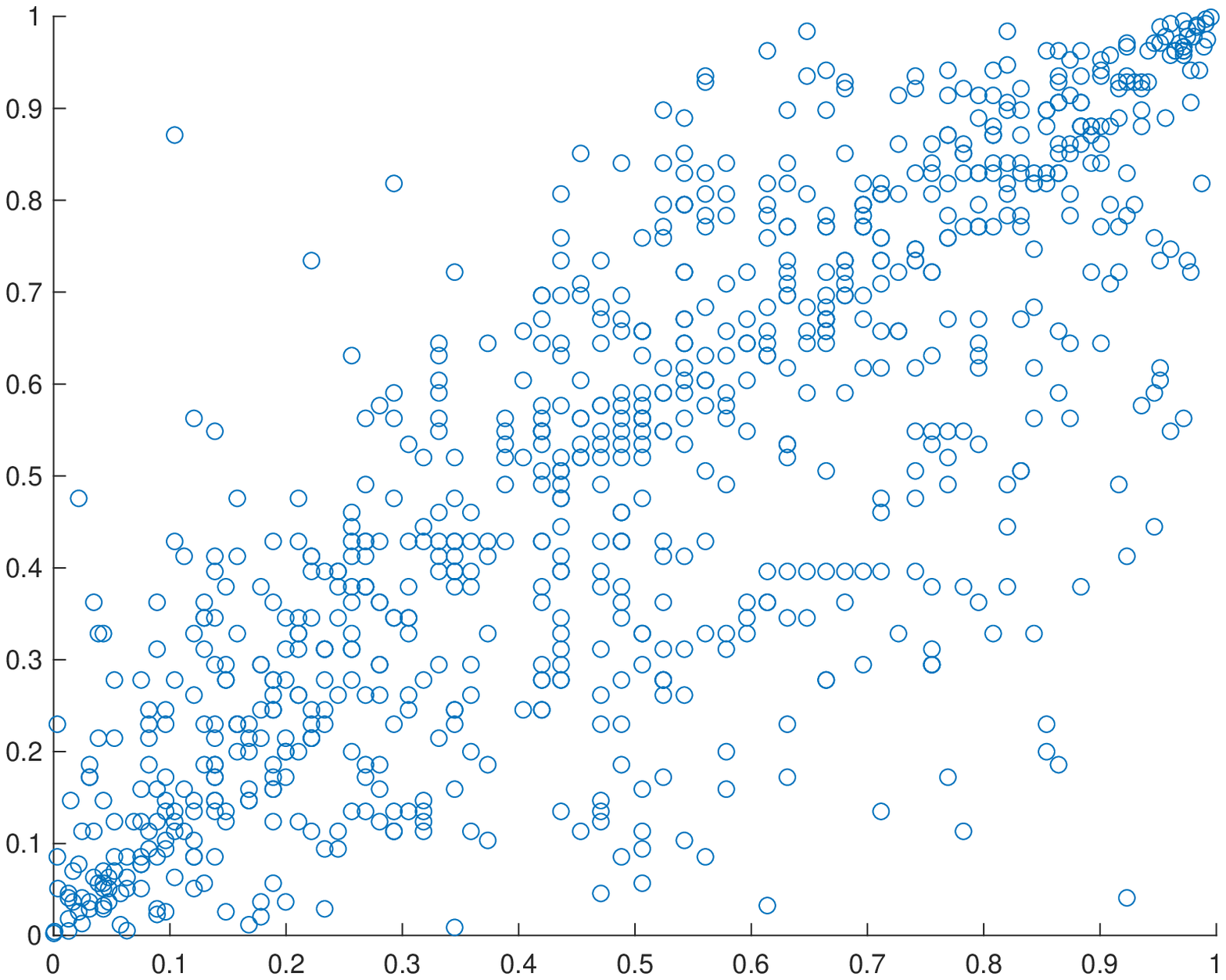}}
      \subfigure[\scriptsize{Predictive sample}]
   {\includegraphics[width=3.5cm]{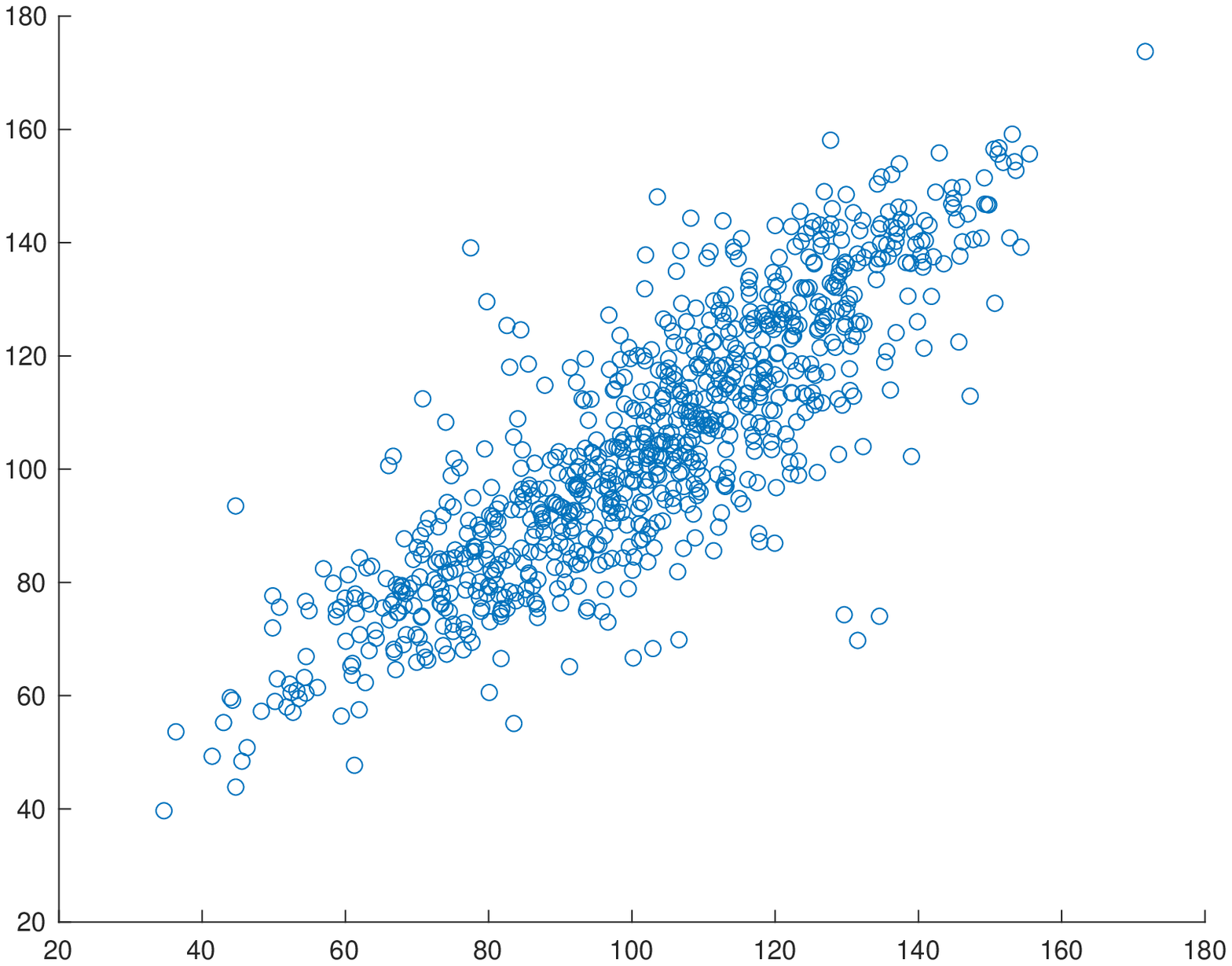}}
    \subfigure[\scriptsize{Predictive transformed sample}]
   {\includegraphics[width=3.5cm]{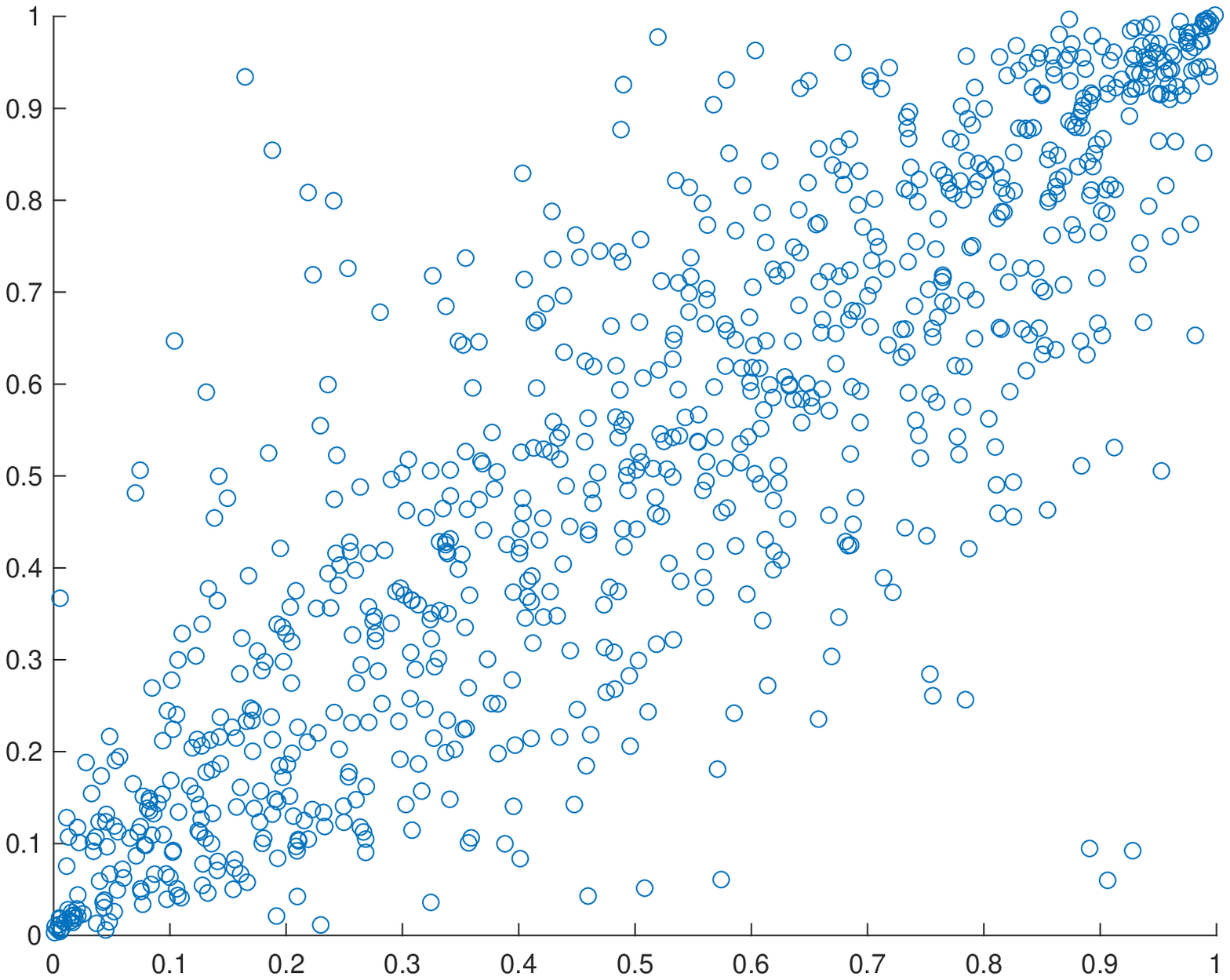}}
   \subfigure[\scriptsize{Histogram of real data}]
  {\includegraphics[width=3.5cm]{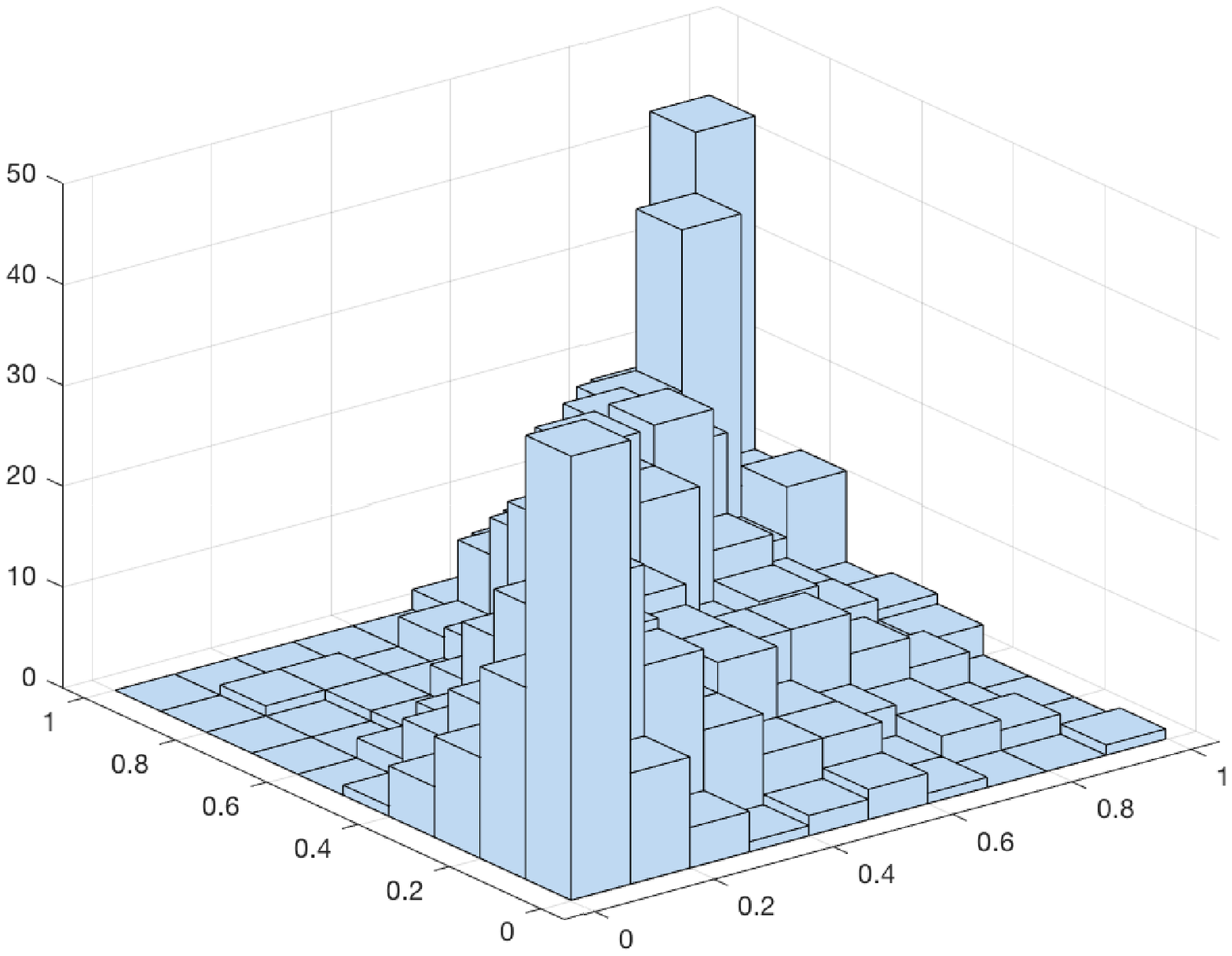}}
    \subfigure[\scriptsize{Predictive histogram}]
  {\includegraphics[width=3.5cm]{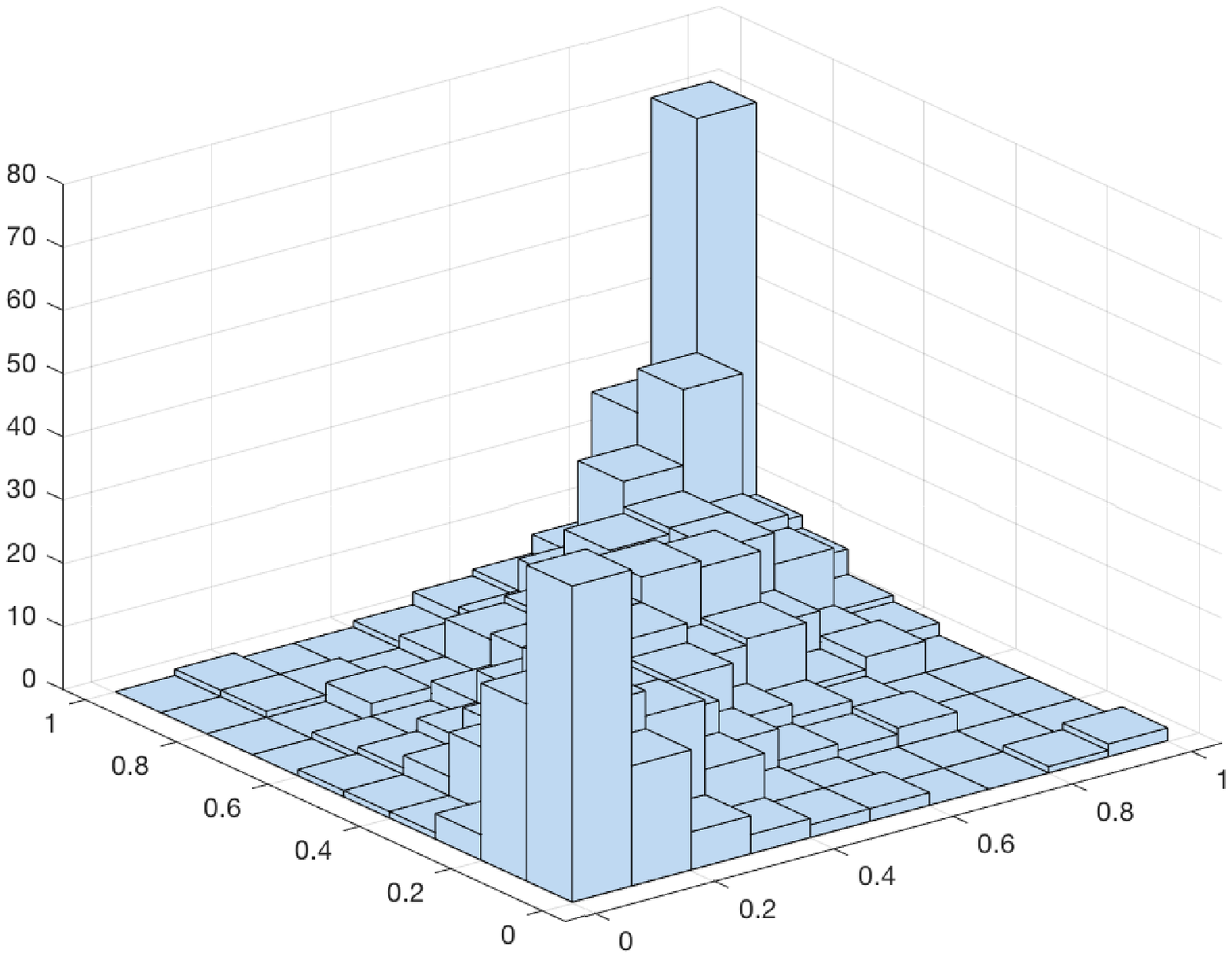}}
\caption{\footnotesize{Panels (a) and (b): scatterplots of the twins' overall scores for the real and pseudo-observations with respect to the father's level of education; panels (c) and (d): scatterplots of the predictive and transformed predictive samples; panels (e) and (f): histograms of the real data and the predictive sample.}}
\label{DataEstimFather}
\end{figure}

As discussed in Section \ref{Intro}, the scatterplots in Figure \ref{DATA} clearly show that there is a positive correlation between the twins' school performance and the strength of dependence varies according to the values of a covariate, which is the mother's (panel (a)) or father's level of education (panel (b)) or the family income (panel (c)).
In Figure \ref{DATA} the effect of the covariates is illustrated by dots of different colours, where we notice that most of the light brown dots are grouped in the upper right corner, while the dark brown dots lie in the bottom left corner. Therefore, the higher the parents' education or family income, the higher the twins' school performance.
In order to model the effect of a covariate, such as the mother's and father's education and family income, on the dependence between the overall scores of the twins, we implement the Bayesian nonparametric conditional copula model.

\begin{figure}[h!]
 \centering
   {\includegraphics[width=8cm]{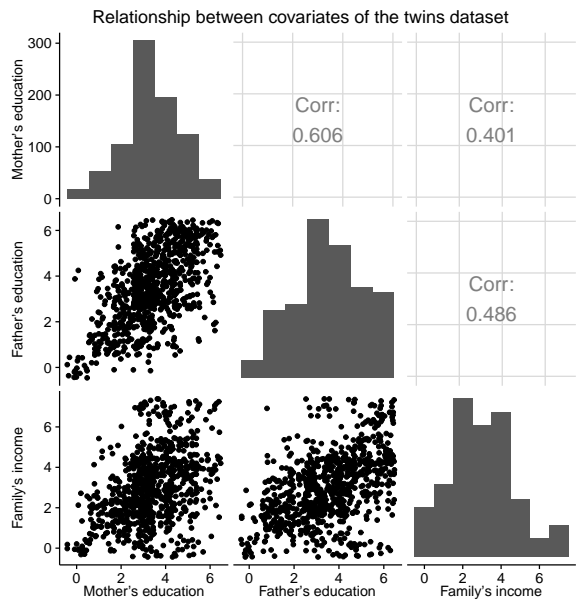}}
\caption{\footnotesize{Relationship between the covariates of the twins dataset. The lower-triangular panels represent pairwise scatterplots, the upper-triangular panels show pairwise Pearson's correlation coefficients, and the diagonal panels represent the histograms of each covariate. (Note that jittering was used in the scatterplots to prevent overplotting).}}
\label{Covariates}
\end{figure}

Figure \ref{Covariates} shows the relationship between the covariates of the twins dataset, where the lower-triangular panels represent pairwise scatterplots, the upper-triangular panels show pairwise Pearson's correlation coefficients, and the diagonal panels represent the histograms of each covariate.  
The scatterplots and Pearson's correlation coefficients in Figure \ref{Covariates} indicate a rather strong positive correlation between each pair of covariates, especially between the mother's and father's level of education.
The high correlations indicate that the data do not contain much information on the independent effects of each covariate, and suggest the inclusion of only one of them in the model. 
For this reason we decided to include only one of the redundant covariates at a time.
Note that, with a different dataset, the methodology may be extended to include more than one covariate. However, model specification issues and increased computational costs must be carefully considered.
    
\begin{figure}[h!]
 \centering
  \subfigure[\scriptsize{Real data}]
   {\includegraphics[width=3.5cm]{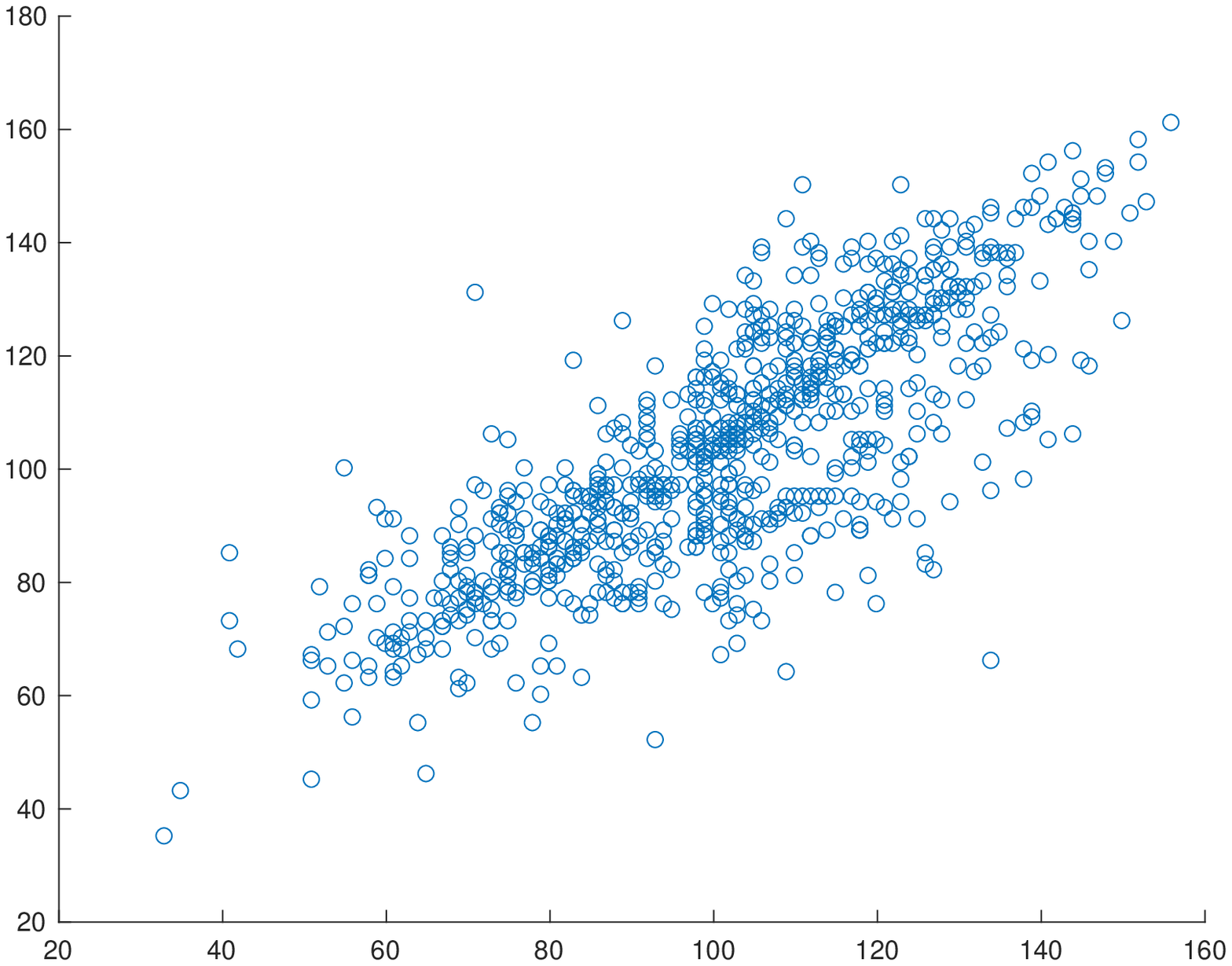}}
    \subfigure[\scriptsize{Transformed pseudo-observations}]
   {\includegraphics[width=3.5cm]{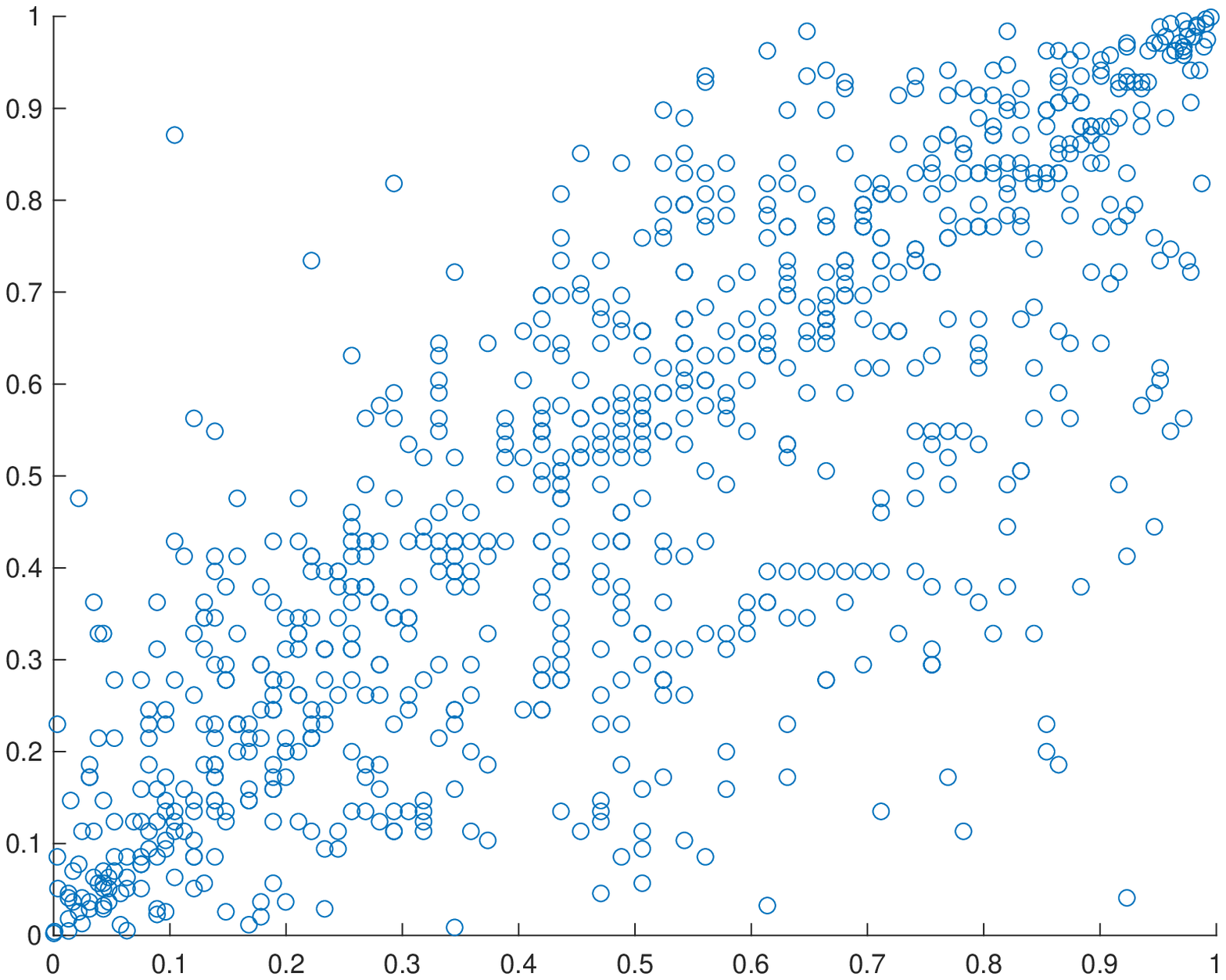}}
      \subfigure[\scriptsize{Predictive sample}]
   {\includegraphics[width=3.5cm]{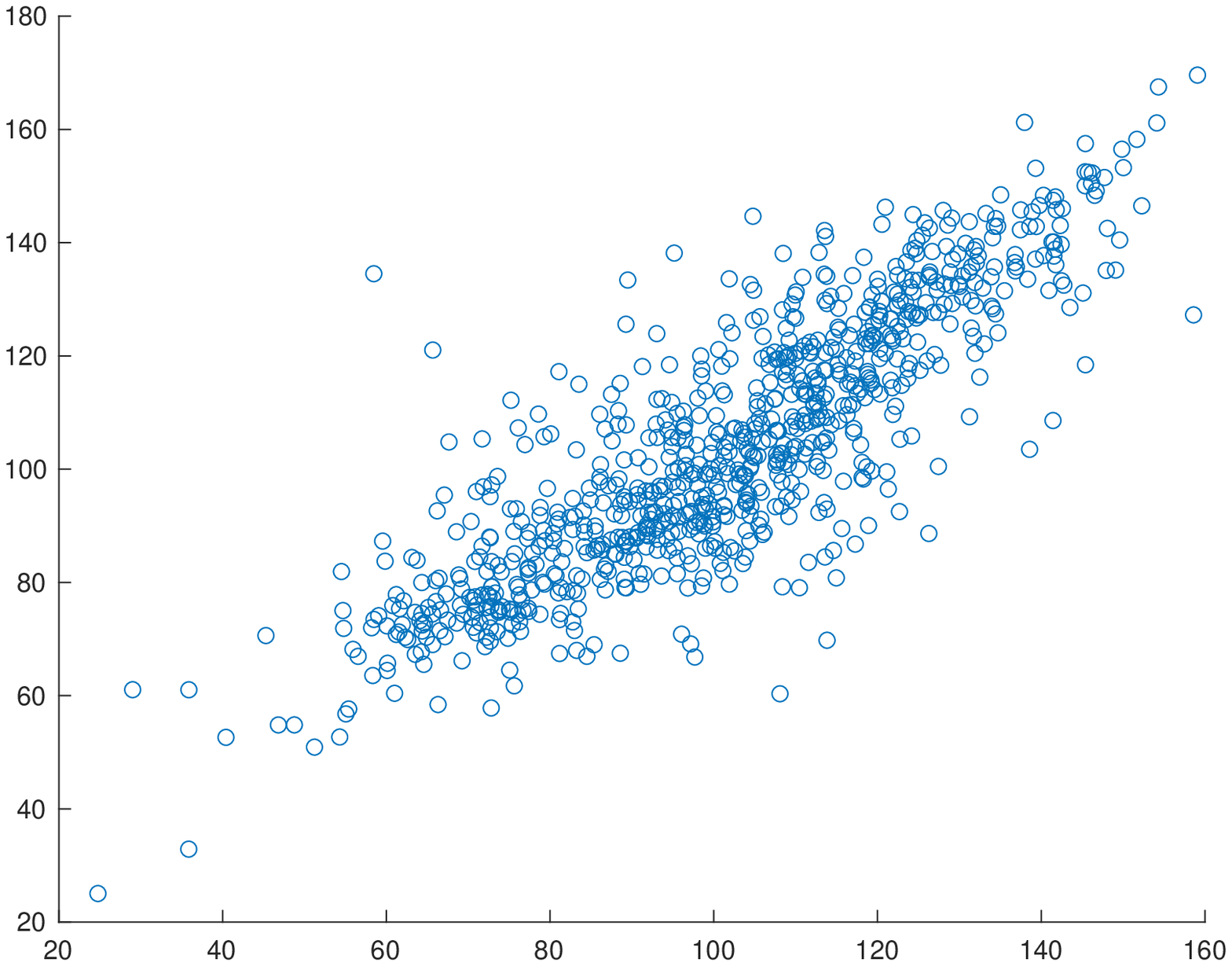}}
    \subfigure[\scriptsize{Predictive transformed sample}]
   {\includegraphics[width=3.5cm]{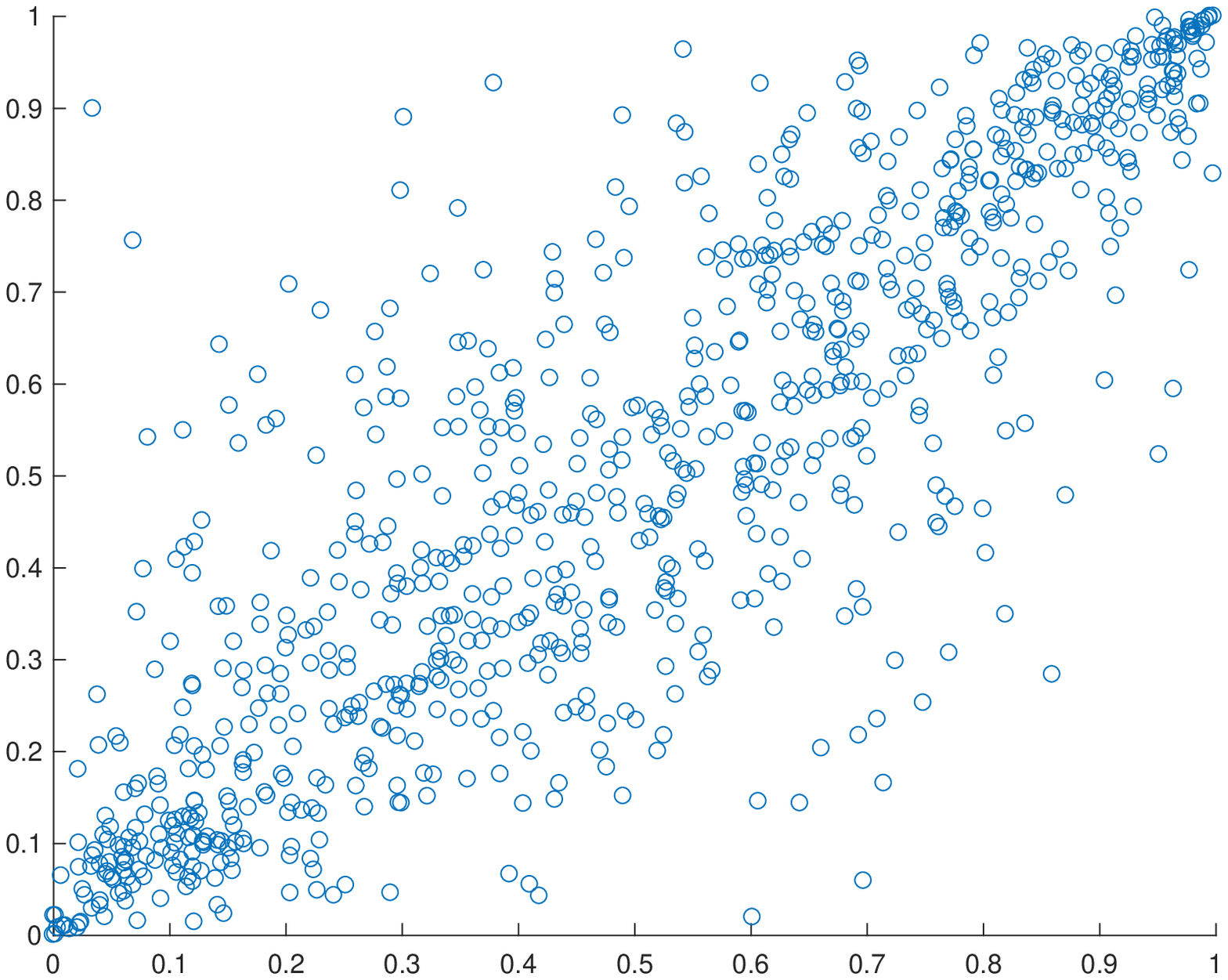}}
   \subfigure[\scriptsize{Histogram of real data}]
  {\includegraphics[width=3.5cm]{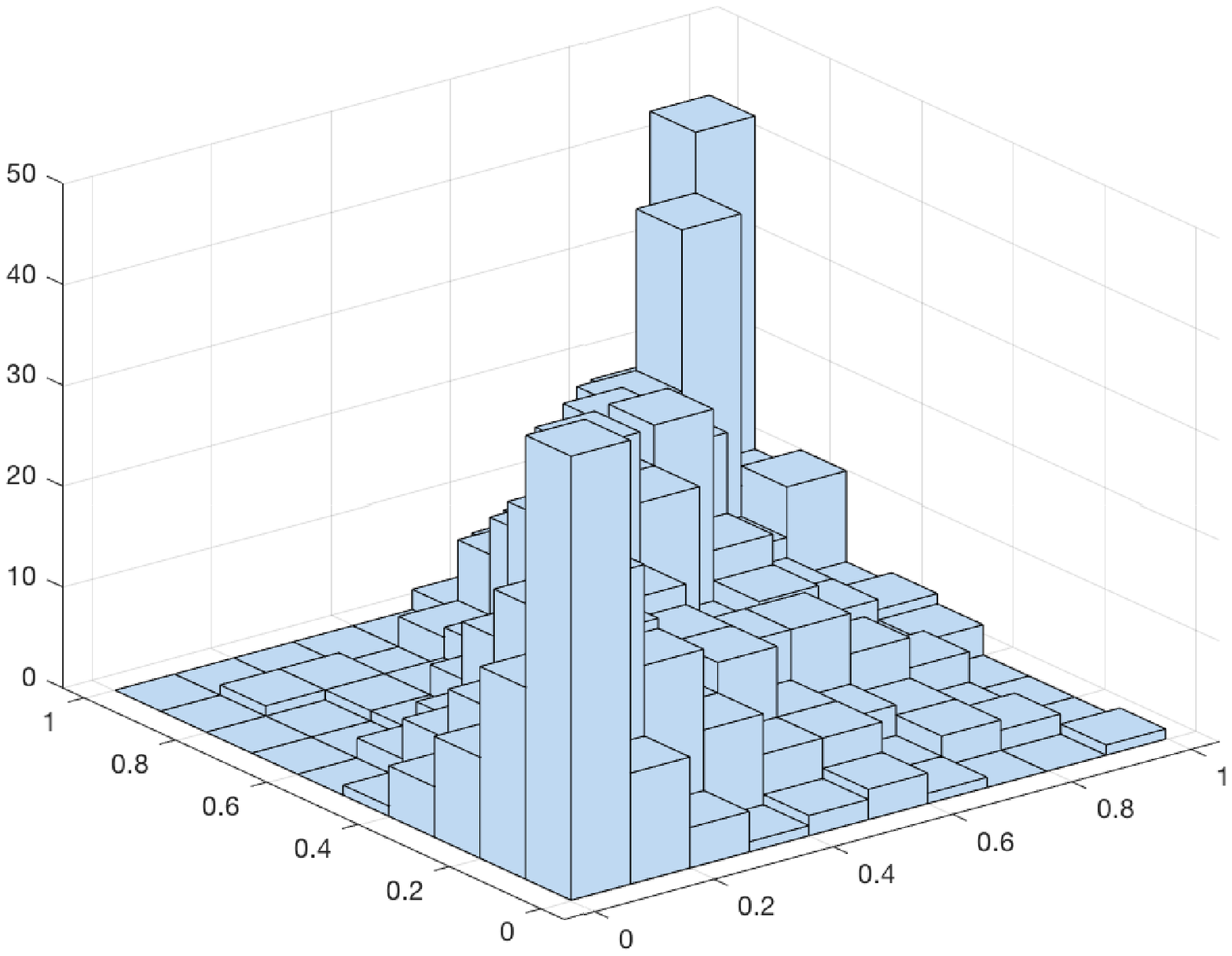}}
    \subfigure[\scriptsize{Predictive histogram}]
  {\includegraphics[width=3.5cm]{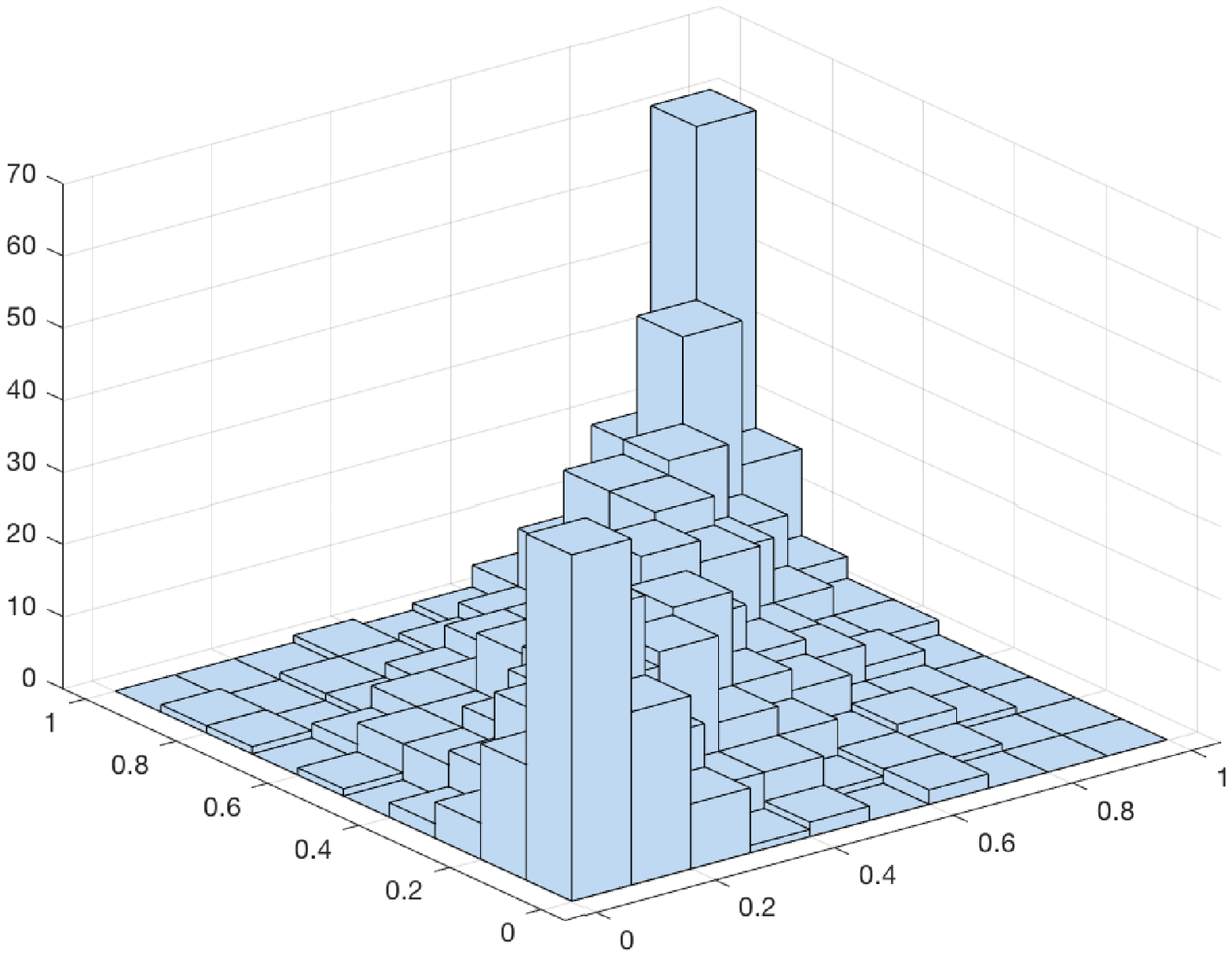}}
\caption{\footnotesize{Panels (a) and (b): scatterplots of the twins' overall scores for the real and pseudo-observations with respect to the family income; panels (c) and (d): scatterplots of the predictive and transformed predictive samples; panels (e) and (f): histograms of the real data and the predictive sample.}}
\label{DataEstimIncome}
\end{figure}

\begin{figure}
\centering
\subfigure[\scriptsize{Mother's level of education}]
{\includegraphics[width=5.5cm]{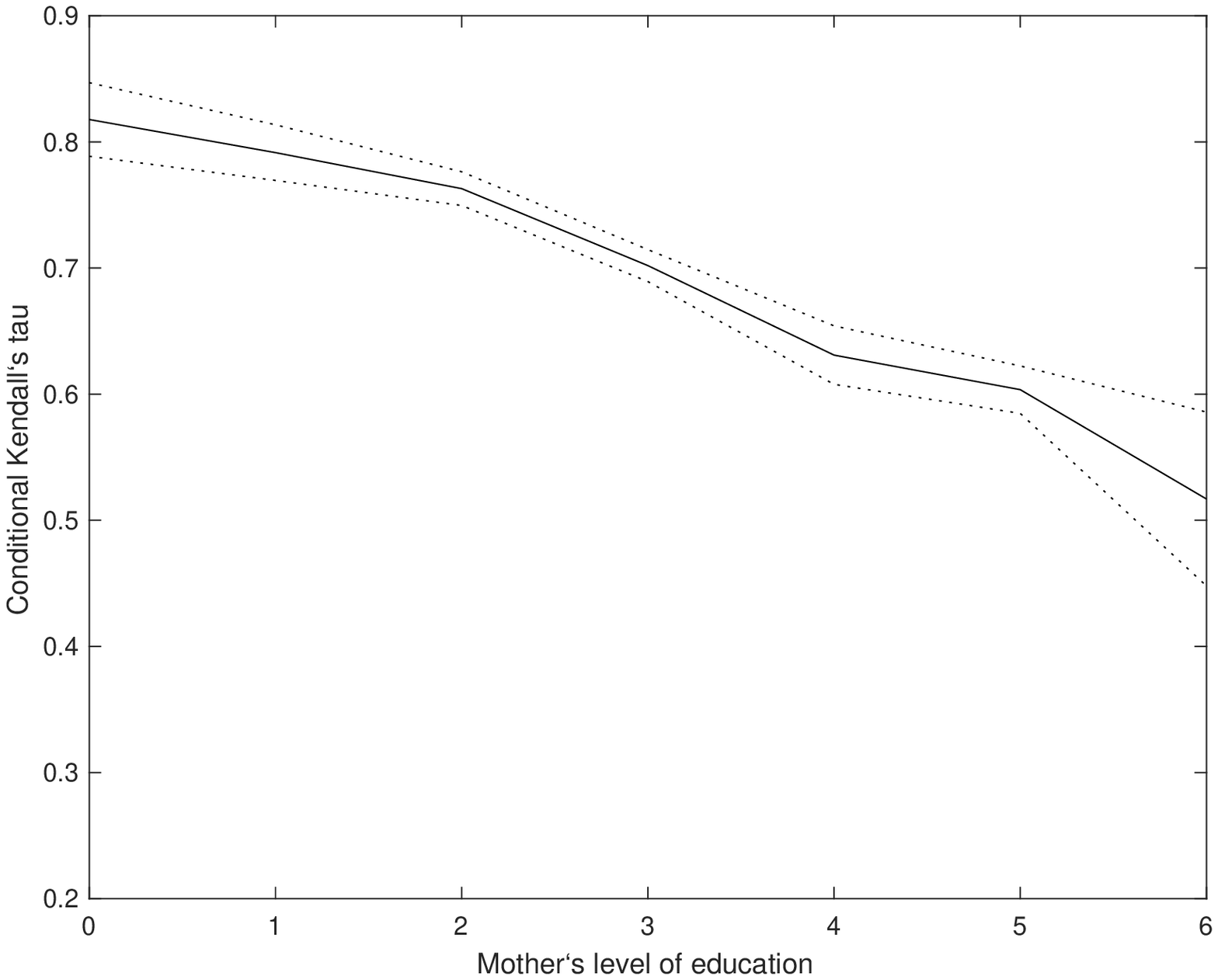}}
\subfigure[\scriptsize{Father's level of education}]
{\includegraphics[width=5.5cm]{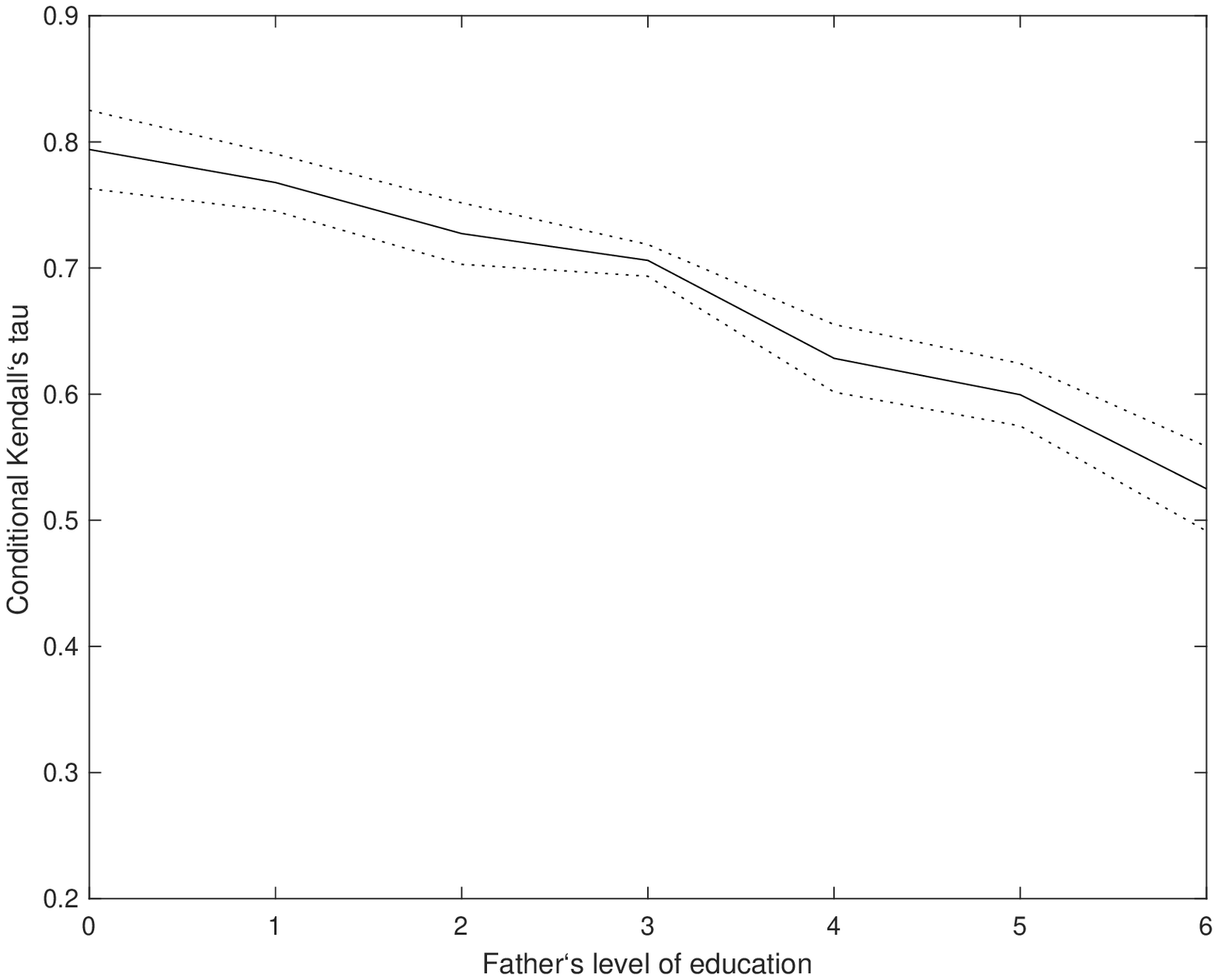}}
\subfigure[\scriptsize{Family income}]
{\includegraphics[width=5.5cm]{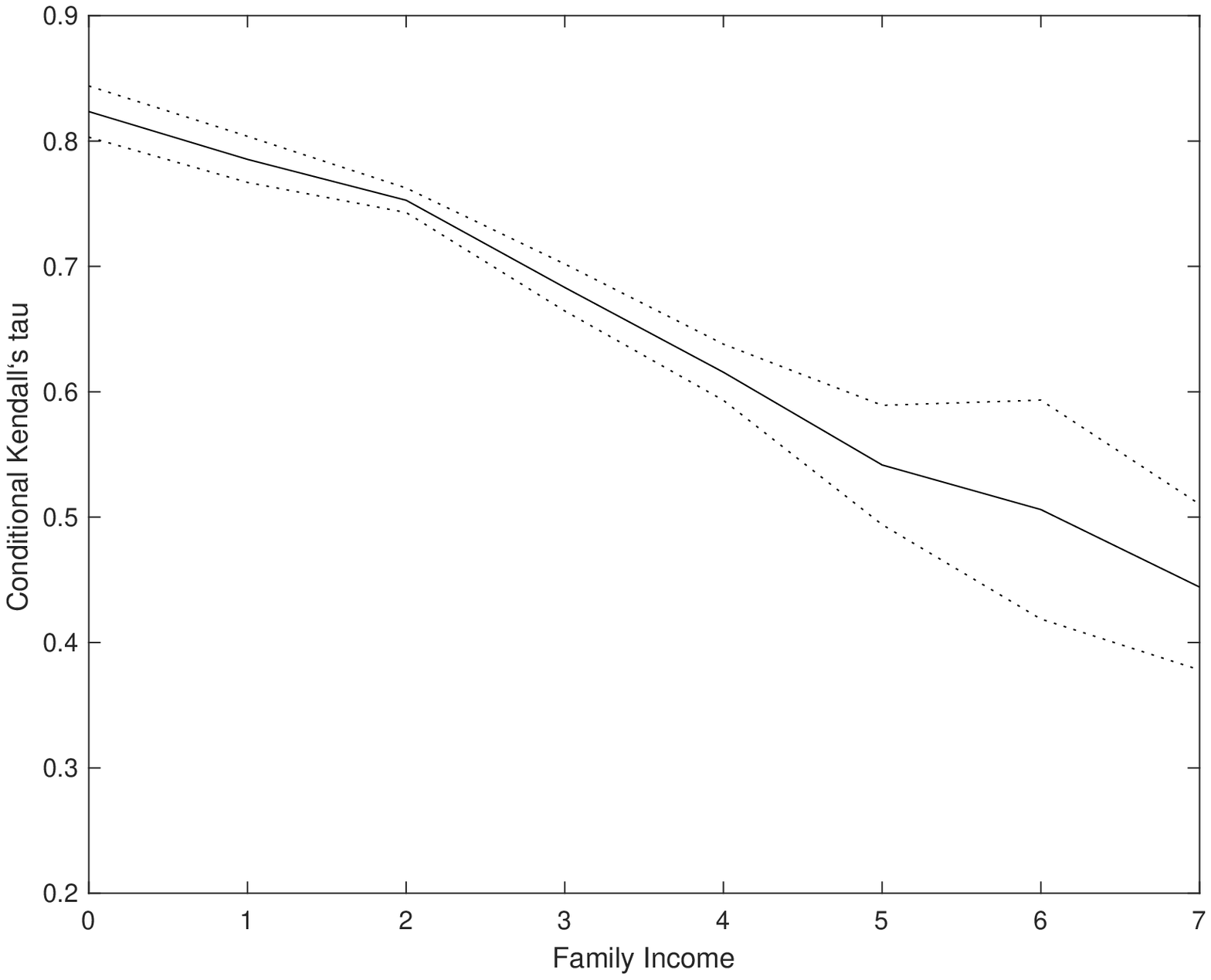}}
\caption{\footnotesize{Estimated Kendall's tau against the mother's (top panel) and father's level of education (middle panel) and the family income (bottom panel) and an approximate $95\%$ credible intervals (dotted lines)}}
\label{Kend}
\end{figure}

Adopting the same priors of the simulation studies, we run the Gibbs sampling algorithm described in Section \ref{Gibbs} for $4000$ iterations.
Figures \ref{DataEstimMother}, \ref{DataEstimFather} and \ref{DataEstimIncome} show, for the mother's and father's education and family income, respectively, the scatterplots of the twins' overall scores using the real and transformed pseudo-observations (panels (a) and (b)),
the scatterplots of the predictive and transformed predictive samples (panels (c) and (d)) and the histograms of the real and the predictive samples (panels (e) and (f)).
Note that the pseudo-observations are obtained using the nonparametric estimation approach described in Section \ref{Gibbs}. 
From the comparison between the scatterplots and histograms of the real and predictive samples obtained with the three different covariates, it emerges that the Bayesian nonparametrics conditional copula model accurately captures the tail structures and the dependence patterns between the twins' overall scores. Moreover, the posterior means of the number of mixture components for the conditional copula are $26.82$, $24.49$ and $27.11$, for the mother's and father's level of education and the family income, respectively, supporting the need for non-Gaussian copulas.
We note that the good performance of this approach in tail modelling makes it suitable to various applications focussing on extremes. To quantify the degree of dependence between the twns' scores, we use the conditional Kendall's tau, which is a nonparametric measures of correlation, known as concordance, between two ranked variables $(Y_1,Y_2)$ with respect to a covariate $X=x$. The conditional Kendall's tau takes the following form:
\begin{equation}
\tau(x)=4\int\int C_x(u_1,u_2)dC_x(u_1,u_2)-1 \notag
\end{equation}
where $C_x$ is the appropriate conditional copula. Figure \ref{Kend} shows Kendall's tau estimated from the model against the mother's (top panel) and
father's level of education (middle panel) and the family income (bottom panel), together with 95\% credible intervals.
The plots clearly illustrate the negative effect of all three covariates on the dependence between the twins' overall scores.
The effect is greater for the family income, where the Kendall's tau decreases from approximately $0.83$ to $0.45$, while for the parents' education levels the Kendall's tau decreases from approximately $0.8$ to $0.6$.
Therefore, the higher the parents' education and family income, the better the socioeconomic status and the higher the differences between the twins' school performances.
The cognitive aptitudes of twins from less advantaged families are more similar to each other than those from high income, highly educated families. 
Families of high socioeconomic status provide supportive and challenging environments, able to offer a wide range of opportunities and choices to their children, and allowing them to express themselves freely. 
Hence, twins raised in wealthy families are encouraged to develop differences in their traits, and may show rather dissimilar cognitive abilities, albeit high on average.
On the contrary, families of low socioeconomic status offer scarce opportunities to their children and may provide limiting and restrictive environments. 
In less advantaged families, twins cannot develop their full potential and individuality, hence both tend to show low cognitive abilities.

This might suggest, as in \cite{LoHaTu09}, an interaction between genetic and environmental factors.
Genes multiply environmental inputs that support intellectual growth such that an increased socioeconomic
status raises the average cognitive ability but also magnifies individual differences in cognitive ability (see \cite{BaLeWe13}).

\section{Conclusion}
\label{Conc}
In this paper we proposed a Bayesian nonparametric conditional copula approach to model the strength and type of dependence between two variables of interest and we applied the methodology to the National Merit Twin Study.
In order to capture the dependence structure between two variables, we introduced two different calibration functions expressing the functional form of a covariate variable. The statistical inference was obtained implementing a slice sampling algorithm, assuming an infinite mixture model for the copula. The methodology combines the advantages of the conditional copula approach with the modeling flexibility of Bayesian nonparametrics.

The simulation studies illustrated the excellent performance of our model with three distinct copula families and different sample sizes.
The application to the twins data revealed the importance of the environment in the development of twins' cognitive abilities and suggests that environmental factors are more influential in families with higher socioeconomic position. On the contrary, other factors, such as genetic causes, may be more dominant in families with lower socioeconomic position.

Although this paper focusses on bivariate copula models, the methodology can be extended to multivariate copulas including more than one covariate. However, the inclusion of multiple covariates needs special attention regarding the choice of variables prior to estimate the calibration functions. Moreover, the increasing computational cost due to the additional covariates should be taken carefully into consideration.



\section*{Acknowledgements}
The authors are thankful to the Associate Editor and the anonymous reviewers for their useful comments which significantly improved the quality of the paper. Fabrizio Leisen was supported by the European Community's Seventh Framework Programme [FP7/2007-2013] under grant agreement no: 630677.
\vspace*{-8pt}


%


%


\bibliographystyle{biom}
\bibliography{BNPCond}

\appendix


\section{Gibbs sampling details}
\label{AppA}
Let $\mathcal{D}_{j}=\{i=1,\dots,n : d_{i}=j\}$ be the set of indexes of the observations allocated to the $j$-th component of the mixture, while $\mathcal{D}=\{j: \mathcal{D}_{j}\ne \emptyset \}$ is the set of indexes of non-empty mixtures components. Let $D^{*}=\sup{\{\mathcal{D}\}}$ be the number of stick-breaking components used in the mixture. As in \cite{walker2011}, the sampling of infinite elements of $\bm{\pi}$ and $\bm{\beta}$ is not necessary, since only the elements of the full conditional probability density functions of $D$ are need.

The maximum number of stick-breaking components to be sampled is:
$$ N^{*}=\max{\{i=1,\dots,n | N_{i}^{*}\}},$$
where $N_{i}^{*}$ is the smallest integer such that $\sum_{j=1}^{N_{i}^{*}} w_{j}> 1- z_{i}$.

\subsection{Update of  $\pi$}\label{V}
We update the stick-breaking components and consequently the weights $w_j$ based on the equation $w_j=\pi_j \prod_{k<j} (1-\pi_k)$. Assuming that $\pi_j$ is distributed as a Beta ($\mathcal{B}e(1,\lambda)$), the full conditional distribution of $\pi_j$ is:
\begin{equation}
\pi_j|\dots \sim \mathcal{B}e(1+ \#\{d_i=j\}, \lambda + \#\{d_i>j\}), \label{UpV}
\end{equation}
where $\# \{d_i=j\}$ are the number of $d_i$ equal to $j$ and $\#\{d_i>j\}$ is the number of $d_i$ greater than $j$ for $j< D^{*}$.

On the other hand, if $j=D^{*}+1,\dots,N^{*}$ we have that
$$\pi_j|\dots \sim \mathcal{B}e(1,\lambda).$$

\subsection{Update of $Z$} \label{Z}
From the full likelihood function (\ref{Lik}), $z_i$ follows a uniform distribution
\begin{equation}
z_i|\dots \sim \mathcal{U}(0,w_{d_i}) \label{UpZ}
\end{equation}
and it is sampled accordingly.

\subsection{Update of $D$} \label{Di}
The allocation variable $d_i$ values lie between $0$ and $N_i$ and the density of $d_i$ satisfies
\begin{equation}
P(d_i=j|\dots) \propto \, \mathbb{I}(z_{i}<w_{d_{i}}) c_{\rho(x_i|\bm{\beta}_{d_i})}(u_i,v_i). \label{UpD}
\end{equation}

\subsection{Update of $\beta$} \label{beta}
The full conditional of the vector of parameters $\bm{\beta}_k$, for $k\ge 1$ is:
\begin{equation}
f(\bm{\beta}_k|\dots) \propto \, \pi(\bm{\beta}_k) \prod_{d_i=k} c_{\rho(x_i|\bm{\beta}_k)}(u_i,v_i),
\label{UpB}
\end{equation}
where $\pi(\bm{\beta}_k)$ is the prior on $\bm{\beta}$.
Since the (\ref{UpB}) is not a standard distribution, we used a Random Walk Metropolis Hastings.

\clearpage

\end{document}